\documentclass[11pt,a4paper]{article}
\usepackage{jheppub}
\usepackage{epsf}
\usepackage{bm}

\def\red{\color{red}}
\def\Mrowczynski{Mr\'owczy\'nski}
\def\putbox#1#2{\epsfxsize=#1\textwidth \epsfbox{#2}}
\def\centerbox#1#2{\centerline{\epsfxsize=#1\textwidth \epsfbox{#2}}}
\def\half{\frac{1}{2}}
\def\Eq#1{eq.~(\ref{#1})}

\def\F{{\bm F}}

\def\B{{\bm B}}
\def\k{{\bm k}}
\def\p{{\bm p}}
\def\q{{\bm q}}
\def\v{{\bm v}}
\def\be{\begin{equation}}
\def\ee{\end{equation}}
\def\bea{\begin{eqnarray}}
\def\eea{\end{eqnarray}}

\def\pmax{p_{\rm max}}
\def\nhard{n_{\rm hard}}
\def\qhat{{\hat{q}}}
\def\qhatinst{\qhat_{\rm inst}}
\def\qhatelast{\qhat_{\rm elastic}}
\def\qhatnew{\qhatinsts}
\def\qhatTelast{\qhatelasts}
\def\qhatTinst{\qhatinsts}
\def\qhatelasts{\qhat_{s,{\rm elastic}}}
\def\qhatinsts{\qhat_{s,{\rm inst}}}
\def\Gsplit{\Gamma_{\rm split}}
\def\Gammael{\Gamma_{\rm el}}
\def\Gsplitjoin{\Gamma_{\mbox{\scriptsize{split/join}}}}
\def\tbroaden{t_{\rm broaden}}

\def\kinst{k_{\rm inst}}
\def\kiso{k_{\rm iso}}
\def\ksplit{k_{\rm split}}
\def\kisosplit{k_{\rm isosplit}}
\def\tcoh{t_{\rm coh}}
\def\tlargeang{t_{\rm large\:angle}}
\def\teq{t_{\rm eq}}
\def\tsplit{t_{\rm split}}

\def\tform{t_{\rm form}}
\def\tiso{t_{\rm iso}}
\def\tisofill{t_{\rm iso}(\pmax)}
\def\tisosplit{t_{\rm isosplit}}
\def\tnewinst{t_{s,\rm inst}}
\def\tTinst{t_{s,{\rm inst}}}
\def\tTelast{t_{s,{\rm elast}}}
\def\tchange{t_{\rm change}}
\def\Tfinal{T_{\rm final}}
\def\OO{{\cal O}}
\def\gsim{\mbox{~{\raisebox{0.4ex}{$>$}}\hspace{-1.1em}
	{\raisebox{-0.6ex}{$\sim$}}~}}
\def\lsim{\mbox{~{\raisebox{0.4ex}{$<$}}\hspace{-1.1em}
	{\raisebox{-0.6ex}{$\sim$}}~}}

\title{Thermalization in Weakly Coupled Nonabelian Plasmas}

\author{Aleksi Kurkela and Guy D.\ Moore}
\affiliation
    {%
    Department of Physics,
    McGill University, 3600 rue University,
    Montr\'eal QC H3A 2T8, Canada
    }%

\emailAdd{guymoore@physics.mcgill.ca}
\emailAdd{kurkela@physics.mcgill.ca}

\date{July 2011}

\abstract{%
      We investigate how relativistic, nonabelian plasmas
      approach equilibrium in a general context.  Our treatment is
      entirely parametric and for small Yang-Mills coupling $\alpha$.
      First we study isotropic systems with an initially
      nonequilibrium momentum distribution.  We consider both the case
      of initially very high occupancy and initially very low
      occupancy.  Then we consider systems which are anisotropic.  We
      consider both weak anisotropy and large anisotropy, and allow the
      occupancy to be parametrically large or small.
      Writing the typical momentum of an initial excitation as $Q$
      and the final temperature as $\Tfinal$, full equilibration occurs
      in a time $\teq \sim \alpha^{-2} \Tfinal^{-1}$ for $\Tfinal > Q$,
      and $\teq \sim \alpha^{-2} Q^{\half} \Tfinal^{\frac{-3}{2}}$
      for $\Tfinal < Q$, unless the initial system is sufficiently
      anisotropic and $\Tfinal > \alpha^{\frac{2}{3}} Q$, in
      which case equilibration occurs somewhat faster,
      $\teq \sim {\rm max}( \alpha^{-2} T^{-1} \,, \,
      \alpha^{\frac{-13}{7}} Q^{\frac 57} \Tfinal^{\frac{-12}{7}} )$.
    }%


\begin{document}

\maketitle

\section{Introduction and summary of results}

Nonabelian plasmas out of equilibrium are a rather generic feature of
early universe cosmology.  For instance, they can arise in the process
of reheating after inflation, whether by perturbative decay of inflatons
\cite{inflaton_decay}, by resonant decay of an inflaton into Standard
Model fields \cite{inflaton_resonant}, or due to other late-decaying
relics.  In some cases the nonabelian plasmas created in these processes
can be locally very anisotropic; for instance, Asaka {\it et al}
\cite{Grigoriev} show that late decays of inflatons can produce
jet-shaped regions of heated plasma which may be important for
electroweak baryogenesis; the interiors of these jets are generically
highly anisotropic.  Nonequilibrium and anisotropic plasmas can also
arise in cosmological phase transitions, such as the electroweak phase
transition \cite{baryo_review}.  In all of these cases, due to
asymptotic freedom in QCD and the weakly coupled nature of weak isospin,
the relevant nonabelian coupling is relatively small.

The early stages of highly relativistic heavy ion collisions also
probably produce a nonabelian (QCD) plasma out of equilibrium.  Probably
in real-world applications the coupling is not small and a perturbative
approach is suspect; but at least in the theoretically clean limit of
extremely high energy collisions the coupling is also weak in this
context.

In any case we think it is a well motivated {\sl theoretical} question
to ask, how in general does a nonabelian plasma approach equilibrium?
For the case of a plasma which is {\sl relatively close} to equilibrium,
we believe that this is well understood.  The dominant processes in this
case are elastic scattering and inelastic, number-changing (splitting
and joining) processes.  These are well described by an effective
kinetic theory \cite{AMY5}.  There is a natural energy scale $T$, and
the time for the system to approach equilibrium is parametrically
$\teq \sim \alpha^{-2} T^{-1}$ up to logarithmic factors which we will
systematically ignore.

However, it is less clear what physics is
most important for a system which starts out very far from equilibrium.
This is particularly the case if the system is also anisotropic.  In
this case, there can be plasma instabilities
\cite{Weibel,Mrow,RRS,ALM,RRS2,AMY7} which may dominate the dynamics
in some cases.  To our knowledge, there has been no comprehensive study
of this case.  The literature is also somewhat fragmentary for the case
of a system which is isotropic, but in which the modes which dominate
the system's energy have typical occupancies much larger or much smaller
than 1.

In this paper we will give a parametric treatment of each of these
cases; systems which are far from equilibrium but isotropic, and
systems which are far from isotropic.  By a parametric treatment we mean
that we will identify the most important physics in each case, as
determined by counting powers of the gauge coupling $\alpha$, and
estimate relevant momentum and time scales in the process of
equilibration as powers of $\alpha$.  We will not attempt to keep
track of factors of the logarithm of the coupling.  Nor will we attempt
to estimate any order-1 coefficients, or determine the range of $\alpha$
values over which our parametric estimates are reliable.
If a more quantitative estimate is
needed in any particular situation, then we have at least identified
what the relevant physics is in making such an estimate.

We will {\sl not} present estimates of the equilibration of systems
which are initially {\sl spatially} nonuniform (inhomogeneous).  This
presents too general a class of situations to allow any comprehensive
study.  However there is a general strategy for using our results to
study these systems as well.  Namely, one can consider the way in which
spatial nonuniformity leads to local momentum-space nonuniformity
through propagation, and then use our results for the evolution of
momentum-space anisotropic systems to determine the local physics which
this causes.  We will use the results of this paper to study one
particularly interesting case, that of a system under 1-dimensional
``Bjorken'' expansion, in a future publication.%
\footnote{We cannot help giving away one central result of our future
  publication.  Assuming the energy density scales as
  $\varepsilon \sim \alpha^{-1} Q^3 \tau^{-1}$ before equilibration,
  and using the estimate for the final equilibration time given in the
  abstract, one finds $\tau_{\rm eq} \sim \alpha^{-\frac{5}{2}} Q^{-1}$,
  slightly different than the conclusions of Baier
  et.~al.~\cite{bottomup}.}

In the main body of the text we will go through each case in detail,
introducing the physics relevant to each situation as the need arises.
However, for the convenience of the reader we will begin with a summary
of our results.

First we consider isotropic plasmas.  For simplicity we only consider
plasmas where all physics is initially dominated by excitations which
have a single characteristic
momentum scale $Q$, with a typical occupancy $f(Q)$ which we write
parametrically as $f(Q) \sim \alpha^{-c}$.  Cases with $c>0$ represent
initially {\sl overoccupied} systems; those with $c<0$ represent
initially {\sl underoccupied} systems.

We find three cases.  For $c>1$ (extreme overoccupancy) the
Nielsen-Olesen instability \cite{Nielsen-Olesen} rapidly converts the
system into one with a larger $Q$ and $c=1$.

For $0<c<1$ (overoccupancy), the dominant processes are elastic
scattering and number-changing effective $2\rightarrow 1$ ``merging''
processes.  Excitations quickly arrange into an $f(p) \propto T_*/p$
distribution with a cutoff scale $\pmax$.  Initially $\pmax\sim Q$
and $T_* \sim \alpha^{-c} Q$, but both scales evolve with time,
see \Eq{pmaxTstar}:
\be
\pmax \sim \alpha^{\frac{2-2c}{7}} Q^{\frac 87} t^{\frac 17} \,, \qquad
T_* \sim \alpha^{\frac{-6-c}{7}} Q^{\frac 47} t^{-\frac{3}{7}} \,,
\qquad
\left(\; t > \alpha^{-2+2c} Q^{-1} \;\right) \,.
\ee
Equilibration occurs when
$\pmax\sim T_* \sim \alpha^{-c/4} Q$,
at $\teq \sim \alpha^{-2+\frac{c}{4}} Q^{-1} \sim \alpha^{-2} T^{-1}$.
The final temperature is determined by the initial energy density;
and $\teq$ is the characteristic equilibration time for a bath of
temperature $T$.

For $c<0$ (underoccupancy), the physics is a little more involved
(though a basically correct exposition was given in
Ref.~\cite{bottomup}, in a slightly different context).  The most important
physics is the formation of a bath of low-momentum ``daughter''
particles, which thermalize and begin to dominate the system's dynamics
at a time scale $t \sim \alpha^{-2+\frac{c}{3}} Q^{-1}$.  The soft
thermal bath then catalyzes the breakup of the hard excitations,
absorbing their energy into the thermal bath.  Both processes are
dominated by number-changing effective $1\rightarrow 2$ processes, and
it is essential to include the modification of the number-changing rate
due to the Landau-Pomeranchuk-Migdal (LPM) effect
\cite{LPM,BDMPS}.  The temperature of the
thermal bath rises with time as $T \sim \alpha^{4-c} Q^3 t^2$.  The hard
excitations are consumed and thermalization completes when
$T= \Tfinal \sim \alpha^{-\frac{c}{4}} Q$ at
$\teq \sim \alpha^{-2+\frac{3c}{8}} Q^{-1}$.  The temperature is again
dictated by the initial energy density; the equilibration time is
$\teq \sim \alpha^{c/8} \alpha^{-2} \Tfinal^{-1}$, longer by $\alpha^{c/8}$
than the characteristic equilibration time of a thermal bath at
temperature $\Tfinal$.
But $\teq$ is much {\sl shorter} than the large-angle
elastic scattering time scale for the initial distribution of
excitations.  Therefore, large angle change through elastic scattering
between the initially present excitations actually plays {\sl no} role
in the system's thermalization.

There is a simple way to understand the equilibration time for an
underoccupied system.
Because of LPM modified bremsstrahlung emission, a hard $p\gg T$
excitation in a thermal bath loses energy at a rate
$dE/dt \sim \alpha^2 T^2 \sqrt{E/T}$ \cite{BDMPS}.  Therefore $\teq$
is the characteristic time scale for an excitation of energy $Q$,
traversing a thermal bath of temperature $\Tfinal$, to
radiate away its energy.  Equilibration could not proceed faster than
this.

Having treated isotropic systems, we then turn to anisotropic ones.
Again we assume that the system starts with excitations possessing a
single characteristic momentum scale $Q$.  But the distribution of
particles is now characterized by two quantities; a typical occupancy,
$f(p) \sim \alpha^{-c}$, and a measure of anisotropy characterized by a
strength $\alpha^{-d}$.  For weakly anisotropic systems with $d<0$, we
define $\epsilon\equiv \alpha^{-d}$ as the relative variation of the
occupancy with direction
(so if $\epsilon=0.1$
then the occupancy in some directions is $10\%$ larger than in other
directions).  For strongly anisotropic systems with $d>0$, we define
$\delta \equiv \alpha^{-d}$, as the angular range within which most of
the excitations reside.  We concentrate on the case of an oblate
distribution, with most excitations' momenta $\p$ lying within
an angle $\delta$ of the $xy$ plane.  In this case, if $\delta =0.1$
then the occupancy is $f(\p)\sim \alpha^{-c}$ if $|p_z| < 0.1 |\p|$, but
$f(\p)$ is small outside this range.  We also consider prolate
distributions with most excitations' momenta within an angle $\delta$ of
the $z$ axis.

\begin{table}
\centerline{\begin{tabular}{|c|c|l|c|} \hline
Region  &  Boundaries  &  timescale for   &  New physics  \\
        &              &  new physics     &  which occurs \\ \hline
  1     &  $c>0$, $3d-c+1>0$, $7d-c+1<0$  &
        $\alpha^{\frac{-d-7+7c}{4}} Q^{-1}$ & joining reduces anisotropy
        \\ \hline
  2     & $c<0$, $3d+c+1>0$, $5d+c+1<0$  &
        $\alpha^{\frac{-7d-7-3c}{4}} Q^{-1}$ & new plasma instabilities
        \\ \hline
  3a    & $c<0$, $5d+c+1>0$, $d<0$, $7d+3c+1<0$ &
        $\alpha^{\frac{-11d-5-3c}{2}} Q^{-1}$ & new plasma instabilities
        \\ \hline
  3b    & $d>0$,  $8d+3c+3>0$, $9d+3c+1<0$ &
        $\alpha^{\frac{-35d-15-9c}{6}} Q^{-1}$ & new plasma instabilities
        \\ \hline
  4a    & $7d-c+1>0$, $7d+3c+1>0$, $d<0$  &
        $\alpha^{\frac{3d-3+3c}{2}} Q^{-1}$ & angle randomization
        \\ \hline
  4b    & $d>0$, $23d+9c+3>0$, $3d+c-1<0$ &
        $\alpha^{\frac{5d-3+3c}{2}} Q^{-1}$ & angle broadening
        \\ \hline
  5     & $3d+c+1>0$, $8d+3c+3<0$ &
        $\alpha^{\frac{-d-1+c}{2}} Q^{-1}$ & new plasma instabilities?
        \\ \hline
  6     & $9d+3c+1>0$, $23d+9c+3<0$, $9d+3c-1<0$ &
        $\alpha^{\frac{-4d-6}{3}} Q^{-1}$ & new, weak instabilities
        \\ \hline
  7     & $3d+c-1>0$, $d-c+1>0$ & $\alpha^{\frac{-d-1+c}{2}} Q^{-1}$
        & angle broadening \\ \hline
  8     & $c>1$, $d-c+1<0$ & less than $Q^{-1}$
        & Nielsen-Olesen instabilities \\ \hline
  9     & $3d+c+1<0$, $c<0$ or $3d-c+1<0$, $0<c<1$
        & as per isotropic & as per isotropic \\ \hline
  10    & $9d+3c-1>0$, $37d+15c+7<0$ &
        $\alpha^{\frac{-7d-25+3c}{12}} Q^{-1}$ & thermal bath elastic scatt
        \\ \hline
\end{tabular}}
\caption[Main results table]
{\label{summary_table}
 Main regions in the occupancy($c$)--anisotropy($d$) plane,
 indicating the time scale on which the physics changes and the nature
 of the new physics.}
\end{table}

\begin{figure}
\centerbox{0.9}{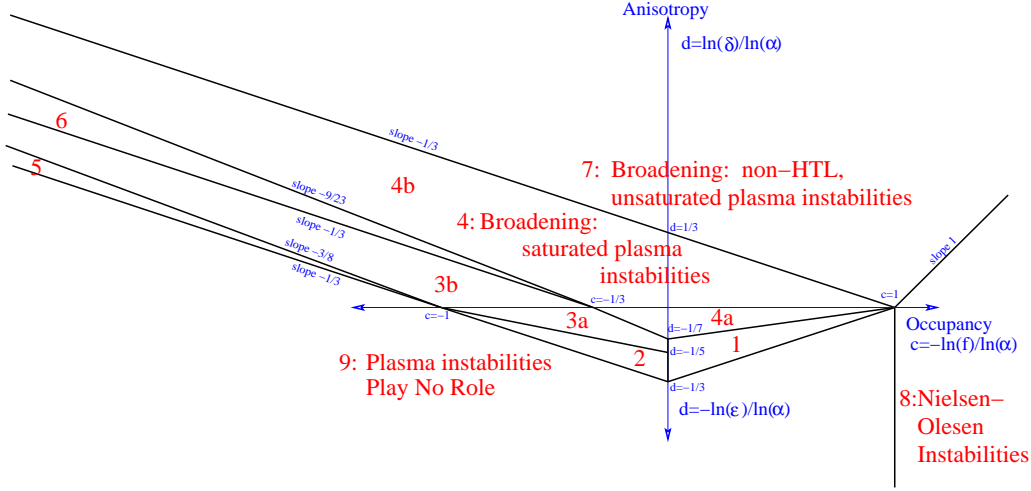}
\caption{\label{summary_fig}
  Main regions in the occupancy-anisotropy plane.  Descriptions of
  each region are in the text, and in Fig.~\ref{fig:cartoon}.}
\end{figure}

Our findings for oblate anisotropic plasmas are summarized in figure
\ref{summary_fig} and table \ref{summary_table}.  Depending on the
values of $c$ and $d$ (occupancy and anisotropy), the system's evolution
can take many different forms.  In almost all regions plasma
instabilities play a key role in the dynamics; but after some time scale
new physics comes to dominate and the behavior of the system changes.
We have only tried to track the evolution until the first significant
change in the dominant physics.  In summary, the regions are:
\begin{enumerate}
\item Plasma instabilities induce joining processes which
  suppress the anisotropy.
\item Plasma instabilities induce splitting processes, split daughters
  are anisotropic and generate new plasma instabilities which come to
  dominate the dynamics.  The most important daughters have occupancies
  limited by re-absorption.
\item Same as 2.\ except re-absorption is irrelevant; the most important
  momentum scale is set by the angle randomization of split daughters.
\item Plasma instabilities randomize the directions of hard excitations.
\item Radiated daughters induce new plasma instabilities before the
  original plasma instabilities finish growing to their ``saturation''
  size.  This region is not fully understood.
\item Radiated daughters form a nearly-thermal bath, but residual
  anisotropy of the bath generates new plasma instabilities.
\item Plasma instabilities broaden the momentum distribution before
  the unstable modes' growth can saturate.  The hard-loop approximation
  does not apply in this region.
\item Nielsen-Olesen instabilities rapidly lower the maximum occupancy.
\item Plasma instabilities play no important role.  Equilibration
  proceeds as in an isotropic system.
\item Plasma instabilities cause splitting which generates a soft
  thermal bath.  The bath comes to dominate the dynamics
  through elastic scattering.  (This region lies off the edge of
  Fig.~\ref{summary_fig}.)
\end{enumerate}
Note that in regions 1, 4, 7, and 8 the processes which redistribute
particle momentum ($\qhat$) become less efficient with time; whereas on
the contrary, in regions 2, 3, 5, 6, and 10 the new physics which
emerges actually makes the redistribution of particle momentum
{\sl more} efficient with time.

Though we have not followed the evolution of anisotropic systems through
to final thermalization, we have an estimate for the final
thermalization time.  First one determines
$\Tfinal \sim \varepsilon^{\frac{1}{4}}$; for $d>0$ it is
$\Tfinal \sim \alpha^{\frac{d-c}{4}} Q$, for $d<0$ it is
$\Tfinal \sim \alpha^{\frac{-c}{4}} Q$.
If $\Tfinal \geq Q$ then $\teq \sim \alpha^{-2} \Tfinal^{-1}$.
If $\Tfinal\ll Q$ then
$\teq \sim \alpha^{-\frac{13}{7}} Q^{\frac 57} \Tfinal^{\frac{-12}{7}}$
if $\Tfinal > \alpha^{\frac{2}{3}} Q$ and
$\teq \sim \alpha^{-2} Q^{\half} \Tfinal^{\frac{-3}{2}}$
if $\Tfinal < \alpha^{\frac 23} Q$.  Both estimates are the time it
takes for an excitation of momentum $Q$ to lose its energy while
traversing a thermal bath of temperature $\sim \Tfinal$; the difference
is that for $\Tfinal > \alpha^{\frac{2}{3}} Q$, the thermal bath
is somewhat anisotropic and plasma instabilities, not elastic
scattering, determine how the $p\sim Q$ momentum excitation loses
energy.  If the resulting $\teq$ is shorter than
$\alpha^{-2} \Tfinal^{-1}$, then the hard $p\sim Q$ excitations break up
before the remaining thermal bath can equilibrate with itself.  But it
still takes time $\sim \alpha^{-2} \Tfinal^{-1}$ for the bath to fully
self-equilibrate, so this is remains a lower bound on the total
thermalization time.

Having summarized our findings, we will now present the details leading
to these results.  Since we presented the main conclusions here,
we will {\sl not} end with a discussion or conclusions section.

\section{Isotropic distributions}
\label{sec:iso}

In this section we will consider systems which are both
homogeneous and isotropic.  Nevertheless they can be very far from
equilibrium.  Such systems might arise rather generically during
preheating after inflation.  We will also re-encounter many of the
physical processes relevant to this case when we consider systems which are
not isotropic.  We begin by reviewing the relevant physical processes,
then we will consider their application to nonequilibrium systems.

\subsection{Elastic and inelastic scattering}
\label{sec:elastic}

We begin by reviewing the basic physical processes relevant for
equilibration in a plasma.
All of the physics discussed here is well known, and is presented in
more complete detail in Ref.~\cite{AMY5}.  At weak coupling and at
sufficiently long length and time scales, the dynamics of the
plasma can be described by that of long-lived weakly interacting quasiparticle
excitations. The perpetual interaction with the medium affects
the quasiparticle's dispersion relation, and to leading order, the
correction looks like an effective mass
\be
E(\p)\sim \sqrt{\p^2+m^2},
\ee
which introduces a new scale, the \emph{screening scale} $m$, which is parametrically
\be
m^2 \sim \alpha \int_ \p \frac{f(\p)}{\p},
\ee
where $\int_\p$ is a shorthand for $\int d^3 p$.

Under certain conditions, the non-equilibrium dynamics of the plasma can be described by the Boltzmann
equations that schematically read
\be
(\partial_t + \v(\p)\cdot \nabla_{\bm x})f({\bm x}, \p,t)= -C[f],
\label{eq:boltzmann}
\ee
where $f({\bm x}, \p,t)$ is the phase space density of quasiparticles in the
plasma, and $\v(\p)$ is the velocity of a quasiparticle with momentum $\p$.
In the following, we will restrict ourselves to spatially homogeneous systems so
that $f$ is only a function of $\p$ and $t$. In the presence of several particle species
(gluons, up- and down-quarks, \emph{etc.}) the distribution function is a multicomponent
vector with a component for each species.  In the absence of external forces, the
quasiparticles change their momentum states by mutual interactions;
$C[f]$ is a spatially local collision
term that represents the rate at which particles get scattered out of state $\p$
minus the rate they get scattered into this state.

The conditions under which the non-equilibrium dynamics of a
theory can be described by the Boltzmann equations of
\Eq{eq:boltzmann} are
\begin{itemize}
\item that the typical size of the wave packets
of quasiparticles is smaller than the mean free path of quasiparticles, and
\item the quantum mechanical formation time of scatterings
is small compared to the mean free time.
\end{itemize}
Generally these conditions will be met for those degrees of freedom in a
plasma with typical occupancy $f(p) \ll \alpha^{-1}$.  An exception is
that certain inelastic processes can have formation times which exceed
the mean free time between scatterings for the excitations involved.
This case requires special treatment (the LPM effect), which we will
return to at length below.

\begin{figure}
\centerbox{0.6}{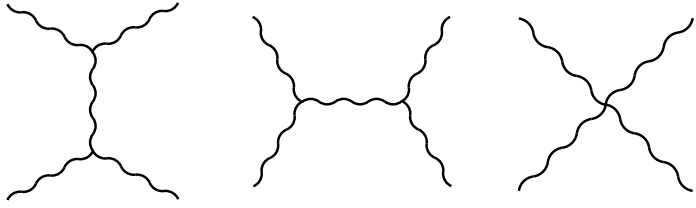}
\caption[Elastic scattering]
{ \label{elastic} Diagrams relevant in elastic scattering of gluons.}
\end{figure}

There are two qualitatively different processes that contribute in leading order
to the collision term, elastic scattering and near-collinear splitting processes.
The collision term for elastic scattering reads
\bea
C^{2\leftrightarrow 2}[f] =
	&&\frac{1}{2}\int_{\k,\p',\k'}
    |M(p,k,p',k')|^2(2 \pi)^4 \delta^{(4)}(P+K-P'-K')\\
&&   \times \big\{ f(\p)f(\k)[1{\pm} f(\p')][1{\pm} f(\k')]
     - f(\p')f(\k')[1{\pm} f(\p)][1{\pm} f(\k)] \big\}.
\eea
Here $P,P',K$, and $K'$ denote on-shell four-vectors. The
$[1{\pm} f]$-factors arise from final state Bose stimulation or Pauli blocking.
$M$ is the elastic scattering
amplitude in non-relativistic normalization, related to the usual relativistically normalized
 matrix element $\mathcal{M}$
by
\be
|M(p,k,p',k')|^2 \sim  \frac{|\mathcal{M}(p,k,p',k')|^2}{(2p_0)(2k_0)(2p'_0)(2k'_0)}.
\ee
In the leading order for gauge bosons, the $\mathcal{M}$ originates from
diagrams in Fig.~\ref{elastic} (from now on we will only consider gauge
bosons, which due to the stimulated interaction rates dominate the
equilibration).%
\footnote{%
    If the typical initial occupancy is large, we must be speaking of
    gauge bosons.  If the initial occupancy is small and the system is
    composed primarily of fermions, then the radiated ``daughter''
    particles are primarily gauge bosons.  The rate of radiation differs
    only by an order-1 group Casimir factor and all parametric estimates
    are identical to the case with primary gauge bosons.  All of these
    remarks also apply to the anisotropic case.%
} In vacuum the matrix element reads
\be
|\mathcal{M}|_{\rm vacuum}^2 \sim \alpha^2 \left( 3-\frac{su}{t^2}-\frac{st}{u^2}-\frac{t u}{s^2}\right),
\ee
where $s$, $t$, and $u$ are the usual Mandelstam variables.
The process is dominated by small momentum exchange, and in this limit
the matrix element becomes
\bea
|M|_{\rm vacuum}^2& \sim &\frac{\alpha^2}{(q_\perp^2)^2},
\label{eq:M^2}
\eea
where $q_\perp$ is the momentum transfer in the elastic scattering. In vacuum the
total cross-section $\sim \int d^2q_\perp |M|^2$ is infrared divergent, but this divergence is regulated by
including the medium dependent self-energy to the exchange line in the diagram of Fig.~\ref{elastic}.
Then the amplitude becomes \cite{Aurenche}%
\footnote{%
    The screening does not affect nearly static magnetic fields,
    which is why the matrix element scales as $1/q_\perp^2$ for very
    small $q_\perp$.  Because of this behavior the total cross-section
    actually remains logarithmically divergent, but the divergence is
    too weak to influence the rate of angle or particle number change.}
\be
|M|^2\sim\frac{\alpha^2}{q_\perp^2(q_\perp^2+m^2)}.
\ee

A particle traveling through the plasma undergoes successive uncorrelated
elastics scatterings that diffuse the momentum of the traveling particle. The
momentum of the particle evolves as a random walk in momentum space, characterized by
the momentum diffusion constant $\qhatelast$; the average squared
momentum transfer $\Delta p^2$ grows linearly in time,
\be
\Delta p^2 \sim \qhatelast t.
\ee
The rate of elastic scatterings is
\bea
\frac{d \Gammael}{d^2 q_\perp}\sim \int dq_z \int_{\p'} |M|^2
        f(\p')[1{+} f(\p'-\q)],
\eea
which for $q_\perp< |\p|$ is
\be
\frac{d \Gammael}{d^2 q_\perp}\sim \frac{\alpha^2}
                                         {q_\perp^2 (q_\perp^2+m^2)}
 \int_{\p'} f(\p')[1{+}f(\p')] \,,
\label{eq:diffrate}
\ee
and the mean squared momentum transfer per unit time is
\be
\qhatelast \sim \int d^2 q_\perp \frac{d \Gammael}
        {d^2 q_\perp}  q_\perp^2 \, .
\label{eq:qhat}
\ee
Collisions with small $q_\perp\sim m$ are the most frequent but change
momentum of the propagating particle the least, while scatterings with
large $q_\perp$ are rare but
produce larger angle deflections. Inserting \Eq{eq:diffrate} into \Eq{eq:qhat},
one sees that all available scales of momentum transfer contribute equally to the
momentum diffusion, such that $\qhatelast$ reads\footnote{That all the available momentum transfer
scales contribute to the momentum diffusion gives rise to a logarithmic enhancement, which is however neglected in our power counting.}
\be
\qhatelast \sim \alpha^2 \int_{\p} f(\p)[1{+}f(\p)].
\label{eq:qhat2}
\ee

\begin{figure}
\centerbox{0.75}{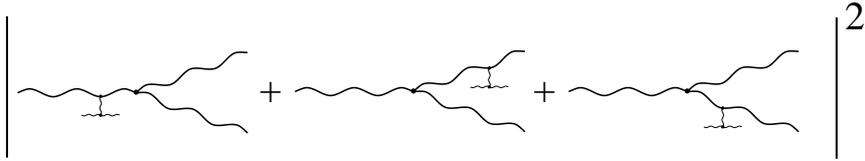}
\caption[inelastic scatterings]
{\label{fig:inelastic}
 Diagrams responsible for inelastic scattering processes in the plasma.
 The thick lines represent an initial particle and its two nearly
 collinear daughters; the thin lines show a scattering process which
 introduces a small amount of momentum, rendering the splitting
 kinematically allowed.}
\end{figure}

Due to the peculiarities of soft and collinear enhancements in QCD,
there is another class of processes which are also leading-order.
These are the inelastic, number changing, collisions \cite{AMY4,AMY5}.
In the vacuum, $1\leftrightarrow 2$ processes are kinematically
disallowed for massless particles, but in the presence of the medium, soft elastic scatterings
can take the particles slightly $\sim m$ off-shell allowing for subsequent
``$1\leftrightarrow 2$'' nearly collinear splitting processes, depicted in Fig.~\ref{fig:inelastic}.
A particle propagating through a plasma experiences soft elastic scatterings with
rate
\be
\label{eq:Gamma}
\Gammael\sim \frac{\alpha^2}{m^2} \int_{\p'}  f(\p') [1{+}f(\p')]
\sim \frac{\qhatelast}{m^2} \, .
\ee
In vacuum, each individual uncorrelated scattering event induces a
collinear radiation with probability $\alpha$ per logarithmic range in
angle, per logarithmic range in energy.  The same applies in-medium
(modulo Bose stimulation factors)
provided that the formation time of the emission process is shorter than
the mean time scale between scattering events.  When the formation time
is longer than the time between scatterings, there is destructive
interference between emission processes, the LPM effect, which we will
discuss in due course.
Assuming the formation time is shorter, and neglecting the log over
emission angles, the particle splitting rate is parametrically
\be
\label{splitrate}
\frac{ d \Gsplit^{\rm BH}}{ d p/p}\sim \alpha \Gammael,
\ee
where $p$ is the energy of the emitted particle.
(The superscript BH stands for Bethe-Heitler, since the regime where
coherence effects are negligible corresponds to the original calculation
of Bremsstrahlung radiation in QED by Bethe and Heitler \cite{Bethe}.)
Including the statistical factors, the effective
collision term for ``$1\leftrightarrow 2$'' processes reads
\bea
C^{\mbox{\scriptsize{``$1\leftrightarrow2$''}}}[f]\sim
 \int d k \frac{ d \Gsplit^{\rm BH}}{ d k}
       \big(
	f(p)[1{+}f(p{-}k)][1{+}f(k)]
	 -f(p{-}k) f(k)[1 {+} f(p)]
	 \big),
\label{eq:inelastic}
\eea
where energy and momentum conservation are implied.

\subsection{$0<c<1$: Cascade by elastic scattering}

\label{high_occ}

We start by considering isotropic and homogeneous systems
whose initial distribution at $t=0$ has occupancies
$\sim \alpha^{-c}$ up to a cutoff scale $Q$
\be
f(p<Q)\sim \alpha^{-c}, \quad \quad f(p>Q)<(Q/p)^4 \label{eq:init}
\ee
so that the high momentum particles ($p>Q$) do not dominate energy density
or anything else and can be neglected in the discussion. For $c>0$,
the soft modes are overpopulated compared to the thermal ensemble with
the same energy density. We expect thermalization to proceed by energy
cascading to higher $p$ modes with lower occupancy, until the typical
occupancy reaches 1 and quantum mechanics cuts off further cascading.
We will see that if $c<1$, both the elastic scattering and inverse
collinear splitting are of comparable effectiveness in transferring
energy to higher momenta. For $c>1$, the occupancies
are so large, that even in the weak coupling limit, the thermalization
is driven by non-perturbative physics. This will be discussed separately in the
next subsection.

The energy density of the system is
\be
\label{energy1}
\varepsilon \sim \int d^3p \,p \, f(p)\sim \alpha^{-c} Q^4,
\ee
and the final temperature is
$\Tfinal\sim \varepsilon^{\frac 14} \sim \alpha^{-c/4}Q$, so the
equilibration cannot occur faster than in time
$\teq \sim \alpha^{-2} \Tfinal^{-1} \sim \alpha^{-2+c/4} Q^{-1}$, which
is the equilibration time for a small deviation from thermal equilibrium in
a system at temperature $\Tfinal$.
 On the other hand systems with high occupancies have Bose-stimulated
interaction rates so we also expect it should not take longer than
this time scale for the system to thermalize.

Let us now estimate the time scale it takes for a particle with $p\sim Q$ to
appreciably change its angle or momentum. For the initial conditions of \Eq{eq:init},
the momentum diffusion due to elastic scattering is dominated by modes with $p\sim Q$
and is parametrically of order
\be
\qhatelast\sim \alpha^2 \int_\p f(\p)[1{+}f(\p)] \sim \alpha^{2-2c}Q^3,
\ee
and the time for large-angle change is
\be
t_{\textrm{large angle}}\sim Q^2/\qhat \sim \alpha^{-2+2c}Q^{-1}.
\ee
On the other hand, for particles with momentum $p \ll Q$ the time scale is
shorter by a factor of $(p/Q)^2$.  Such ``soft'' particles of low
momentum will equilibrate quickly into an infrared tail with
occupancy $f(p\ll Q) \propto p^{-1}$ -- the functional form which
is in quasi-equilibrium in the sense that the ``gain''
and ``loss'' terms in the Boltzmann equation cancel.  Such a soft
tail does not dominate either screening, elastic scattering, or
the energy density of the system, so it plays no role in directing
the dynamics of the typical excitations.  Such $f(p) \propto 1/p$ tails
will be a common feature in the sections which follow, so we introduce
some notation to describe them.  In our case, at and after the time
$t_{\textrm{large angle}}$  the particles have had
time to organize themselves into such a $1/p$ tail.  The distribution is
characterized by two scales; the maximum momentum $\pmax$ below which the
occupancy behaves as $\propto 1/p$, and an ``effective temperature''
determining the occupancy below $\pmax$:
\be
f(p)\sim \frac{T_*}{p} \Theta(\pmax-p) \, .
\label{eq:fproto}
\ee
Of course the Heaviside function is an oversimplification of the
transition from $\propto T_*/p$ behavior to rapid falloff; but for
parametric estimates it will be sufficient.
The parameter $T_*$ is solved from energy conservation
\be
\label{energy2}
\varepsilon \sim T_* \pmax^3 \,.
\ee
The subsequent equilibration is then controlled by the evolution of the
scale $\pmax$. Before thermalization $\pmax< \Tfinal < T_*$,
and when $\pmax$ increases $T_*$ decreases. When
$\pmax\sim \Tfinal\sim T_*$, the typical occupancies
at scale $\pmax$ reach 1, and all scales below $\Tfinal$ have thermalized.
After that, the cascade will continue (without Bose stimulation) to higher momentum
scales.  However,  in thermal ensemble the energy density, screening,
and elastic scattering are dominated by scale $\Tfinal$, so that even though
the ultraviolet tail of the distribution has not completely adjusted to the Maxwell-Boltzmann
form, the system can be considered thermalized.

The scale $\pmax$ can grow either by elastic scatterings
or inverse splitting processes.  The rate for large angle elastic collisions is
\bea
\qhatelast&\sim& \alpha^2 T_*^2 \pmax, \label{eq:qhat3}\\
\Gamma_{\textrm{large angle}}&\sim& \frac{\qhatelast}{\pmax^2}
\sim \alpha^2 \,T_* \left(\frac{T_*}{\pmax}\right),
\eea
and is enhanced compared to the equilibrium one $\sim \alpha^2 T$.

The rate for a hard joining process, in which two particles with
$p\sim \pmax$ merge, so that the momentum of the resulting particle
has changed by order 1, is given by the \Eq{eq:inelastic} with
$p\sim k \sim (p{-}k) \sim \pmax$ so that
\be
\partial_{t}f(\pmax) \sim \Gamma_{\textrm{hard merge}} f(\pmax)\sim \alpha \Gammael f(\pmax)[1{+}f(\pmax)]\\,
\ee
and inserting the elastic scattering rate,
\be
\Gamma_{\textrm{hard merge}}\sim \alpha^2\, T_*
 \left(\frac{T_*}{\pmax}\right).
\ee
The rates for large momentum transfer either from elastic scattering or hard joining
are the same (up to logarithms) so both the two processes compete, but all
parametric estimates made using either one will give the same results.


\begin{figure}
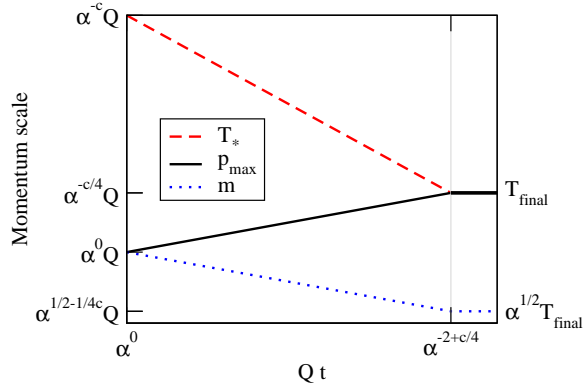

\centerbox{0.5}{high_occ.eps}
\caption{
Evolution of the momentum scales in a log-log plot during the cascade
with initial distribution of \Eq{eq:init} with $c=1$.
}\label{fig:high_occ}
\end{figure}


During the evolution of $\pmax$ the total particle number changes, and we have
assumed that the number changing processes can keep up with the change of $\pmax$ and
hence $T_*$. We expect that this is true because number-changing
$2\leftrightarrow 1$ splitting processes are parametrically as efficient
as elastic scattering.  But even if they are numerically less efficient,
particle number is still efficiently changed, because the particle
number changing rate is large in the infrared.  Specifically, the total
particle number changing rate is of order
\be
\Gamma^{\mbox{\scriptsize{``$1\leftrightarrow 2$''}}}
  \sim \int d k \frac{ d \Gsplit^{\rm BH}}{dk}
  [1{+}f(k)] \sim \alpha \Gammael\, f(m)\\,
\ee
which is greater than the hard merging or the large angle elastic
scattering rate by a power of
$(\pmax/m) \sim \sqrt{\pmax/\alpha T_*}$
\cite{ArnoldDoganMoore}.

The time evolution of $\pmax$ and $T_*$ can be determined by self-consistently solving
\be
\pmax^2 \sim \qhatelast t,
\ee
together with \Eq{eq:qhat3} and energy conservation, \Eq{energy1}
and \Eq{energy2}:
\be
\pmax \sim \alpha^{\frac{2-2c}{7}} Q^{\frac 87} t^{\frac 17} \,, \qquad
T_* \sim \alpha^{\frac{-6-c}{7}} Q^{\frac 47} t^{-\frac{3}{7}} \,.
\label{pmaxTstar}
\ee
These expressions hold for all times $t$ such that $\pmax > Q$,
that is, $t > \alpha^{\frac{2c-2}{7}} Q^{-1}$, until
equilibration is complete, which occurs when $\pmax \sim T_*$, at time
\be
\teq\sim \alpha^{-2+\frac{c}{4}} Q^{-1}
  \sim \alpha^{-2} \Tfinal \,.
\ee
These results are summarized in Fig.~\ref{fig:high_occ}.

\subsection{Nielsen-Olesen Instability}
\label{sec:NO}

In the last subsection we considered the case $f(p\sim Q) \sim
\alpha^{-c}$ with $0<c<1$.  Now we briefly comment on what happens in
the case $c>1$, that is, the case of extremely high initial occupancy.
The physics in this regime is nonperturbative;
large occupation numbers
of order $\gtrsim \alpha^{-1}$ can cancel the suppression
of interactions coming from the small coupling constant.
Specifically, the nonperturbative physics of the Nielsen-Olesen
instability \cite{Nielsen-Olesen} will dominate in this regime.

Very large occupancy fields can be viewed as effectively classical.  In
our case, we have classical color-magnetic fields of coherence length
$l_{\rm coh} \sim Q^{-1}$ and of field strength
$B^2 \sim \alpha^{-c} Q^4 \rightarrow B \sim \alpha^{-c/2} Q^2$.
The Lamor radius of a particle propagating in such a field is
$L\sim \alpha^{-1/4}B^{-1/2} \sim \alpha^{(c-1)/4} Q^{-1}$, which is
shorter than the coherence length of the field.  Therefore we can
neglect the spatial variation of the magnetic field.  In an intense and
uniform magnetic field, relativistic excitations split up into Landau
levels with energy spacing $\sim L^{-1}$.  But as pointed out by Nielsen
and Olesen \cite{Nielsen-Olesen}, the way the spin-magnetic interaction
splits the Landau levels means that, for spin-1 particles, one spin
state of the lowest Landau level
is actually exponentially unstable, with a growth time
$\gamma \sim L^{-1} \sim \alpha^{\frac{1-c}{4}} Q$.
In a time scale%
\footnote{%
    The timescale also depends on the size of the initial seed
    fluctuations in $p\sim L^{-1}$ modes.  But at minimum there are
    vacuum fluctuations, which behave as $f(p)\sim 1/2$.  So the time
    for the fluctuations to grow from $f\sim 1$ to $f\sim \alpha^{-1}$
    is $\sim L^{-1} \ln(\alpha^{-1})$, longer than $L^{-1}$ by a
    logarithm.
}
$t\sim L^{-1}$
these unstable modes grow to absorb the energy density of the
magnetic fields, transferring it to modes of typical wave number
$p \sim L^{-1} \sim \alpha^{\frac{1-c}{4}} Q \gg Q$.
So the instability transfers energy towards ultraviolet
and works towards thermalizing the distribution.  Using energy
conservation, $\alpha^{-c} Q^4 \sim f_{\rm new}(p) p^4$, we find
the final occupancy of these modes to be $f(p) \sim \alpha^{-1}$.
Therefore the final state has
\be
Q_{\rm new} \sim \alpha^{\frac{1-c}{4}} Q \,, \qquad
f_{\rm new} \sim \alpha^{-1} \,.
\ee
So the end product of the Nielsen-Olesen instability is a distribution
with $c=1$.  The equilibration of this ``final'' state will then
proceed as described in the previous subsection.

\subsection{Landau-Pomeranchuk-Migdal suppression}
\label{sec:LPM}

We will turn next to dilute systems, $c<0$ or $f(\p) \ll 1$.  In this
case the screening scale $m^2$ gets smaller, but surprisingly the
elastic scattering rate does
not; while there are fewer scattering ``targets,'' this is compensated by
the weaker screening, so the explicit factor of $f(\p)$ in \Eq{eq:Gamma}
is canceled by the implicit factor in $1/m^2$.  But the individual
elastic scatterings are by smaller and smaller angles.  If the
scattering is by a sufficiently small angle, any emitted radiation stays
``on top'' of the emitter for so long that it can still interfere with
an emission from the next scattering, which is called the
Landau-Pomeranchuk-Migdal  (LPM) effect \cite{LPM}, after the scientists
who understood the effect in the context of QED.  This effect was
first understood correctly in the context of QCD by Baier {\it et al}
(``BDMPS'') \cite{BDMPS}.
These interference effects will prove to play an important role in the
case of a dilute (low initial occupancy) system, so
we will review the physics of the LPM effect before we discuss
the low-occupancy case.

Consider the quantum mechanical formation time $\tform$ of a nearly collinear
splitting process, defined as the time it takes to separate the wave packet of the
emitted daughter particle from the emitting mother particle in the transverse direction.
Take $k$ to be the momentum of the daughter particle and $k_\perp$ its transverse
momentum with respect to the mother particle.
The transverse size of the wave packet is then $\Delta x_\perp \sim 1/k_\perp$,
and its transverse velocity $v_\perp \sim k_\perp/k$, so that the daughter's wave
packet overlaps with the mother particle until the time
\be
\label{tform1}
\tform\sim \frac{\Delta x_\perp}{v_\perp} \sim \frac{k}{k_\perp^2}.
\ee

The daughter particle acquires transverse momentum by random elastic scatterings
with the other particles in the medium.  If $\tform$ is longer
than the mean time between such scatterings, we can replace individual
scatterings with transverse momentum diffusion and estimate that
$k_\perp^2 \sim \qhatelast \tform$, giving the formation time%
\footnote{%
  In nonabelian theories, both the mother and daughter particles are subject
  to elastic scattering as the gauge bosons are charged. In an abelian
  theory the photon is not charged and the formation time depends on the
  momentum of the mother particle instead of that of the daughter. The
  analogous estimate in the abelian case is
  $\tform(p)\sim \sqrt{p^2/k \qhatelast}$,
  where $p,k$ are the momenta of the mother and daughter respectively.
}
\be
\label{tform2}
\tform(k)\sim \sqrt{\frac{k}{\qhatelast}}.
\ee

\begin{figure}
\centerbox{0.5}{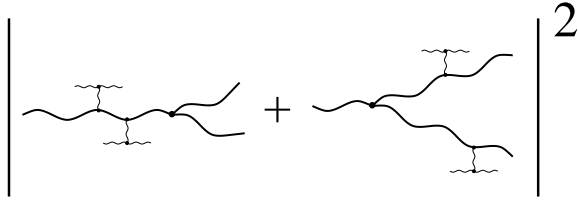}
\caption[LPM effect]
{\label{fig:LPM}
 Interference between diagrams in which an emission occurs either before
 or after multiple scattering events.  When they occur on the final
 states, the scattering events can involve either external leg.
 These interference effects are  responsible for the LPM effect.}
\end{figure}

If the formation time is longer than the time between elastic collisions $\Gammael^{-1}$, the
kinetic theory treatment of \Eq{eq:boltzmann} is no longer adequate as
the diagrams depicted in Fig.~\ref{fig:LPM} interfere coherently, an effect
not taken into account in the
kinetic treatment.  Physically, the emitting particle can no longer resolve individual kicks
originating from individual collisions, and effectively sees only the net deflection from all
collisions during the formation time; the particle's trajectory bends smoothly.
Then the rate for splitting processes reads\footnote{
	Note that the
	formation time of an inverse splitting process in the previous subsection
	(overoccupied case) is $\tform\sim  (\alpha T_*)^{-1}$. This is of the
	same order of magnitude as the time between elastic collisions
	 $\Gammael^{-1}\sim  (\alpha T_*)^{-1}$, so  the LPM suppression does not affect
	 the parametric estimates.
	}
\bea
\label{eq:LPM}
\Gsplit^{\rm LPM}(k) &\sim& \alpha \, t^{-1}_{\rm form}(k),\\
\Gsplit(k)&\sim& \min[\Gsplit^{\rm BH}(k),\Gsplit^{\rm LPM}(k)].
\eea

Equivalently, the spectrum of split particles changes from Bethe-Heitler $\log(k)$ to LPM type $k^{-1/2}$ spectrum
at the momentum scale $k_{\rm LPM}$ where the formation time is of the same order of magnitude
as the time between elastic scatterings:
\bea
\label{kLPM}
t^2_{\rm form}(k_{\rm LPM}) &\sim& \Gammael^{-2} \nonumber \\
\frac{k_{\rm LPM}}{\qhatelast} & \sim &
             \left( \frac{m^2}{\qhatelast} \right)^2 \nonumber \\
k_{\rm LPM}\sim \frac{m^4}{\qhatelast}.
\eea

\subsection{$c<0$: Low occupancy and ``Bottom-up'' thermalization}
\label{sec:bottomup}

Next we consider the equilibration of an isotropic homogeneous system, whose
initial distribution is given again by \Eq{eq:init} with typical
occupancies $\alpha^{-c}<1$ ($c<0$),
so that the initial distribution consists of underoccupied hard particles at a typical scale $Q>\Tfinal$.
The thermalization  proceeds in close analogy to the ``bottom-up'' scenario described in
 Ref.~\cite{bottomup}: First, a population of soft modes with momenta of order
$\sim m$ is generated by soft splitting, induced by small angle elastic
collisions between the hard particles. The soft sector subsequently
thermalizes in a time $Q t \sim \alpha^{-2-c/3}$. By this time,
the soft sector carries only a small portion of the total energy of the system,
but it draws energy from the hard modes by hard LPM suppressed splitting. The system
becomes fully thermalized when the hard sector has lost all its energy to the
soft sector at $Q t\sim \alpha^{-2}\alpha^{-3c/8}$.

\subsubsection{$Qt<  \alpha^{-2-c}$}
For the underoccupied initial condition, the screening scale, momentum diffusion
constant due to elastic scattering, and small angle elastic scattering rates read
\bea
\label{qhat_h}
\qhatelast&\sim& \alpha^2 n_h \sim \alpha^{2-c} Q^3 \,,\\
\label{m_h}
m_h^2&\sim& \alpha n_h/Q \sim \alpha^{1-c} Q^2 \,,\\
\label{Gamma_h}
\Gammael&\sim& \frac{\qhatelast}{m_h^2} \sim \alpha Q \,,
\eea
where $n_h\sim \alpha^{-c}Q^3$ is the number of initial hard particles.
The index $h$ indicates that the screening is due to the original hard modes. If
the equilibration would proceed via elastic scatterings as in the overoccupied
case, the estimate for the equilibration time would be dictated by the large angle
collision time for the hardest modes $\teq\sim Q^2/\qhatelast \sim \alpha^{-2+c}Q^{-1}$.
However, as we will see, inelastic scatterings provide a faster route to
the thermal distribution.

The hard particles radiate soft gluons creating a new population of particles at small
momentum. The production rate of gluons with soft momentum $k$ by
hard particles with momentum $p$ is Bose stimulated.  The production
rate of soft particles of momentum $k$, from \Eq{eq:inelastic}, is
\be
\Gamma_{\rm prod}(k) \sim
\int_p \Gsplit(k) f(p)[1{+}f_s(k)][1{+}f(p)]
 \sim \int_p \Gsplit(k)f(p)[1{+}f_s(k)]\,,
\ee
where $f_s$ is the distribution of the new soft modes.
But the stimulation factor $[1{+}f_s(k)]$ is
cancelled by the inverse process where the soft gluon merges onto a
hard particle
\be
\Gamma_{\rm absorb} \sim
\int_p \Gsplit(k) f_s(k) f(p)[1{+}f(p)]\sim
\int_p \Gsplit(k) f_s(k) f(p)\,,
\ee
so that the total production rate of soft particles becomes
\be
\label{prod_minus_absorb}
(\Gamma_{\rm prod}-\Gamma_{\rm{absorb}}) \sim
\int_p \Gsplit(k) f(p) \,.
\ee
The spectrum of the emitted particles is then
\be
\label{eq:spectrum}
f_s(p)\sim n_h \Gsplit(p) t/p^3,
\ee
where $p^{-3}$ arises from the momentum space element.
Using \Eq{splitrate} and \Eq{eq:LPM},
\bea
\label{spec_BH}
f_s(p)&\sim& \alpha \,n_h  \Gammael t /p^3,\quad \hspace{0.65cm}\textrm{for } p< k_{\rm LPM},\\
f_s(p)&\sim&\alpha \,n_h \sqrt{\qhatelast} t /p^{7/2},\quad \textrm{for }p > k_{\rm LPM},
\label{spec_LPM}
\eea
with (see \Eq{kLPM}, \Eq{qhat_h} and \Eq{m_h})
\be
k_{\rm LPM}\sim \alpha^{-c}Q.
\ee
As long as $c>-1$ so that $k_{\rm LPM} > m_h$,  the production of the softest modes is described by
the Bethe-Heitler rate.%
\footnote{%
  For extremely dilute initial conditions $c<-1$, the production of even the
  softest modes is LPM suppressed. In this case the first soft gluons
  emerge at the time $t\sim \tform(m_h)\sim \alpha^{-3/4+c/4}Q^{-1}$,
  after which the evolution will follow the normal $c>-1$ case discussed
  in the next sub-subsection.
}
However, interactions can change the actual spectrum from this
production rate.  In particular, an $f(p)\sim p^{-3}$ spectrum rises
more steeply than a thermal spectrum.  Below some
scale $\pmax$ the spectrum will collapse into a thermal-like form,
\be
f_s(p) \sim T_*/p, \quad \textrm{for } p < \pmax
\label{spec_sat}
\ee
in close analogy to the overoccupied case discussed in
subsection~\ref{high_occ}.  The soft sector's contribution to
the screening and elastic scattering are dominated by the scale $\pmax$,
as they were in the overoccupied case.
The major difference is, however, that the growth of the soft sector is driven by
soft splittings from the hard particles. In addition to interacting
among themselves, the soft particles can scatter with the hard particles.

Let us now determine what physics dominates the evolution of $\pmax$.
There are four mechanisms which compete in redistributing the soft sector:
\begin{itemize}
\item Elastic scattering with hard particles, described by $\qhatelast$: The scale
which has had enough time to experience order 1 momentum changes is
\be
\pmax^{\textrm{hard scat}}\sim (\qhatelast t)^{1/2}\sim
\alpha^{1-c/2}Q(Qt)^{1/2} \,.
\ee
All scales below $\pmax^{\textrm{hard scat}}$ have had time to
redistribute, via elastic scattering, into an $f(p)\propto 1/p$ thermal-like
tail.

\item Elastic scattering among soft particles: That the integral in
\Eq{eq:qhat} diverges for $\propto 1/p^3$ already shows the
inconsistency of such a soft distribution. While the total number of soft
particles is initially smaller than that of the hard particles,
their interaction rates are Bose stimulated, so it could be possible
that elastic scattering is dominated by the soft sector.

Assuming that the  evolution of $\pmax$ is dominated by the elastic scatterings between
soft particles, we can estimate the momentum diffusion constant arising from the soft sector
$\qhatelasts\sim \alpha^2 T_*^2 \pmax^{\rm soft\:scat}$. This
momentum diffusion coefficient re-arranges the soft particles out to
$\pmax \sim (\qhatelasts t)^{\frac 12}$.  The energy density
of the soft sector is dominated by the particles emitted at scale $\pmax$,
so the energy density of the soft sector is
$\varepsilon_s \sim n_h \Gsplit(\pmax^{\rm soft\: scat})t
			\sim T_* \pmax^3$.
Combining and solving self-consistently, we get
\be
\pmax^{\textrm{soft scat}} \sim \alpha^{6/5-2c/5}Q(Q t)^{3/5} \,.
 \ee

\item Merging at the $\pmax$ scale: The particle number in the overoccupied
regime can decrease by merging pairs of particles to create fewer particles at higher
momentum, analogously to the hard merging in the overoccupied case.
All modes up to a limiting scale $\pmax^{\textrm{hard merge}}$ have had time to undergo
repeated merging to push the extra particle number to the scale $p^{\textrm{hard merge}}_{\max}$.

To estimate the scale  $\pmax^{\textrm{hard merge}}$ we find the
hardest modes which have had time to have an order 1 probability of merging
 \be
 \Gamma_{\rm{merge}}(p^{\textrm{hard merge}}_{\max}) t \sim 1
 \ee
 with the Bose stimulated rate
 \be
 \Gamma_{\rm merge}(p)\sim \Gamma_{\rm{split}}(p)[1{+}f(p)]\,,
 \ee
so that
\be
p^{\textrm{hard merge}}_{\max}\sim \alpha^{4/3-c/3}Q(Q t)^{2/3}\,.
\ee

\item Saturation:
For a given soft momentum $k \ll Q$, emissions will cause $f_s(k)$ to grow
until it reaches a size where absorption (joining) processes shut off
the production.  Let us find the occupancy where this occurs.
The production and absorption rates per hard particle $f(p)$ are
\bea
\label{eq:saturation}
f(p) \left( \Gamma_{\rm prod} - \Gamma_{\rm absorb} \right)
 &\!\sim\! & \Gsplit(k) \Big( f(p) [1{+}f(p{-}k)] [1{+}f_s(k)]
      - f(p{-}k) f_s(k) [1{+}f(p)] \Big)
\nonumber \\
& \!\sim\! & \Gsplit(k) \left(
  k f_s(k) \frac{df}{dp} + f(p) [1{+} f(p)] \right) \,.
\eea
This rate, integrated over $\p$, goes to zero when
\be
f_s(k) \sim \frac{\int_\p f(p)[1{+}f(p)]}{\int_\p -k\: df/dp} \,.
\ee
Since the phase space
opens up as $p^2 dp$, it must be that $df(p)/dp$ is on average
negative.  So we can estimate $df/dp \sim -f(p) / p$.
Using this estimate and using that $p\sim Q$ is the dominant scale, we
find
\be
f_s(k) \sim \frac{p}{k} [1{+}f(p)] \sim \frac{Q}{k} (1+\alpha^{-c}) \,.
\label{fk}
\ee
(The $1+\alpha^{-c}$ term is included so the formula can also be applied
for overoccupancy.)

Whenever the occupancy of soft modes reaches this value, soft modes are
absorbed as quickly as they are emitted.  The production
mechanism shuts off and the occupancy grows no further.  If this shutoff
is responsible for deciding where the momentum tail goes from
$k^{-3}$ to $k^{-1}$ behavior, then the scale $\pmax$ is determined by the
condition that the emitted spectrum, $f_s(\pmax)$ from \Eq{eq:spectrum},
has $f_s(\pmax) \sim Q/\pmax$.  Inserting \Eq{spec_BH}, we find
\be
\pmax^{\rm{saturation}}\sim \alpha^{1-c/2}Q(Q t)^{1/2},
\ee
which turns out to be the same as the scale set by elastic scatterings
with hard particles.

\end{itemize}

At early times the hierarchy of these scales is
\be
\pmax^{\textrm{hard scat}} \sim
\pmax^{\rm{saturation}} >
\pmax^{\textrm{hard merge}} >
\pmax^{\textrm{soft scat}} \label{eq:hierarchy}
\ee
so that the evolution of $\pmax$ is set by elastic scatterings with the hard particles (as well as
saturation of soft particle production):
\be
\pmax\sim \pmax^{\textrm{hard scat}} \sim \alpha^{1-c/2} Q (Qt)^{\frac 12}\,.
\ee
The occupancies in the infrared tail can be solved again by estimating
how much energy density the soft sector has had time to eat. The energy
density in the infrared tail is dominated by the scale $\pmax$ so that
\bea
\varepsilon_s &\sim& T_* \pmax^3  \sim \pmax n_s(\pmax)\,,\\
n_s(\pmax) & \sim & \Gsplit^{\rm BH}(\pmax) t n_h
   \,, \nonumber \\
T_* &\sim& \Gsplit^{\rm BH}(\pmax) t n_h/\pmax^2 \sim Q,
\eea
showing that $T_*$ is a constant scale.

There are a number of assumptions we have made whose breaking may lead to
a revision of the time evolution of $\pmax$ and $T_*$. First, we assumed
that the hard sector
dominates elastic scattering, screening and energy density. Of these,
screening is most sensitive to soft modes. The contribution of the soft
sector to screening is dominated by the scale $\pmax$,
\be
\label{msoft}
m_s^2 \sim \int_\p \frac{f_s(\p)}{\p}\sim \alpha T_* \pmax\sim \alpha^{2-c/2} Q^2 (Qt)^{1/2} ,
\ee
which becomes comparable to hard screening at the time
(compare \Eq{msoft} to \Eq{m_h})
\be
Q t \sim \alpha^{-2-c}.
\ee
Second, we also made the assumption of the hierarchy given in \Eq{eq:hierarchy},
which also breaks down at the same time $Qt \sim  \alpha^{-2-c}$, when all
four scales reach
\be
\pmax^{\textrm{hard scat}} \sim \pmax^{\rm{saturation}}
\sim \pmax^{\textrm{hard merge}} \sim \pmax^{\textrm{soft scat}}
     \sim k_{\rm LPM}.
\ee

\subsubsection{$\alpha^{-2-c} < Q t < \alpha^{-2+c/3}$}

In the second stage, the scale $\pmax$ continues its growth, but
as it enters the LPM suppressed regime, the increase of energy to
the soft sector is no longer enough to keep $T_*$
a constant scale. At the same time, the soft sector grows large enough
to dominate screening, so that the rate of elastic scattering reduces
and the new LPM scale follows $\pmax$.

Assuming that the evolution of $\pmax$ is still set by the
elastic scattering off the hard particles, we get the same estimate
as before,
\be
\pmax\sim (\qhatelast t)^{1/2}\sim \alpha^{1-c/2} Q (Qt)^{\frac 12}\,.
\ee
Then, from the energy density that the infrared tail has had time to eat we get
\be
T_*  \sim  \Gsplit^{\rm LPM}(\pmax) t n_h/\pmax^2 \sim \alpha^{-1/2-c/4}(Q t)^{-1/4}Q.
\ee

The contribution of the soft screening and elastic scattering from soft particles are then
\bea
m_s^2 &\sim& \alpha^{3/2-3c/4}Q (Q t)^{1/4},\\
\qhatelasts&\sim& \alpha^{2-c}Q^3,
\eea
so that the soft and hard sectors give comparable contributions to
elastic scattering but the soft sector
dominates screening. Hence, the assumption that
$\pmax$ can be estimated from hard elastic scattering is selfconsistent.
Furthermore, as the dynamics of the soft sector is set by the
soft sector itself, we know already from the overoccupied case that the
hard merging scale will equal $\pmax$.

The growth of $m_s$ decreases the elastic scattering rate, and so the
LPM scale starts to grow. The new LPM scale is then
\be
k_{\rm LPM}\sim m_s^4/\qhatelast\sim \pmax,
\ee
so that using either the BH rate or the LPM rate will give the correct
estimate for the energy density of the infrared tail.

When $\pmax\sim T_*$, the occupancies at the scale $\pmax$ become
of order 1 and the soft sector thermalizes. This happens at $Q t\sim \alpha^{-2+c/3}$,
and at later times, the estimates again need to be revised.

\subsubsection{$\alpha^{-2+c/3} < Q t < \alpha^{-2+3c/8}$}
\label{sec:bath}

In the final stage of the equilibration the soft particles dominate
screening, particle number and $\qhat$ but the hard sector still carries
most of the energy.  The soft sector is thermalized and is
described by a single scale, its
temperature $T$, with $k_{\rm LPM}\sim T$,
$n_s\sim T^3$, $\varepsilon_s \sim T^4$, and
\be
\label{qhat_therm}
\qhatelast  \sim \alpha^2 T^3.
\ee

The hard sector loses its energy to the soft sector by splitting off
daughters, which split further, cascading down in momentum to join the
thermal bath.  There is a subtlety, first identified by Baier {\it et al}
\cite{bottomup}.  As discussed, the spectrum of excitations emitted by a
hard $p\sim Q$ particle scales as $k^{-1/2} dk/k$.  In terms of energy
lost, the spectrum is dominated by the largest $k$; in terms of number
of excitations, it is dominated by small $k$.  But in a given amount of
time $t$, a sufficiently high-energy daughter will not actually
give its energy to the thermal bath, but will remain intact.  In terms
of energy gain by the thermal bath, the most important emitted daughters
are therefore the highest-energy daughters which can break down
completely and yield their energy to the thermal bath within time $t$.
These are precisely those excitations which can hard-split in less than
time $t$.  The rate for a particle of momentum $k$ to hard-split is of
order the same as the rate for a much harder $p\sim Q$ particle to emit
a particle of momentum $k$.  Therefore the most important daughter
momentum scale $k$ is that, such that each hard particle emits
$\OO(1)$ such daughters in time $t$.  Call this momentum scale
$\ksplit$; parametrically (see \Eq{tform2}, \Eq{eq:LPM} and
\Eq{qhat_therm})
\bea
\label{def:ksplit}
\Gsplit^{\rm LPM}(\ksplit)t \sim 1,
  \quad \quad \ksplit\sim \alpha^4 T^3 t^2 \,.
\eea
The particles with momentum $\ksplit$ then undergo multiple splittings, in
a time scale which is much shorter than the formation time of the initial emission,
depositing all their energy to the soft thermal bath.
The energy density and the temperature of the soft sector are then
\bea
\varepsilon_s &\sim& n_h \ksplit \,, \qquad
T \sim \varepsilon_s^{\frac 14} \,,
\eea
and solving self consistently we get for the temperature and for the splitting scale
\bea
T&\sim & \alpha^4 n_h t^2 \sim \alpha^{4-c} Q^3 t^2 \,, \\
\ksplit &\sim&\alpha^{16-3c}Q (Q t)^8  .
\eea

By the time the splitting scale reaches Q, all of the hard modes have had time
to undergo hard splitting and join the thermal bath, and the system thermalizes
when
\be
Q t \sim \alpha^{-2+3c/8} \,,
\qquad
\Tfinal \sim \alpha^{-c/4} Q \,.
\ee
The final equilibration time should be compared to the characteristic
time scale $t \sim \alpha^{-2} T^{-1} \sim \alpha^{-2+c/4} Q$ for the
thermal bath to self-equilibrate.  The equilibration time is longer but
only by a factor of
$\sqrt{Q/T} \sim \alpha^{c/8} \sim f_{\rm initial}^{-\frac{1}{8}}$.
In particular, equilibration is much faster than the naive large-angle
scattering time for the original hard excitations,
$\teq \sim \alpha^{-2+c} Q^{-1}$.
Physically, the factor $\sqrt{Q/T}$ arises because the time it takes a
particle of momentum $Q\gg T$ to radiate away its energy in a thermal
bath of temperature $T$ is $t \sim (\alpha^2 T)^{-1} \sqrt{Q/T}$.  So
the final equilibration time is just the time for the hard excitations
to break up in a thermal bath at the final temperature.


\begin{figure}
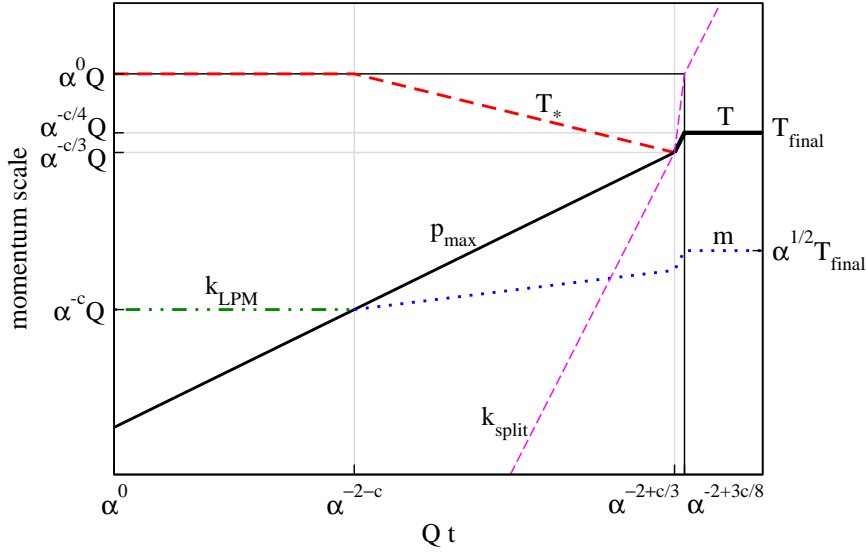

\centerbox{0.75}{low_occ.eps}
\caption{
Evolution of the momentum scales in a log-log plot
with initial distribution of \Eq{eq:init} with $c=-1$.
}\label{fig:low_occ}
\end{figure}


\section{Weakly anisotropic systems}

In some applications we expect a nonabelian plasma to form, in which,
rather than being evenly distributed in direction, particle occupancy is
larger in some directions than in others.  In some cases
the excess occupation may be modest.  For instance, a nonabelian plasma
which is experiencing shear flow develops an anisotropic particle
distribution.  This has been discussed by Asakawa {\it et al}
\cite{anom_visc}, who argue that in a range of shear flow strengths, the
plasma remains nearly isotropic but only because of the physics of
plasma instabilities.  Weakly anisotropic plasmas should also be rather
generic in the presence of phase interfaces during early Universe phase
transitions, such as the electroweak phase transition (if it exists).

After an overview of the physics of plasma instabilities, we will first
treat this case of weakly anisotropic plasmas, and then consider plasmas
with a very high degree of anisotropy in the next section.
Since anisotropic plasmas are rather complex and because the ``parameter
space'' describing them is larger than for the isotropic case, we will
not attempt to follow the dynamics through every stage to equilibrium.
Instead we will identify the relevant dynamics {\sl initially,} and then
determine the time scale before the dynamics change appreciably and what
the new relevant dynamics becomes.

\label{sec:weak}

\subsection{Plasma instabilities}

\label{sec:weibel}

So far we have only considered systems which are statistically locally
isotropic.  A new complication enters when we consider systems which are
locally {\sl anisotropic}.  Namely, certain long-wavelength magnetic
gauge field excitations are then generically unstable to exponential
growth, in a process called the Weibel instability (see
\cite{Weibel} for the electromagnetic case and
\cite{Mrow,RRS,ALM,RRS2} for an overview of the nonabelian case and
\cite{ALMY,RRS3,others} for some nonabelian numerical studies).  At weak
coupling, in many circumstances these fields are expected to grow large
enough that they come to dominate much of the dynamics of the system.
In particular, the ``hard'' excitations which dominate the energy
density of the system may predominantly change direction by deflection
in these magnetic fields.  These deflections may then dominate the
splitting or joining processes these excitations undergo.  The key
physics involved in plasma instabilities is the physics of screening.
So far we have only needed a very crude description of plasma screening,
but plasma instabilities involve finer details of the screening process.
Therefore we begin with a more careful discussion of the physics of
screening.

\begin{figure}
\putbox{0.45}{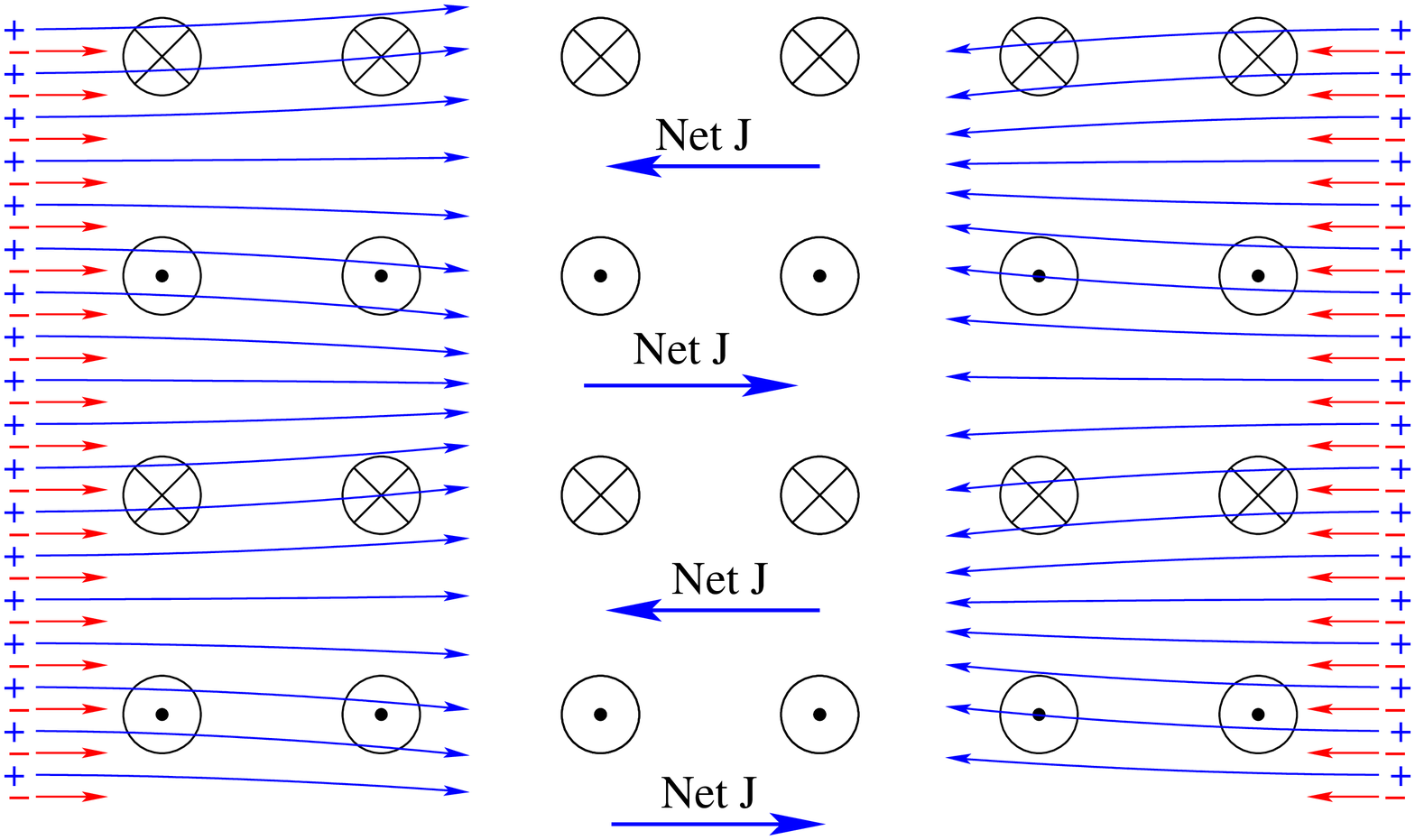} \hfill
\putbox{0.45}{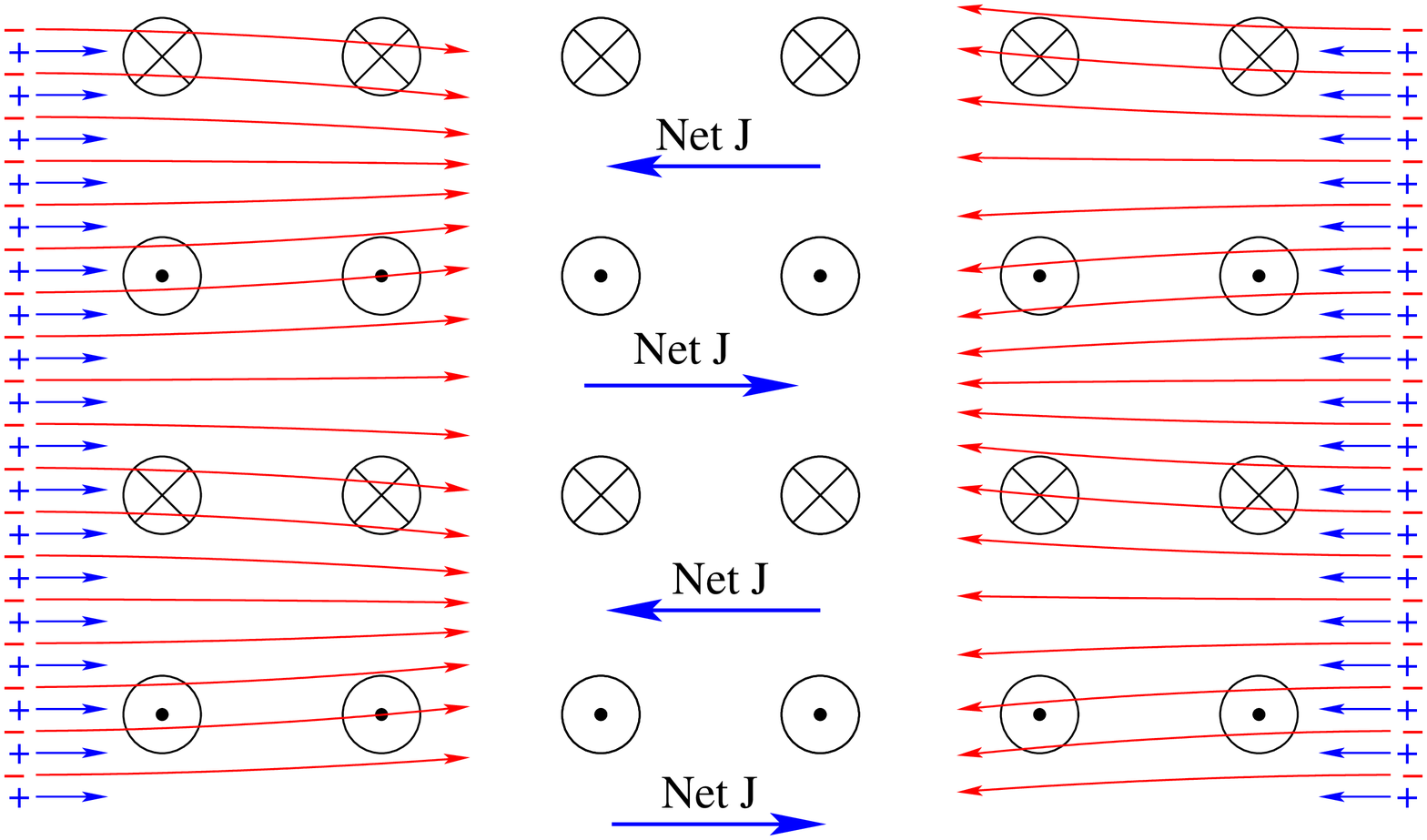}
\caption[Weibel instabilities]
{\label{fig:weibel}
Illustration of the Weibel instability.  Particles flowing in the $z$
direction are deflected by a magnetic field with both polarization and
wave-vector at right angles to the particle flux.  Left:  the deflection
of positive charged particles (in blue) focuses them into some regions
and defocuses them from others.  Right:  opposite charged particles
(red) are focused in different regions but generate the same-sign net
currents.  Both are focused such that their currents amplify the
original magnetic field.}
\end{figure}

There are already nice, physical and quantitative discussions of plasma
instabilities in the literature, see for instance section II of
Ref.~\cite{ALM} and
Refs.~\cite{RRS,RRS2}.  Nevertheless we will give
a quick qualitative overview
to capture certain salient points for our discussion.  We start by
explaining how to think about the screening scale $m$ we previously
introduced.  Consider the case where many particle-like excitations%
\footnote{%
    By particle-like excitations we mean, excitations which can be
    described in terms of wave packets which are of smaller extent than
    the coherence length of the classical fields in which they move; and
    yet are built of a narrow enough distribution of momenta that they
    move nearly non-dispersively.  If the typical particle momentum is
    $p$, the range of momenta particles carry (and therefore the range
    of momenta out of which wave packets are built) is $\Delta p$, and
    the characteristic wave number of the background fields they move in
    is $k$, this requires $p \gg \Delta p \gg k$.  We will encounter a
    case where these criteria cannot be met in subsection \ref{sec:high},
    but in most of our discussion the scale ordering will be true
    parametrically.%
}
all move along the $\pm x$ axis, in the presence of a magnetic field
with wave-vector $\k$ in the $y$ direction and $\B$ field in the $z$
direction, as illustrated in Fig.~\ref{fig:weibel}.
The force on a particle is
\begin{equation}
\F = q^a \v \cdot \B^a \,.
\label{eq:F}
\end{equation}
If the particle trajectories are undeflected at time $t=0$, then by time
$t$ their velocities and positions have changed, to linear order in $B$,
by
\begin{equation}
v_y(t) = \frac{q^a B^a t}{p} \,, \qquad
\delta y(t) = \frac{q^a B^a t^2}{2p} \,.
\label{eq:vy}
\end{equation}
This particle motion concentrates particles in some regions and depletes
them in others.  The concentration fraction is $k \delta y(t)$, and the
net current associated with this concentration is
\begin{equation}
J^a \sim \int d^3 \p \; f(\p) q^a k \delta y(t)
\sim \int d^3 \p \;
f(\p)\frac{q^a q^b}{2p} k B^b t^2 \,.
\label{eq:J}
\end{equation}
Here $\int d^3 \p f(\p)$ is just counting the density of particles.
We recognize the combination
\begin{equation}
\int d^3 \p \; f(\p) \frac{q^a q^b}{p}
= \delta_{ab}\; g^2 T_r \int \frac{d^3 \p\; f(\p)}{p}
\sim \delta_{ab} m^2
\label{eq:m_again}
\end{equation}
so the induced current is
\begin{equation}
J^a \sim k B^a m^2 t^2 \,.
\label{J_induced}
\end{equation}
This current enters the equations of motion for the magnetic field,
where it competes with terms of form $\nabla \times B \sim kB$.
Therefore the current term becomes important in the evolution of a
magnetic field whenever $m^2 t^2 \sim 1$.  So the correct physical
interpretation of the thermal mass $m$ is that it is the inverse of the
time scale for particle deflections to add up to enough that they
influence the gauge field dynamics.  That is, if particles remain
coherently in a region of electric or magnetic field of one sign for
time scales of order $1/m$ or longer, then the induced currents are
large enough to significantly influence the field evolution.
In the case illustrated, since the
particles enhance the magnetic field which deflects them, they lead to
an exponential growth in the strength of the magnetic field
$B(t) \sim B(t=0) \exp(\gamma t)$, with a growth rate $\gamma \sim m$.

For the case considered in Fig.~\ref{fig:weibel} the currents from the
particle excitations strengthen the magnetic field, which destabilizes
the plasma towards magnetic field growth.  Particles
flying in certain other directions contribute currents which oppose the
magnetic field.  For an {\it isotropic} distribution of particle
momenta, a magnetic field will not induce any current.  That is because
a magnetic field induces an overall rotation of the distribution of
particles; but if the particle distribution is isotropic then a rotation
does not change it.  Therefore an isotropic distribution is neutral in
the sense that it neither stabilizes nor destabilizes magnetic fields.
For an anisotropic distribution of particles, we can instead ask about
the average over directions and polarizations of magnetic fields.  This
is equivalent to averaging over the particles' momentum distribution.
Hence, for anisotropic systems, the plasma's impact on magnetic fields,
averaged over directions and polarizations of magnetic fields, is
neutral.  That means that
there will be some directions and polarizations where the particles
stabilize the magnetic fields (generate currents which oppose the
magnetic field), while for other directions and polarizations they will
destabilize the magnetic fields (generate currents which enhance the
magnetic field).  Hence, in an anisotropic plasma, magnetic
instabilities will {\sl always} be present in {\sl some} field directions
and polarizations \cite{ALM}.

In practice we will be interested in magnetic fields which change with
time, which always means there are also electric fields present.
Electric fields are stabilized by an isotropic distribution of
particles, so time-changing fields tend to be more stable than static
magnetic ones.  Also, if the electromagnetic fields vary on too short of
a length scale, particles will not have trajectories which stay in a
coherent region of field strength for the requisite $t \gsim 1/m$ needed
for currents to dominate the Yang-Mills equations of motion.  To
make somewhat more precise statements about which modes are unstable
under which circumstances, we will sketch a diagrammatic evaluation.

At the linearized level, an instability will occur if the retarded
propagator has a pole for some wave number $k$ and pure imaginary
frequency $\omega = i \gamma$, corresponding to exponential growth.
The transverse part of the inverse retarded propagator is
\begin{equation}
\label{inv_propagator}
G^{-1}_{ij}(\k,\gamma) =
          (k^2 + \gamma^2) (\delta_{ij} - \hat{k}_i \hat{k}_j)
             -\Pi_{ij}(\k,\gamma) \,.
\end{equation}
If the self-energy loop momentum $p$ is large compared to the external
momentum $k$, then the self-energy behavior becomes spin independent and
we can treat the simplest case, which is a scalar in the loop.
For this case,
\bea
\Pi^{\mu\nu}(k) & \sim & g^2 \int d^4 p \left[ -2g^{\mu\nu} \delta(p^2) f(p)
  +(2p+k)^\mu (2p+k)^\nu \frac{\delta(p^2) f(p)}{(p+k)^2}
  \right. \nonumber \\ && \hspace{0.76in} \left. {}
  +(2p-k)^\mu (2p-k)^\nu \frac{\delta(p^2) f(p)}{(p-k)^2}
   \right] \,.
\label{eq:fullpi}
\eea
Assuming $p\cdot k \gg k^2$, we can expand in $2p\cdot k \gg k^2$.
Together with the spin-independent form for the self-energy, this
constitutes the {\sl hard-loop} approximation for the self-energy.
In this approximation, the self-energy is
\bea
\Pi_{ij} & = & -g^2 T_r \int_{\p} \frac{f(\p)}{p}
\left[ \delta_{ij} - \frac{k_i v_j + k_j v_i}{\v\cdot \k -i\gamma}
    + \frac{(k^2+\gamma^2) v_i v_j}{(\v\cdot \k -i\gamma)^2}
    \right] \nonumber \\
(\delta^{ij} - \hat\k^i \hat\k^j) \Pi_{ij} & = & - g^2 T_r
 \int_{\p} \frac{f(\p)}{p} \: \left[ 2 +
   \frac{(k^2+\gamma^2)(1-(\v\cdot \hat\k)^2)}
        {(\v\cdot \k - i \gamma)^2} \right]\,. \;\;
\label{Pi_inst}
\eea
Positive-sign contributions inside the square brackets contribute
positively to \Eq{inv_propagator}, stabilizing the field.  Negative
contributions destabilize the field, and if they are large enough -- if
the RHS of \Eq{Pi_inst} equals or is larger than $2(k^2+\gamma^2)$ --
then they allow an exponential instability.  We see that those particles
in the angular range $\v \cdot \k < \gamma$ are the ones contributing to
the instability; all others stabilize.

The above discussion is based on linearizing in the size of the magnetic
field, and possibly in other interactions.  It breaks down when higher
loop contributions to the self-energy become as large as the one-loop
self-energy discussed above.  Most of the time, the dynamics controlling
the equilibration of the full system take place on a time scale long
compared to $1/\gamma$ the growth rate of the instabilities.  In this
case, the instabilities will grow until some nonlinear physics limits
their size.  After this, the relevant magnetic fields will remain large,
evolving but in a ``quasi-stationary'' fashion so that statistical
properties of their distribution evolve only on the time scale of the
system's approach to equilibrium.  The saturation of instabilities and
quasi-stationary evolution have been verified in numerical studies
\cite{AMY7,ArnoldMoore1,ArnoldMoore2007}.

\subsection{Instabilities for weak anisotropy}
\label{sec:weak_inst}

Consider a system along the lines of what we discussed in the previous
sections; that is, a distribution of ``particles'' with a typical
momentum scale $p \sim Q$, sufficiently fast falloff in $f(p)$ above
this scale, and typical occupancies
$f(p) \sim \alpha^{-c}$ for $p \sim Q$.  We will consider both the
possibility $c>0$ (overoccupancy) and $c<0$ (underoccupancy).  But now
we will consider the case where $f(\p)$ is also a function of the
direction $\hat{\p}$.  In this section we will take this angular
dependence to be weak,
\be
f(\p) = f(p)  \times \left( 1 + \epsilon F(\hat{\p})
                                  \right)
\label{eq:fweak}
\ee
with $F(\hat{\p})$ some order-1
combination of spherical harmonics of number 2 and higher, but dominated
by low numbers,%
\footnote{%
    That is, the $Y_{00}(\hat{\p})$ component defines $f(p)$.  Any
    $Y_{1m}(\hat\p)$ component can be removed by a choice of rest frame.}
and $\epsilon$ characterizing how small the non-spherical component of
$f(\p)$ is.  We also define $d$ as $\epsilon \equiv \alpha^{-d}$ (note
that $d$ is negative) to make it easier to mix powers of $\alpha$ and of
$\epsilon$.

Inevitably the distribution we consider here is not the most general we
could consider.  While there is no need to allow $Y_{1m}(\hat{\p})$
content in $F(\hat{\p})$
 we could have considered cases where $f(\p)$ does
not factorize into a radial and an angular function, that is, where the
typical particle energy is larger in some directions than in others.

First we compute the screening scale and elastic scattering $\qhat$,
\be
\label{eq:msq_weak}
m^2 \sim \alpha Q^2 f(Q) \sim \alpha^{1-c} Q^2
\,, \qquad
\qhatelast \sim \alpha^2 Q^3 f[1{+}f] \sim
    \left\{ \begin{array}{ll} \alpha^{2-2c} Q^3 & c>0\,, \\
                              \alpha^{2-c} Q^3  & c<0\,. \\
     \end{array} \right.
\ee
To determine whether plasma instabilities play any role in this system,
we need to determine what gauge field modes become plasma-unstable, how
large the amplitudes of the associated magnetic fields grow, and
how much deflection these magnetic fields cause in the hard particles.
That is, we need to compute $\qhatinst$ arising from
soft-unstable magnetic fields and compare it to
$\qhatelast$ to
determine whether the instabilities play an important role.

First let us see how many modes are unstable and how fast they grow.
As discussed in the last section, a mode is unstable and grows if the
inverse retarded propagator, \Eq{inv_propagator}, vanishes for an imaginary
frequency $\omega = i\gamma$.  $-\Pi_{ij}$ gets more positive as
$\gamma$ is increased, so the range of $\k$ which are unstable is
determined by considering $\gamma=0^+$.
The isotropic part of the integral vanishes for $\gamma=0^+$ and
gives $\pi \gamma/k$ for small $\gamma$.  The anisotropic part averages to
zero over $\hat\k$ directions, but is positive in some directions and
negative in other directions, {\it e.g.}\ in unstable directions
\be
-\Pi_{ij} \sim \frac{m^2}{2} \left( \frac{\pi \gamma}{k} - \epsilon
 \right) \,.
\label{Pi_weak}
\ee
Applying this we find instabilities in about half of
directions, with maximum $k$ of
\be
\kinst^2 \sim \epsilon m^2 \,.
\label{kinst_weak}
\ee
A simpler way to get this scale would be to compute $m^2$ {\sl just}
from the anisotropic piece of the particle distribution, which gives
$m^2_{\rm inst} \sim \epsilon m^2$.  The unstable modes are those with
$\kinst^2 \sim m^2_{\rm inst}$.  So long as $\epsilon \leq 1$ and
$c \leq 1$, the scale $\kinst \ll Q$ and so the ``hard-loop''
approximation used to establish these estimates is self-consistent.

The maximal growth rate is roughly $\gamma$ such that the
$m^2 \pi \gamma/2k$ term cancels the negative term, giving
\be
\gamma \sim \epsilon \kinst \sim \epsilon^{\frac{3}{2}} m \,.
\label{gamma_weak}
\ee
This is {\em not} the answer we would get by just considering the
contribution of the unstable modes.  That is because a time-varying
gauge field always has an electric component, and an isotropic
distribution of particles exerts a stabilizing effect on electric
fields.  So while the isotropic part of the particle distribution does
not change which modes are unstable, it can slow down the growth of the
unstable modes.  (This is an example of the general phenomenon that
scales $k \ll m$ have slow dynamics \cite{ASY}).

\begin{figure}
\centerbox{0.6}{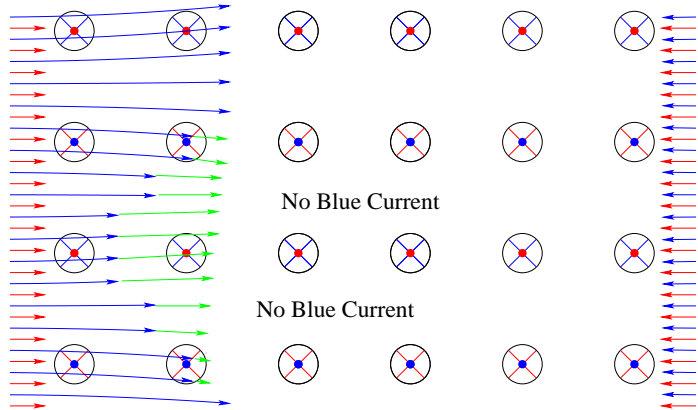}
\caption[Color randomization]
{\label{fig:color-rotate} A cartoon illustrating how color rotation
  could inhibit plasma instabilities.  A blue-anti-red magnetic field
  deflects blue and red color charges.  However where color rotation
  occurs, the induced current will be in another color direction (such
  as green).  This does not amplify the inducing magnetic field because
  it is in the wrong color direction; hence plasma instabilities are
  inhibited.}
\end{figure}

\begin{figure}
\centerbox{0.6}{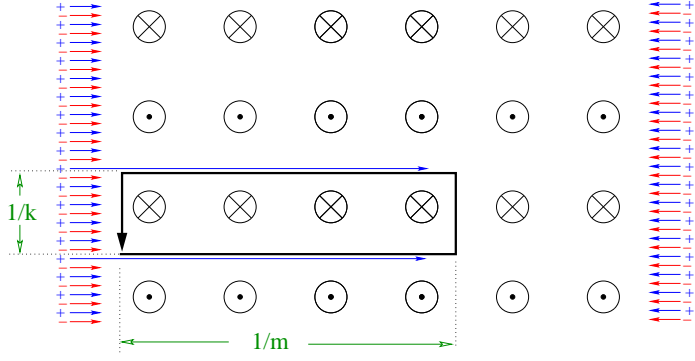}
\caption[Wilson loop]
{\label{fig:Wilson}
 Setup for the Weibel instability as in Fig.~\ref{fig:weibel}, but with
 a Wilson loop indicated.  When the phase around the Wilson loop is
 order-1, then in any gauge one or both of the indicated particle
 trajectories suffers a large color rotation.  Choosing the Wilson loop
 to be of length $\sim 1/m$ and width $\sim 1/k$, an order-1 phase in
 the Wilson loop is sufficient to ensure that the type of color rotation
 illustrated in Fig.~\ref{fig:color-rotate} will occur over the length
 scale relevant for the development of the instability.}
\end{figure}

This analysis of which modes are unstable is perturbative.  The mode
amplitudes grow until the perturbative treatment breaks down, which will
occur when some nonabelian term becomes important somewhere in the above
calculation.  The evaluation of the hard loop assumes particles
can propagate a distance $1/\kinst$ without color rotating.  Color
rotation destroys the growth of plasma instabilities, as illustrated in
figure~\ref{fig:color-rotate}:  when particles color-rotate, the current
they induce is in a different color direction than the magnetic field
which generated the current.  Hence the current does not enhance the
generating magnetic field.  Color rotation is important when a Wilson line
$U={\rm Pexp}\, ig \int dl \cdot A$ of length $l \sim \kinst^{-1}$
generically contains an order-1 phase factor.
A gauge-invariant criterion, for such phase factors to be large, is that
a Wilson loop of size $1/\kinst$ on a side have an order-1 phase factor,
see Fig.~\ref{fig:Wilson}.  The
phase factor of such a Wilson loop is $g^{-1} B/\kinst^2$.  So color
randomization is important if $B^2 \sim \alpha^{-1} \kinst^4$.
This translates to an occupancy for the unstable modes of
$\OO(\alpha^{-1})$.  In other words, the growth of plasma instabilities
is shut off when plasma-unstable modes have nonperturbatively large
occupancies.  We would arrive at the same conclusions by investigating
where mutual interactions between such modes significantly influence
their evolution.  We would also reach the same conclusions by asking
where two-loop self-energy corrections grow to be as large as the
one-loop self-energy correction we studied.

\subsection{Momentum diffusion from the instability}

Next we compute the mean squared rate at which these unstable modes
deflect the excitations of momentum $\sim Q$.  Such an excitation feels
a time varying force of order $F \sim g B$.  This force remains
coherently in the same direction for the characteristic coherence length
of the magnetic field,
$\tcoh\sim \kinst^{-1}$.  (This is also the typical time
scale it takes the particle to rotate in color, which would also
randomize the direction of the force.)  Provided $c<1$ and $\epsilon<1$,
the momentum accumulated in one coherence length,
\be
\Delta p \sim \tcoh g B \sim \kinst^{-1} g \frac{\kinst^2}{g}
    \sim \kinst \sim \epsilon^{\frac 12} \alpha^{\frac{1-c}{2}} Q
\label{eq:deltap}
\ee
is small compared to the typical particle momentum $Q$.  Therefore the
change in particle momentum will accumulate as many small incoherent
``kicks'' and can be well characterized by a momentum diffusion
coefficient $\qhatinst$,
\be
\qhatinst \sim F^2 \tcoh \sim \alpha B^2 / \kinst \,.
\label{qhat_weak1}
\ee
We already saw that the magnetic field saturates at an amplitude
$B^2 \sim \alpha^{-1} \kinst^4$.  Therefore
\be
\qhatinst \sim \alpha B^2 \tcoh
     \sim \frac{\alpha B^2}{\kinst} \sim \kinst^3
     \sim \epsilon^{\frac 32} m^3
     \sim \epsilon^{\frac{3}{2}} \alpha^{\frac{3}{2} (1-c)} Q^3 \,.
\label{qhat_inst}
\ee
The instability is important if this $\qhat$ dominates the scattering
from elastic processes, but we expect it to be irrelevant if it is
smaller than the scattering already occurring due to normal elastic
processes.  So the domain where the instability is important is
\bea
\qhatinst & > & \qhatelast \,, \nonumber \\
\epsilon^{\frac{3}{2}} \alpha^{\frac{3}{2} (1-c)} Q^3 & > &
       \alpha^{2-(2,1)c} Q^3
\label{q_compare1}
\eea
leading to
\be
\mbox{Instability important if } \left\{ \begin{array}{ll}
      \epsilon > \alpha^{\frac{1+c}{3}}  &  c < 0\,, \\
      \epsilon > \alpha^{\frac{1-c}{3}}  &  c > 0\,. \\
     \end{array} \right.
\label{result1}
\ee

When the instability is important, it can change two things.  One thing
it can do is to enhance angle change, erasing $\epsilon$.  The other
thing it can do is for this angle change to enhance splitting/joining
processes.  This can cause $c$ and $\epsilon$ to evolve, and in the case
of low occupancy it can create a ``cloud'' of soft excitations which can
dominate the anisotropy of screening even if they do not replace the
hard particles numerically.
It cannot lead to an evolution of $c$
through elastic scattering because soft unstable fields are primarily
magnetic, and only electric fields change particle energy.  To estimate
the electric field subdominance, note that $E \sim \partial_t A$ while
$B \sim \partial_x A$, so $E \sim (\gamma/\kinst)B \sim \epsilon B$.
Therefore the diffusion of particle energy $\sim E^2 \tcoh$ is slow
compared to the diffusion of particle direction.  The time
scale for particle energy to change is correspondingly longer than the
time scale for $\epsilon$ to be erased, ending the instability.

The rate for $\epsilon$ to change appreciably due to scattering is the
rate for large angle change, which is
\be
\tlargeang^{-1} \sim \frac{\qhatinst}{Q^2}
  \sim \epsilon^{\frac{3}{2}} \alpha^{\frac 32(1-c)} Q \,.
\label{large_angle}
\ee
This time scale is larger (rate smaller) than the growth rate of the
unstable modes, $\gamma \sim \epsilon^{\frac{3}{2}} \alpha^{\frac{1-c}{2}} Q$,
so the treatment is self-consistent in that the unstable modes have time
to grow before they significantly modify $\epsilon$.

To find the rate at which splitting/joining processes occur, we need
to determine the formation time of emitted particles.  Moving through the
soft-unstable fields, an emitted particle picks up a kick of $\kinst^2$
every length $1/\kinst$, which is the effective inter-scattering
distance.  The formation time for radiation of a particle of energy $k$
is therefore
\be
\tform(k) \sim \frac{k}{k_{\perp}^2} \sim
        \frac{k}{\qhat \tform} \quad \Rightarrow \quad
\tform(k) \sim k^{\half} \kinst^{-\frac{3}{2}}
        \sim k^{\half} \epsilon^{-\frac 34}
        \alpha^{-\frac{3}{4} (1-c)} Q^{-\frac 32}
\label{tform_weak}
\ee
which is always longer than $1/\kinst$.  Therefore split/join
processes are always in the LPM regime.  The rate of these processes
is a factor of $\alpha[1{+}f(k)]$ over the formation time;
\be
\Gsplitjoin(k) \sim \alpha [1{+}f(k)] \tform^{-1} \sim
      k^{-\half} \epsilon^{\frac 34}
      \alpha^{\frac 74 -\frac{3}{4} c} Q^{\frac 32} [1{+}f(k)]\,.
\label{t_split}
\ee
To determine whether this rate is physically important, we have to
consider the physics of the particles generated.

\subsection{Joining and $\epsilon$ suppression for overoccupancy}
\label{eps_over}

For the case $c>0$ ($f(Q)>1$) we need not worry about splitting
processes; we saw in section \ref{high_occ} that $f(p<Q)$ scales at most
as $1/p$, which is too soft to dominate screening.
We also see, combining \Eq{t_split}, \Eq{large_angle} and \Eq{result1},
that $\Gsplitjoin \tlargeang \ll 1$.  Therefore the
particles randomize direction, and $\epsilon$ is degraded, before
particle number can change appreciably.

Nevertheless it is possible that $\epsilon$ is significantly affected by
joining processes.  The rate of this process is
$\Gsplitjoin(Q) \sim \epsilon^{\frac 34}
          \alpha^{\frac{7}{4}(1-c)} Q$
(\Eq{t_split} with $k \sim Q$ and $f \sim \alpha^{-c}$).
But what this expression disguises is that this
rate is anisotropic in direction, at the order one level.  The reason is
that $\qhat$ arising from plasma instabilities is in general strongly
anisotropic, and peaked for those particles which are overoccupied.  So
what can happen is the following.  Only an $\OO(\epsilon)$ fraction of
particles are anisotropic; but they generate instabilities, which cause an
order-one anisotropic $\qhat$.  This makes the rate of joining processes
faster, by an order-one amount, in the directions which have an excess
of particles.  Joining reduces the number of particles moving in those
directions, which is precisely a reduction in $\epsilon$.

The anisotropy is erased by the joining when an $\epsilon$
fraction of particles in the overoccupied direction have had time
the join, creating an $\epsilon$ change in $f(\p)$. This takes place at
the time
\be
t_{\epsilon }\sim \epsilon\, \Gsplitjoin^{-1}
\sim \epsilon^{\frac 14} \alpha^{\frac{-7+7c}{4}} Q^{-1} \,.
\label{t_epsilon}
\ee
This must be compared to the rate at which $\epsilon$ is
degraded by angular deflection; the anisotropy is also erased in
a time it takes to redistribute the directions of particles
$t_{\textrm{large angle}}$.
Whichever of these processes
is faster dominates, so that the dominant mechanism
bringing down $\epsilon$ is particle joining if
\bea
\label{eq:join}
t_{\epsilon} & < &\tlargeang
\nonumber \\
\epsilon^{\frac{1}{4}} \alpha^{-\frac{7}{4}(1-c)} Q^{-1}
& < &
       \epsilon^{-\frac 32} \alpha^{-\frac{3}{2}(1-c)} Q^{-1}
\nonumber \\
\epsilon & < & \alpha^{\frac{1}{7}(1-c)} \,.
\eea
Otherwise anisotropy is mostly removed by angle change due to $\qhat$.
The time scale for isotropization is
\be
\tchange \sim \left\{ \begin{array}{cl}
  t_{\epsilon} \sim  \epsilon^{\frac 14} \alpha^{\frac{-7+7c}{4}} Q^{-1}
  \,, &  \epsilon < \alpha^{\frac{1}{7}(1-c)} \,, \\
  t_{\textrm{large angle}} \sim \epsilon^{\frac 32}
                \alpha^{\frac{3}{2} (1-c)} Q^{-1}
  \,, &  \epsilon > \alpha^{\frac{1}{7}(1-c)} \,. \\
  \end{array} \right.
\ee
In either case, after time $\tchange$ the plasma has become nearly
isotropic, and equilibration proceeds as described in
subsection \ref{high_occ}.  Note that, in every case, $\tchange$ is
short compared to $\alpha^{-2} \Tfinal^{-1}$, the full equilibration
time found in subsection \ref{high_occ}.

\subsection{Soft emissions and $\epsilon$ for underoccupancy}

Next consider the case $c<0$, so $f(Q) < 1$ (underoccupancy).  In this
case the rate of hard particles to split off softer particles is larger
than the rate for them to join.  Soft particles are more efficient at
screening than hard particles.  The splitting rate is also highly
anisotropic, because $\qhat$ is.  Therefore, more soft particles get
generated in the directions where there is already a particle number
excess.  This tends to increase the anisotropy of screening, potentially
enhancing the strength of plasma instabilities.
In fact we will see that this happens over most, but not all, of the
parameter range of interest.

The soft particle production is in the LPM regime and,
analogously to the isotropic case, the spectrum of the soft particles
is given by \Eq{spec_LPM}, \Eq{spec_sat} (replacing $\qhatelast$ with
the dominant $\qhatinst$),
\bea
\label{fs_big}
f_s(p)&\sim&\alpha \,n_h \sqrt{\qhatinst} t /p^{7/2},\quad \textrm{for }p > \pmax,\\
\label{fs_small}
f_s(p)&\sim& T_*/p, \hspace{2.09cm}  \quad \textrm{for }p < \pmax\,.
\eea
 In this case the evolution of $\pmax$ is dominated by saturation,
\Eq{fk}, so that $f_s(p)$ is limited from above by $f_s(p)\lesssim Q/p$
and $T_*\sim Q$.  We find $\pmax$ by equating \Eq{fs_big}, \Eq{fs_small}
and using \Eq{qhat_inst}:
 \bea
 T_* &\sim& Q,\\
 \pmax &\sim& \qhatinst^{1/5}(\alpha n_h/Q)^{2/5} t^{2/5}
 \sim \epsilon^{\frac{3}{10}}
      \alpha^{\frac{7}{10}(1-c)} Q (Q t)^{2/5}\,.
      \label{pmaxsat}
 \eea

As the $\qhatinst$ is $\OO(1)$ anisotropic the soft particles are created
with an $\OO(1)$ anisotropy. The modes are then driven towards isotropy by momentum diffusion
dominated by $\qhatinst$. The anisotropy of the soft modes is described by $\epsilon_s(k)$,
and how anisotropic they are depends on the momenta of the soft particles and how large $t$ is.
The momentum scale which is just becoming isotropic is
\be
\label{kiso}
\kiso \sim \sqrt{\qhatinst t}
     \sim \epsilon^{\frac 34} \alpha^{\frac{3}{4}(1-c)}
     Q (Q t)^{\frac 12} \,.
\ee
At early times $\pmax$ is larger than $\kiso$, while for large $t$
the larger of the two scales is $\kiso$. In particular, the time when
$\kiso$ catches up with $\pmax$ is
\be
\label{eq:tiso}
\tisofill \sim \frac{m_h^8}{\qhatinst^3}
\sim \epsilon^{-\frac{9}{2}}\alpha^{-\frac{1}{2}(1-c)} Q^{-1}.
\ee
 For $k>\kiso$,
the soft particles have the order 1 anisotropy
they where created with, $\epsilon_s(k>\kiso)\sim 1$.  Below $\kiso$,
the particles that were created less than $\tiso(k) \sim k^2/\qhatinst$
ago are still anisotropic; particles created longer ago have had time to
undergo large angle change and isotropize.  The fraction of anisotropic
particles, and hence $\epsilon_s$, at the scale $k<\kiso$ is then
\bea
\epsilon_s(k) &\sim& \tiso(k)/t \sim k^2/ \kiso^2 \quad \quad {\rm{for}} \quad k < \kiso,\\
\epsilon_s(k) &\sim& 1 \hspace{2.97cm} \quad \quad {\rm{for}} \quad k > \kiso.
\eea

The new anisotropic population of soft particles can subsequently cause
new plasma instabilities which contribute to $\qhatinst$, and may eventually
dominate it.
The contribution  to $\qhatinsts$
from the new soft particles at the scale $k$ is
\be
\qhatinsts(k) \sim \epsilon_s^{3/2}(k)m_s^3(k),
\ee
where $m_s(k)$ is the screening from particles at scale $k$.
The scale $\pmax$ always
dominates soft-particle contributions to screening: for $k<\pmax$,
the distribution only grows as $1/k$ and screening depends
on  $\int k dk f(k)$. On the other hand for $k>\pmax$ the soft
distribution falls as $k^{-\frac 72} dk$, so that screening
falls as $\int k^{-\frac 52}dk$, and gets its dominant contribution
again from the scale $\pmax$.
Then, if $\pmax > \kiso$ so that the particles which dominate
screening are order-1 anisotropic, then also $\qhatinsts$ is dominated by the
particles at the scale $\pmax$. In this case, the new
$\qhatinsts$ is
\be
\qhatinsts\sim m_s^3(\pmax) \sim \alpha^{3/2}\pmax^{3/2}T_*^{3/2}
\sim \epsilon^{\frac{9}{20}}\alpha^{\frac{3}{20}(17-7c)}Q^3 ( Q t)^{3/5},
 \quad \textrm{for } \pmax> \kiso.
\ee
On the other hand, if $\kiso > \pmax$, then the scale
$\pmax$ may dominate screening but it will not dominate soft
contributions to $\qhat$ as $\epsilon_s(k)\propto k^{2}$ decreases too
fast with decreasing $k$.  So the contribution to $\qhat$ scales as
$\int^{\kiso} k^{-\frac 52} k^2 dk \sim \int^{\kiso} k^{-\half} dk$
which is dominated by $\kiso$, and
\bea
\label{newqhat}
\qhatinsts\sim m_s^3(\kiso)
	&\sim& \left[\alpha \kiso^2 f_s(\kiso) \right]^{3/2}
	\sim \alpha^3 n_h^{3/2} \qhatinst^{-3/8}t^{3/8}\\
	&\sim& \epsilon^{-\frac{9}{16}} \alpha^{\frac{39-15c}{16}}Q^3 (Q t)^{\frac{3}{8}},
	 \quad \quad \quad \quad \quad \quad\quad
      \textrm{for } \pmax< \kiso\,.
\eea

\begin{figure}
\centerbox{0.9}{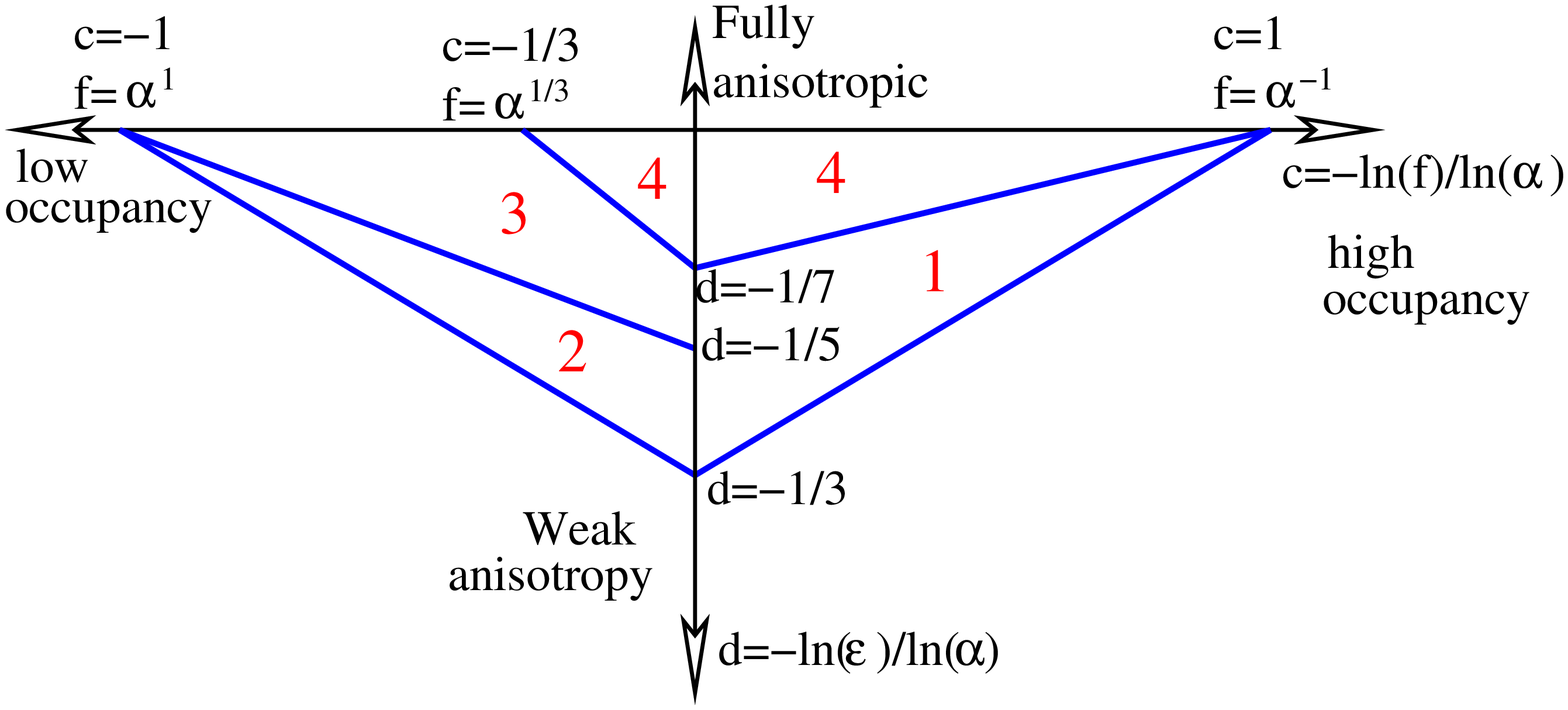}
\caption[Weak anisotropy]
{\label{fig:weak}
 For weak anisotropy, the $d$, $c$ plane (representing the level
 of anisotropy and of over- or underoccupancy) divides into four
 regions.
 In region 1, plasma instabilities induce particle joining processes
 which erase the anisotropy.
 In regions 2 and 3, particle splitting
 creates soft particles which are anisotropic and generate new plasma
 instabilities. In
 region 2 this occurs because of particles which have saturated
 occupancies.  In region 3 it occurs later, when the saturated particles
 have become isotropic and higher energy, still-anisotropic particles
 are responsible for the new anisotropy.
 In region 4, the particles isotropize due to instability-induced angle
 change.
 Below the blue triangle, angle
 change is dominated by ordinary elastic scattering rather
 than by plasma instabilities.
}
\end{figure}

As the soft sector grows, the new $\qhatinsts$ increases and it may
start to dominate over the $\qhatinst$ from the hard particles; we define
$\tchange$ as the time when
\be
\qhatinsts(\tchange) \sim \qhatinst \,.
\ee
We then have three possibilities:
\begin{itemize}
\item The soft $\qhatinsts$ starts to dominate $\qhat$ at early times
when particles at $\pmax$ are still fully anisotropic, $\pmax(\tchange)>\kiso(\tchange)$.
In this case the soft sector takes over $\qhatinst$ at time
\be
\label{eq:tchange}
\tchange \sim \epsilon^{\frac 74} \alpha^{\frac{-7-3c}{4}} Q^{-1}\,.
\ee
This is consistent with the assumption for (see \Eq{eq:tiso},
\Eq{eq:tchange})
\be
\tchange < \tisofill \qquad \Rightarrow \quad \epsilon < \alpha^{\frac{1+c}{5}}.
\ee
This is the Region 2 in Fig.~\ref{fig:weak}.

\item The new plasma instabilities start to dominate after the scale $\pmax$
has isotropized, but before time $\tlargeang$ which is when $\kiso \sim Q $
and momentum diffusion has isotropized all modes.
In this case the soft sector begins to dominate $\qhatinst$ by the time
(see \Eq{newqhat})
\be
\qhatinst \sim \qhatinsts(\tchange) \quad \Rightarrow \quad
\tchange \sim \alpha^{-8 }\qhatinst^{\frac{11}{3}}n_h^4
\sim \epsilon^{\frac{11}{2}}\alpha^{-\frac{5}{2}-\frac{3c}{2}}Q^{-1}.
\ee
This occurs if
\be
\tisofill < \tchange < \tlargeang\qquad \Rightarrow
\quad \alpha^{\frac{1+c}{5}}< \epsilon < \alpha^{\frac{1+3c}{7}}.
\ee
This is the Region 3 in Fig.~\ref{fig:weak}.

In this regime the evolution of $\pmax$ may cease to be controlled
by saturation before $\tchange$. In particular if
$\epsilon>\alpha^{\frac{1+c}{15}}$, elastic scattering among the
soft particles will start to push $\pmax$ to higher scales
(reducing $T_*$ as in the isotropic case) before the new instabilities
take over momentum diffusion. However,
$\qhat$ remains dominated by particles at the scale $\kiso$, so
this can have no effect on $\tchange$.

\item The third possibility is that soft particles never grow to dominate
$\qhatinst$ and the large angle changes erase the anisotropy, which occurs if
\be
\epsilon > \alpha^{\frac{1+3c}{7}} \,.
\ee
This is the Region 4 in Fig.~\ref{fig:weak}.
In this case the anisotropy shrinks appreciably at
$\tchange \sim \tlargeang$ given in \Eq{large_angle}.  When $\epsilon$
falls below $\alpha^{\frac{1+3c}{7}}$ the system will enter region 3.
\end{itemize}
These conclusions are displayed in Fig.~\ref{fig:weak}.

All we have computed above is the time scale before some new physics
significantly changes the dynamics; we have not computed the final
equilibration time $\teq$.  We will postpone doing so until
the end of the next section.

As a final aside, note that the total screening of soft particles
actually dominates the screening arising from hard particles before
$\tlargeang$ if there are enough particles at the scale
$\pmax$ at that time.  The total screening from such particles is
$m^2_s \sim \alpha \, \pmax T_*
                \sim \alpha \, \pmax Q$.
This is to be compared with $m^2 \sim \alpha^{1-c} Q^2$ from hard
particles.  Using \Eq{pmaxsat} and \Eq{large_angle},
this occurs if $\epsilon < \alpha^{\frac{1+9c}{3}}$.
However this effect is not important because even if soft particles
dominate screening, if they are isotropic they have no influence on
which modes are unstable to plasma instabilities and how large the
plasma instabilities grow.  Therefore this large contribution to $m^2$
does not actually influence the physics of plasma instabilities and can
be ignored.

\section{Large anisotropy}
\label{large_aniso}

A nonabelian plasma with a violent creation mechanism, or under strong
anisotropic expansion or compression, can be very far from equilibrium
in a strongly anisotropic fashion.  In particular, we expect very strong
anisotropy at early times in heavy ion collisions, at least in the
theoretically clean case of extremely high energy nuclei (so that the
energy density after the collision represents a scale large compared to
$\Lambda_{\rm QCD}$ and asymptotic freedom ensures perturbative
couplings) \cite{bottomup}.  Therefore we will consider very anisotropic
systems, which may also possess typical occupancies very far from 1.

Since we cannot study everything, we will concentrate on two of the
simplest possibilities; an initial momentum distribution with a single
characteristic scale $Q$ and a strongly oblate (planar) momentum
distribution, and a similar distribution but with a very prolate
(linear) momentum distribution.  We will try to determine the evolution
of such a system, starting with the growth of the plasma instabilities,
and continuing until some other piece of physics takes over as the
dominant mechanism driving thermalization.  We will not follow the
system's evolution all the way to thermalization for every case we
introduce, but at the end we will give an estimate of the final
equilibration time scale $\teq$.

\subsection{Instabilities of an oblate distribution}

Consider first a distribution of excitations with a characteristic
momentum $p \sim Q$, but with a very oblate angular distribution.
Namely, the occupancy is $f(p) \sim \alpha^{-c}$ if $p \lsim Q$
{\sl and} $p_z \lsim \delta Q$, but $f(p)$ is small outside this range.
That is, the excitations are concentrated in an angular range of width
$\sim \delta$ about the $p_z=0$ plane (so in spherical coordinates,
$|\cos(\theta)| \lsim \delta$ for most of the excitations).  We take the
angular range $\delta$ to be small, characterizing it when convenient as
$\delta \sim \alpha^{d}$.  Thus, $d<0$ (last section) represents
distributions which are nearly isotropic, while $d>0$ represents
distributions which are very far from isotropic.

The energy density of this system is
\be
\varepsilon \sim \int d^3 p \; p f(p) \sim Q^4 \delta \alpha^{-c}
\label{E_extreme}
\ee
where the factor $\delta$ is because the phase space volume which is
occupied is only $Q^3 \delta$, not $Q^3$.  If the energy density is
larger than $\alpha^{-1} Q^4$ then there will be Nielsen-Olesen
instabilities, see subsection~\ref{sec:NO}.  This occurs if
$\delta \alpha^{-c} > \alpha^{-1}$ or $c-d > 1$.
We have already discussed the Nielsen-Olesen instability so we will not
consider this region further.

The screening scale is
\be
m^2 \sim \alpha \int \frac{d^3 p}{p} f(p) \sim
\alpha^{1-c} \delta Q^2 \sim \alpha^{1-c+d} Q^2
\label{eq:msq}
\ee
and magnetic modes are unstable even in a larger range of $k_z$.
According to Ref.~\cite{ArnoldMoore2007}, the range of wave numbers
which exhibit plasma instabilities is $\k$ such that
$k_x,k_y \lsim m$ and $k_z \lsim \delta^{-1} m$, and the growth rate is
$\gamma \sim m$.  We can quickly re-derive these results by turning
again to \Eq{inv_propagator} and \Eq{Pi_inst}.
If $k_x/k_z \lsim \delta$ and $k_y/k_z \lsim \delta$ then
$\v \cdot \k \lsim \delta k$ for most of the excitations in the plasma.
If $\gamma > \v\cdot \k$ over most of this range then the contribution
from the second term in \Eq{Pi_inst} is negative and dominant, and
$\Pi_{ij} \sim m^2 k^2/\gamma^2$.  The largest this can be,
given $\gamma > \v\cdot \k \sim \delta k$, is
$\Pi_{ij} \sim m^2 \delta^{-2}$.  Substituting into \Eq{inv_propagator},
the largest value of $k$ which can be unstable is
$k^2 \sim \Pi \sim \delta^{-2} m^2$ or $k \sim \delta^{-1} m$.
Since we required $k_x,k_y \lsim \delta k$ and $\gamma \sim \delta k$ in
finding this largest unstable $k$ value, we find that the transverse
components of $k$ must be $\lsim m$, and the growth rate is
$\gamma \lsim m$.

\begin{figure}
\centerbox{0.6}{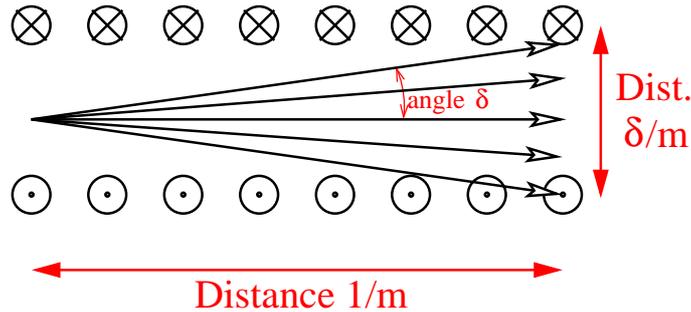}
\caption{\label{fig:coherent}
Illustration of how a magnetic field can vary rapidly in the $z$
direction, and yet particles will have long free paths in same-sign
regions of magnetic field because they propagate in a narrow range of
angles.}
\end{figure}

Let us develop a clearer understanding of this result.  We saw
previously that the physical meaning of the scale $m$ is that
excitations must remain coherently within a region of magnetic field of
stable sign and magnitude for a time scale $1/m$ if the current is to
grow big enough to modify field evolution.  It is immediately clear that
$\gamma \lsim m$, since $\gamma$ is directly an inverse time scale for
the magnetic field to change.  Now $k$ determines the spatial
nonuniformity of the magnetic field, and $\v \cdot \k$ tells how fast a
particle explores this nonuniformity.  If the field varies only in the
$z$ direction but a particle moves purely in the $xy$ plane, it will
remain in a region of coherent magnetic field forever.  Since particles
propagate almost purely in the $xy$ plane, the $k_x$ and $k_y$
components must be $\lsim m$ so that in-plane propagation remains in a
region of coherent magnetic field.  Since the $z$ component of an
excitation's propagation is small, the field can vary faster in the $z$
direction without the particle feeling different signs of $B$ field.
This is illustrated in figure~\ref{fig:coherent}.

There is an assumption we have built into the above discussion.  We have
assumed that the excitations which give rise to these plasma
instabilities can be treated as point-like entities.  This only makes
sense if they can be described using wave packets with spatial extent
$\Delta z < 1/k_z$, requiring momenta $\Delta p_z > k_z$.  Clearly
$\Delta p_z$ cannot be larger than $\delta Q$, the width of the $p_z$
distribution of our excitations.  Therefore our treatment is only
self-consistent if $\delta^{-1} m < \delta Q$.  Using \Eq{eq:msq},
this requires
\be
m < \delta^{2} Q \quad \mbox{and} \quad
m \sim \delta^{\frac 12} \alpha^{\frac{1-c}{2}} Q \quad \Rightarrow \quad
\delta^3 > \alpha^{1-c} \quad \mbox{or} \quad
c < 1-3d \,.
\label{non-htl}
\ee
Above this line, we must start from \Eq{eq:fullpi} and re-derive what
modes are unstable and how fast they grow.  We will return to the case
$c > 1-3d$ at the end of this section.  In the remainder of this section
we will assume $c < 1-3d$, so the hard-loop treatment is valid.

As in the previous section, plasma instabilities give rise to large
infrared magnetic fields, which can dominate the dynamics by deflecting
the hard excitations, with a characteristic momentum diffusion strength
$\qhatinst$.  We want to determine $\qhatinst$, which requires
understanding how large the instabilities grow.
The plasma instabilities will grow until either
\begin{enumerate}
\item
\label{case1}
they cause some physics which changes the distribution of excitations or
otherwise changes the dynamics in an essential way, or
\item
\label{case2}
they get large enough that nonlinear effects limit their growth.
\end{enumerate}
Case \ref{case2} occurs first over almost all the parameter space we
explore, so we will consider it in more detail.  There will be a narrow
region where case \ref{case1} actually occurs first, and we will handle
it when it arises.

The growth of instabilities will be cut off when any mechanism either
interferes with the process which causes growth, or provides an
efficient mechanism to remove energy from the unstable modes.  Color
randomization is a mechanism to cut off the growth mechanism, just as in
the weak-anisotropy case.  Refer again to Fig.~\ref{fig:Wilson}.
When the Wilson line
of a propagating particle randomizes its phase over a length scale
$1/\v\cdot \k \sim 1/m$, then color randomization occurs and instability
growth is shut off.  This occurs if the typical
amplitude for a gauge field (associated with the plasma-unstable $B$
fields) is $A \sim m/g$.  The magnetic field is
$B \sim \nabla \times A \sim \kinst A \sim m^2 \delta^{-1} g^{-1}$,
corresponding to
\be
\label{maxB}
B^2 \sim \alpha^{-1} \delta^{-2} m^4\, .
\ee
Equivalently, we need a Wilson loop of size $1/m \times 1/k$ to contain
an order-1 phase, requiring $B \sim km/g$, returning the same estimate.
Other estimates for the
limiting size of the magnetic fields, such as the scale where nonlinear
interactions between soft $B$ fields become important, give the same
estimate.  This estimate is also supported by lattice studies
\cite{ArnoldMoore2007}.

The limiting size of the instabilities again corresponds to occupancies
$f(k) \sim \alpha^{-1}$ for the unstable modes.  To see this, note that
the energy density in soft magnetic fields is
\be
\label{eq:occupancy}
\varepsilon \sim B^2 \sim \int^{\kinst} dk_z \int^{\sim m} d^2 k_{\perp}
 \: k f(k)
\sim \kinst^2 m^2 f(k)
\sim m^4 \delta^{-2} f(k)
\ee
requiring the typical unstable field occupancy to be
$f(k) \sim \alpha^{-1}$.

These magnetic fields give rise to momentum diffusion, but now the
momentum diffusion coefficient $\qhat$ is strongly direction dependent.
For a particle moving at an angle $\theta>\delta$ with respect to the $xy$
plane (that is, with $p_z / p = \sin\theta$), the magnetic field changes
orientation on a length scale $\kinst / \theta$, and so
\be
\label{qhat_theta}
\qhat(\theta) \sim F^2 \tcoh
  \sim \alpha B^2 \frac{\theta}{\kinst}
  \sim \delta^{-2} m^4 \frac{\theta}{\delta m}
  \sim \frac{m^3}{\delta \theta} \,.
\ee
For generic angles this is $m^3/\delta$, while for very planar
directions it is $m^3/\delta^2$.  The latter case is relevant for the
typical excitations with $p\sim Q$, $p_z\sim \delta Q$.  So for the
typical excitations we have
\be
\qhatinst(\delta) \sim \delta^{-2} m^3 \sim \delta^{-\frac 12}
    \alpha^{\frac{3}{2}(1-c)} Q^3 \,.
\label{qhat_extreme}
\ee

The situation becomes still more complicated if we consider relatively
low-momentum excitations, with $p \lsim m/\delta^2$.  Specifically, for
excitations with $p_z < m/\delta$, the extent of the wave packet is
larger than the inhomogeneity scale of the magnetic field and the
particle-approximation (WKB) treatment of propagation, implicit above,
breaks down.  Therefore we must consider $p \theta < m/\delta$ as a
special case.

The key to understanding the response of such a low-momentum particle is
to remember that the momentum transfer has to be a multiple of the
wave-number of the nonabelian field.  But if a particle of momentum
$p \ll m/\delta^2$ picks up a single kick of momentum
$k_z \sim m/\delta$, then its angle will instantly change to be
$\theta \sim k_z/p \sim m/(p\delta) \gg \delta$.  In this case, one
should estimate the rate of momentum accumulation using this larger
value of $\theta$.  Using this estimate, we find
$\qhat \sim m^3 / \delta \theta$ but $\theta \sim m/p\delta$, so
$\qhat \sim p m^2$.  This estimate is valid for $p > m/\delta$.  For
$p<m/\delta$, there are $k$-values which the momentum-$p$ particle
simply cannot pick up, since the magnetic field is nearly static and
does not change the particle energy.  Hence the largest $k$ which can be
picked up is $k \sim p$.  Again, such a $k$ would cause large-angle
change to the particle, so the coherence time is only $1/k$.  This again
gives the estimate
\be
\label{qhat_smallp}
\qhat \sim pm^2 \,, \qquad
\mbox{when } pm^2 < \frac{m^3}{\delta \theta} \,.
\ee

\subsection{Momentum broadening}

Now we are ready to determine the range of momenta where plasma
instabilities are more important than ordinary elastic scattering.
The rate of ordinary elastic scattering is
\be
\label{qhatelast}
\qhatelast \sim \alpha^2 \int d^3 k f[1{+}f]
  \sim \alpha^{2-c} \delta Q^3 [1{+}f] \sim \left\{ \begin{array}{ll}
     \alpha^{2-2c} \delta Q^3 & c > 0 \,, \\
     \alpha^{2-c}  \delta Q^3 & c < 0 \,. \\
  \end{array} \right.
\ee
Later we will often find it convenient to work only in terms of the
variables $\alpha,\delta,m,Q$.  Using that
$m^2 \sim \alpha \int f(p)\, d^3 p \sim \alpha f \delta Q^3$, we can
eliminate the factors of $f$ in favor of factors of $m^2$ and write
\be
\label{elast_msq}
\qhatelast(c<0) \sim \alpha m^2 Q \,.
\ee
Now we need to compare $\qhatelast$ with $\qhatinst$ to determine where
plasma instabilities dominate the dynamics.
Within the range we are considering, $d>0$ and $c<1-3d$, we find
$\qhatinst > \qhatelast$ whenever $c>0$, while for $c<0$ the instability
dominates provided that
\bea
\label{qhat_compare}
\qhatinst &>& \qhatelast \nonumber \\
\alpha^{\frac 32 - \frac{3c}{2} - \frac{d}{2}} Q^3 & > &
   \alpha^{2 - c + d} Q^3 \nonumber \\
c & > & -1-3d
\eea
or equivalently $d>\frac{-1-c}{3}$.  Outside this region, elastic
scattering is more efficient than plasma instabilities.  Note that the
region where elastic scattering wins out corresponds to systems with
extremely low occupancy, $c<-1$ or $f(p) < \alpha$.  We will not try to
determine the physics in this region.%
\footnote{%
    One might expect that the physics here will be analogous to
    isotropic, low occupancy systems studied in section~\ref{sec:bottomup}.
    This may be true but it is not obvious, since low-momentum radiated
    daughters may be anisotropic and may generate important plasma
    instabilities.  We leave this problem as an exercise to the
    reader.}
Also note that, for excitations at generic angles $\theta \sim 1$, the
plasma instabilities give a smaller $\qhat$, $\qhat\sim m^3/\delta$ from
\Eq{qhat_theta}.  For such particles, plasma instabilities dominate in
the narrower region $c > -1-d$.  But this fact will turn out not to play
a role in our discussion below.

Having identified the region where plasma instabilities dominate $\qhat$
and having found the relevant $\qhat$, it remains to determine what
physics first leads to a significant change to the dynamics.  We will
take as our criterion that some new physics causes $\qhat$ for our
typical excitations to change appreciably.  The simplest possibility is
that the $\qhat$ we just found causes the typical excitations' angular
distribution to get broader, increasing the value of $\delta$ and
reducing the value of $c$.  Since the mean squared change in
$p_z$ is
\be
\Delta p_z^2(t) \sim \qhat t \,,
\label{Dpz}
\ee
this occurs when $\Delta p_z^2(t) > p_z^2 \sim \delta^2 Q^2$, which
requires a time
\be
\label{tbroaden}
\tbroaden \sim \frac{\delta^2 Q^2}{\qhatinst}
  \sim \frac{\delta^4 Q^2}{m^3}
  \sim \delta^{\frac 52} \alpha^{-\frac{3}{2}(1-c)} Q^{-1} \,.
\ee
Self-consistency requires $t > m^{-1}$.  We see that this is satisfied
but is marginal for the boundary case $c=1-3d$, see \Eq{non-htl}.

\subsection{Splitting in highly anisotropic plasmas}

Just as for isotropic and weakly anisotropic systems, we also expect
joining (for $c>0$) and splitting (for $c<0$) may be important.  As for
weak anisotropy, these processes will generically be in the LPM regime,
so an analysis in terms of $\qhat$ and formation times is relevant.
And the split daughters can again dominate the dynamics, either through
elastic scattering or through new plasma instabilities.%
\footnote{%
    In \cite{ArnoldMoore1,ArnoldMoore2} the authors argued based on numerical
    simulations of plasma instabilities that $k\gsim m$
    excitations are also produced by re-scattering of the unstable
    fields (the ``nonabelian cascade'').  While true, we find that this
    production method is less efficient than bremsstrahlung emission, a
    process which the lattice techniques employed in \cite{ArnoldMoore1}
    do not incorporate.  Therefore we will neglect the nonabelian
    cascade in the remainder of this paper.}

Assuming $\qhatinst$ is the dominant momentum exchange process, and
estimating the rate of LPM modified splitting or joining in the same way
we did in previous sections, the formation time to split or join to the
momentum scale $p \lsim Q$ is
\be
\tform^2(p) \sim \frac{p}{\qhatinst} \sim
\left\{ \begin{array}{ll}
  \frac{p \delta^2}{m^3} & p>\delta^{-2} m \,,  \\ & \\
  \frac{1}{m^2}          & p<\delta^{-2} m \,.  \\
\end{array} \right.
\label{tform}
\ee
The different value when $p< \delta^{-2} m$ happens both
because the momentum broadening rate is smaller for such soft momenta,
and because these emissions are not actually LPM suppressed; $1/m$ is
the mean time between scattering events so the emission rate will be the
same as if we assumed each scattering [every time $1/m$] provides the
opportunity for an emission.  The rate for a hard excitation to emit an
excitation of momentum $p$ is
\be
\frac{d\Gamma}{dtdp/p} \sim \alpha \tform^{-1} [1{+}f(Q)]
\equiv \Gsplit(p) \,.
\label{tsplit}
\ee
Here $[1{+}f(Q)]$ is a Bose stimulation factor for the case that the
``hard'' $p\sim Q$ excitations are at high occupancy.  In this case, the
final-state hard particle is Bose enhanced.  (Bose enhancement for the
emitted, soft particle is canceled by absorption, as discussed
in subsection \ref{sec:bottomup}, see \Eq{prod_minus_absorb}.)

The time scale on which a typical hard excitation undergoes {\sl hard}
splitting or joining is
\be
\tsplit(Q) \sim \alpha^{-1} \alpha^{0,c} \tform(Q)
\sim \frac{\alpha^{-1+(0,1)c} Q^{\frac 12}}{\delta^{-1} m^{\frac 32}}
\sim \alpha^{-\frac{7}{4} +\left( \frac{3}{4},\frac{7}{4}\right)c}
     \delta^{\frac 14} Q^{-1} \,.
\label{tsplit2}
\ee
Here $(0,c)$ refers to the low-occupancy or high-occupancy case
respectively.  This time scale is shorter than $\tbroaden$ if
\bea
\label{split_condition}
\delta^{\frac 14}
 \alpha^{-\frac{7}{4} +\left( \frac{3}{4},\frac{7}{4}\right)c} &<&
\delta^{\frac 52} \alpha^{-\frac{3}{2}(1-c)}
\nonumber \\
\delta^{-\frac 94} & < & \alpha^{\frac{1}{4}
               + \left( \frac{3}{4}, \frac{-1}{4}\right) c} \,.
\eea
If $1>c>0$, in which case we use the $-c/4$ case, this expression is
never satisfied and broadening always happens before hard joining.
Splitting will also play no important role when $c>0$; just as in
subsection \ref{eps_over}, soft occupancies will become an
$f(p) \sim \alpha^{-c} Q/p$ tail, which however
never dominates screening, scattering, or
plasma instabilities.  Therefore if $c>0$ then the first new physics
is always momentum broadening at time $\tbroaden$.

However if $c<0$ then in the region $\delta > \alpha^{(-1-3c)/9}$,
or $c< -3d-1/3$, hard splitting will occur before the plasma
instabilities can broaden the hard particle distribution.  Therefore
splitting will play a role in a region {\sl at least} this large.  The
actual region will be larger, because soft $p\ll Q$ split-off daughters
can dominate the dynamics before hard splitting becomes important.

\subsection{Scales $\kiso$ and $\ksplit$}

The analysis of {\sl what} physics takes over from the ``ordinary''
instabilities will be a little intricate.  In particular, one
complication is that, as we have seen, the momentum broadening rate
$\qhatinst(\theta)$ is angle-dependent.  Since we will often be
concerned with whether or not the radiated daughters are isotropic, this
is a complicating effect.  We summarize how the daughters fill in small
momentum scales, including this complication but neglecting splitting of
daughters, in Appendix \ref{the_appendix}.

However, in the end we are always interested
in what is happening ``at the last minute'' when some other piece of
physics comes to dominate $\qhat(\theta\sim \delta)$.  In every case we will
find that this new piece of physics has a $\qhat$ which is not
{\sl parametrically} anisotropic.  That is, $\qhat$ will be anisotropic
by an $\OO(1)$ amount, but not by a power of $\delta$ or $\alpha$.
Therefore, at the time scale when a
new $\qhat$ is coming to dominate, we may treat $\qhat$ as approximately
{\sl isotropic}.  Similarly, in \Eq{tform} we can ignore the
$p<\delta^{-2} m$ case, which occurred because of the anisotropy of
$\qhat$, and always take $\tform$ to comply to the first expression.
It is most important to get this last time scale
correct.  If we also treat $\qhat$ as isotropic at all times, while we
will mis-represent physics at early times, we always get the final
stage, the time scale and physical mechanism which replace the original
plasma instabilities correct.

To understand how radiated daughters can come to dominate $\qhat$, we
need to study the time development of the population of such
daughters.  We will distinguish two important scales.  The first is the
scale $\kiso(t)$, introduced in \Eq{kiso}.  This is the time-dependent
momentum scale such that
particles at this scale have had just enough time for transverse momentum
diffusion to drive them to an isotropic distribution at the order-1
level.  It is determined by
\be
\label{kiso2}
\kiso^2 \sim \qhatinst t \qquad \rightarrow \qquad
\kiso \sim \delta^{-1} m^{\frac 32} t^{\frac 12}
      \sim \alpha^{\frac{3-3c-d}{4}} t^{\frac 12}Q^{\frac 32}\,.
\ee
Excitations at the scale $\kiso$ can grow to dominate $\qhat$ by
providing a new source for plasma instabilities.
Unlike in the case for weak anisotropy, the scale $\pmax$
lies below $\kiso(t)$ for all $t>m^{-1}$.  The scale $\pmax$ can only
play a role by dominating $\qhat$ through elastic scattering.  However,
if elastic scattering ever provides $\qhatelast \sim \qhatinst$, then
all excitations with $k \lsim \kiso$ thermalize and $\qhatelast$ is
controlled by the scale $\kiso$.  Later we will check whether elastic
scattering from the scale $\kiso$ can dominate, and we find that this is
not the case in the regime of interest.

The other important scale is $\ksplit(t)$, defined in \Eq{def:ksplit}:
$\Gsplit(\ksplit) t \sim 1$.  As discussed in subsection \ref{sec:bath},
a daughter with
$k\lsim \ksplit$ in turn has time to re-split and fragment completely.
Therefore, all particles with $k \lsim \ksplit$ redistribute into a
thermal bath or thermal tail.
Applying \Eq{tform} and
remembering $c<0$ so $[1{+}f(Q)] \simeq 1$, we find
\be
t \sim \Gsplit^{-1} \sim \alpha^{-1} \tform \sim \alpha^{-1}
             \frac{\ksplit^{\frac 12} \delta}{m^{\frac 32}}
\qquad \Rightarrow \qquad
\ksplit \sim \delta^{-2} \alpha^{2} m^3 t^2
        \sim \alpha^{\frac{7-d-3c}{2}} Q^3 t^2 \,.
\label{ksplit}
\ee

Both $\kiso$ and $\ksplit$ increase with time; but $\ksplit$ rises
faster, as $t^2$ rather than as $t^{\frac 12}$.  The scales cross at a
momentum $\kisosplit$ and time $\tisosplit$ such that
\bea
\label{tisosplit}
\ksplit(\tisosplit) & = & \kiso(\tisosplit) \equiv \kisosplit
\nonumber \\
\delta^{-2} \alpha^{2} m^3 \tisosplit^{2}
& \sim &
\delta^{-1} m^{\frac 32} \tisosplit^{\frac 12}
\nonumber \\
\tisosplit^{\frac 32}
& \sim &
\delta \alpha^{-2} m^{-\frac 32}
\nonumber \\
\tisosplit & \sim &
\delta^{\frac 23} \alpha^{-\frac 43} m^{-1}
\sim \alpha^{-\frac{11}{6} +\frac{1}{2}c + \frac{1}{6}d} Q^{-1} \,,
\\
\kisosplit & \sim & \delta^{-2} \alpha^2 m^3 \tisosplit^2
           \sim \delta^{-\frac 23} \alpha^{-\frac 23} m
           \sim \alpha^{\frac{-1-3c-d}{6}} Q \,.
\label{kisosplit}
\eea
At {\sl all} times, modes with $k>\kisosplit$ split before $\qhat$
drives them to isotropy, while modes with $k<\kisosplit$ become
isotropic before splitting.


\begin{figure}
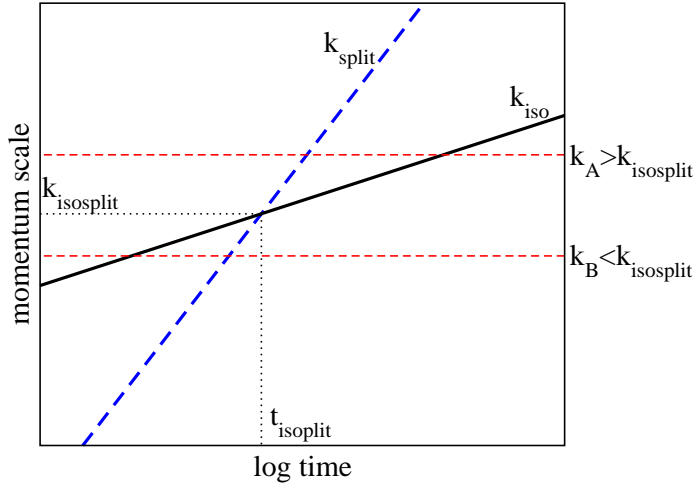

\centerbox{0.6}{isosplit.eps}
\caption{\label{summary2_fig}
  Scales $\kiso$ and $\ksplit$ intersect at time $\tisosplit$ at momentum scale $\kisosplit$.
  All modes above $\kisosplit$ (e.g. the scale $k_A$ in the plot) split before they isotropize, while
  modes with $k<\kisosplit$ (such as the scale $k_B$) become isotropic before splitting.}
\end{figure}


\subsection{$t<\tisosplit$:  new instabilities}

Before the time $\tisosplit$, $\kiso > \ksplit$, and we can effectively
ignore re-splittings of daughters.  The daughter particles are
generically anisotropic and can give rise to ``new'' plasma
instabilities.  We will now show that the most important scale for any
``new'' plasma instabilities is the scale $\kiso$.  Due to the
nonabelian LPM effect, the number of particles varies with scale as
$f(k) d^3k \sim k^{-1/2} dk/k$ which is IR dominated;
the contribution to screening scales as $f(k)d^3k/k \sim k^{-3/2} dk/k$
and is more IR dominated.  Define the contribution to screening from
momenta around $k$ to be $m^2(k)$; we see $m^2(k) \propto k^{-3/2}$.
The level of anisotropy also varies with
scale.  For $k>\kiso$ the particles are strongly anisotropic but are few
in number.  Since the mean squared transverse momentum they can
accumulate is $\kiso^2$, they will lie in an angular range
$\theta(k) \sim \kiso/k$.  These daughters will give rise to new
instabilities with a $\qhat$ of order
\be
\label{k_gt_kiso}
\qhat(k>\kiso) \sim \theta^{-2} m^3(k) \propto (k/\kiso)^2 k^{-9/4}
\propto k^{-1/4}
\ee
which is dominated by small $k$.  Meanwhile, for $k<\kiso$, the
excitations are nearly isotropic.  Only the fraction which were
introduced too recently to isotropize are anisotropic, and this fraction
is of order $(k/\kiso)^2$.  Therefore, for $k<\kiso$, we have weak
anisotropy with $\epsilon \sim (k/\kiso)^2$.  Then we find
\be
\label{k_lt_kiso}
\qhat(k<\kiso) \sim \epsilon^{\frac 32} m^3(k)
   \sim (k/\kiso)^3 k^{-\frac 94} \propto k^{\frac 34}
\ee
which is large $k$ dominated.  Hence the most important scale is
$\kiso$, where the anisotropy is order-1.

To determine the value of $\qhat(\kiso(t))\equiv \qhatnew$ due to plasma
instabilities at the scale $\kiso(t)$, we need to determine
$m^3(\kiso)$. We have that $\tform^2(\kiso) \sim \kiso \qhat^{-1}$.  But
$\qhat \sim \kiso^2 t^{-1}$ so
$\tform^2 \sim t/\kiso \sim \delta t^{\frac 12} m^{-\frac 32}$.
The number of daughters produced is
$n(\kiso) \sim \nhard \alpha t/\tform$.
And $m^2 \sim \alpha \nhard/Q$ so
$\alpha \nhard \sim m^2 Q$.  Therefore
\be
\label{Nkiso}
n(\kiso) \sim \frac{\alpha \nhard t}{\tform(\kiso)}
 \sim \frac{m^2 Q t}{t^{\frac 12} \kiso^{-\frac 12}}
 \sim m^2 \kiso^{\frac 12} Q t^{\frac 12}
\ee
and $m^2(\kiso) \sim \alpha n(\kiso)/\kiso$, so
\be
\label{msq_iso}
m^2(\kiso) \sim \alpha m^2 \kiso^{-\frac 12} Q t^{\frac 12}
 \sim \alpha \delta^{\frac 12} m^{\frac 54} Q t^{\frac 14}
\qquad \Rightarrow \qquad
\qhatnew \sim m^3(\kiso) \sim \alpha^{\frac 32}\delta^{\frac 34}
  m^{\frac{15}{8}} Q^{\frac 32} t^{\frac{3}{8}} \,.
\ee
The strength of plasma instabilities from daughters grows with time as
$t^{\frac 38}$.

These ``new'' plasma instabilities will catch up with
$\qhatinst \sim \delta^{-2} m^3$ at the time scale
\bea
\label{t_newinst}
\alpha^{\frac 32}\delta^{\frac 34}
  m^{\frac{15}{8}} Q^{\frac 32} \tnewinst^{\frac{3}{8}}
&\sim& \delta^{-2} m^3
\nonumber \\
\tnewinst^{\frac 38} & \sim &
\alpha^{-\frac{3}{2}} \delta^{-\frac{11}{4}} m^{\frac{9}{8}}
     Q^{-\frac 32}
\nonumber \\
\tnewinst & \sim & \alpha^{-4} \delta^{\frac{-22}{3}} m^3 Q^{-4}
\nonumber \\
  & \sim & \alpha^{-\frac{5}{2} - \frac{3}{2}c -\frac{35}{6}d} Q^{-1} \,.
\eea
Of course, we assumed $\tnewinst<\tisosplit$, so this answer is only
self-consistent if
\be
\label{t_vs_isosplit}
\tnewinst < \tisosplit \quad \Rightarrow \quad
 \alpha^{-\frac{5}{2} - \frac{3}{2}c -\frac{35}{6}d} Q^{-1} <
 \alpha^{-\frac{11}{6} + \frac{c}{2} + \frac{d}{6}} Q^{-1}
\quad \Rightarrow \quad
c < -3d - \frac{1}{3} \,.
\ee
Below this line in the $c,d$ plane, new instabilities from split
daughters dominate $\qhat$ before $p\sim Q$ particle angle-change
becomes important.
This happens to be {\sl the same} as the line we found earlier, when we
determined where hard splitting happens before $\tbroaden$.

We should also check that the time scale we found in \Eq{t_newinst} is
longer than the time scale $1/m$.  After all, the instabilities grow on
this time scale, and it is the shortest formation time of any daughter
induced by scattering in the instabilities.  Therefore, if we find
$\tnewinst < 1/m$, all of our calculations are inconsistent.  The
condition is
\be
\label{t_vs_m}
\tnewinst > m^{-1} \quad \Rightarrow \quad
\alpha^{-4} \delta^{\frac{-22}{3}} m^3 Q^{-4} > m^{-1}
\; \Rightarrow \;
\alpha^{-2 -2c -\frac{16}{3} d} > 1
\quad \Rightarrow \quad
d > \frac{-3(1+c)}{8} \,,
\ee
which is a line starting at $d=0,c=-1$ and rising with slope $-3/8$.
Between this line and the line $d=-(1+c)/3$, daughters become important
and control $\qhat$ via plasma instabilities in a time scale $\sim 1/m$,
possibly before the plasma instabilities from the original hard
particles have finished growing.  We have {\sl not} completely worked
out the physics in this region, and we suspect that the physics in the
region immediately below the $d=-(1+c)/3$ line may also be nontrivial.
However we will be lazy and leave the complete understanding of this
region for future work.

\subsection{Elastic scattering versus new instabilities}

Let us return to the region with $d < -(1+3c)/9$ but $d>-3(1+c)/8$.  We
just found that new plasma instabilities, associated with the scale
$\kiso$, dominate $\qhat$ before $\tisosplit$ in this region.  However
there is one more thing we should check.  The modes at and below the
scale $\kiso$ could in principle also dominate $\qhat$ via elastic
scattering.  Before continuing, we should pause to make sure that this
does not occur.

Suppose to the contrary that $\qhatelast \gsim \qhat(\kiso)$ which we
just computed.  Then $\kiso$ due to elastic scattering will be at least
as large as that from plasma instabilities.  But
as discussed in subsection \ref{sec:bottomup}, elastic scattering will
fully thermalize all $k<\kiso$, either into a thermal bath or an
$f(p)\propto T_*/p$ type tail.  We will consider separately the case
$f(\kiso) \gg 1$, in which case we get a $T_*/p$ type tail, and the case
$f(\kiso) \ll 1$, in which case we get a thermal bath.

Begin with the case $f(\kiso) \gg 1$, and compare the sizes of
$\qhatelast$ with $\qhatnew$:
\bea
\qhatelast \sim \alpha^2 \int^{\kiso} f[1{+}f]d^3 k & \mbox{ whereas } &
\qhatnew \sim m^3
\nonumber \\
\qhatelast
\sim \alpha^2 f^2 \kiso^3 &\mbox{ whereas }& \qhatnew \sim
 (\alpha f \kiso^2)^{\frac 32}
\sim \alpha^{\frac 32} f^{\frac 32} \kiso^3 \,.
\label{est:f>1}
\eea
We see that $\qhatelast < \qhatnew$ at time $\tnewinst$ provided that
$f(\kiso(\tnewinst)) < \alpha^{-1}$.
We can also quickly check that $\qhatelast(t)$ was not larger at some
earlier time; we have $\kiso \sim t^{\frac 12}$, and a quick calculation
shows $f(\kiso) \sim \kiso^{-\frac 72} t \sim t^{-\frac 34}$, and
therefore $\qhatelast \sim t^0$.

On the other hand, if $f(\kiso,\tnewinst) < 1$ and again assuming
that $\qhatelast \gsim \qhatnew$, then elastic scattering
re-organizes the momentum distribution of $k<\kiso$ particles into a
thermal distribution.  The temperature is set by the energy density,
$T^4 \sim f(\kiso) \kiso^4$.  Meanwhile, particles with $k$ just above
$\kiso$ have not ``collapsed'' and still provide $\qhat\sim\qhatnew$ via
plasma instabilities.  Making the comparison again, we find
\bea
\label{est:f<1}
\qhatelast \sim \alpha^2 T^3 & \mbox{ whereas } &
\qhatnew \sim m^3
\nonumber \\
\qhatelast \sim \alpha^2 (f \kiso^4)^{\frac 34} &\mbox{ whereas }&
\qhatnew \sim (\alpha f \kiso^2)^{\frac 32}
\nonumber \\
\qhatelast \sim \alpha^2 f^{\frac 34} \kiso^3 &\mbox{ whereas }&
\qhatnew \sim \alpha^{\frac 32} f^{\frac 32} \kiso^3 \,.
\eea
In this case, $\qhatnew$ is larger provided
$f(\kiso(\tnewinst)) > \alpha^{\frac 23}$.
In this case $T$ grows with time, and so $\qhatelast$ was smaller at all
earlier times.  Combining the two cases, $\qhatnew$ dominates provided
that $\alpha^{-1} \gg f \gg \alpha^{\frac 23}$.

Now let us check the actual value of $f(\kiso,\tnewinst)$.
Using $f \sim n(\kiso) / \kiso^3$ and combining \Eq{kiso2}, \Eq{Nkiso},
and \Eq{t_newinst}, we find
\be
\label{f_actual}
f(\kiso,\tnewinst) \sim
\frac{n(\kiso)}{\kiso^3} \sim m^2 \kiso^{-\frac 52} Q \tnewinst^{\frac 12}
\sim \delta^{\frac 52} m^{-\frac 74} Q \tnewinst^{-\frac 34}
\sim \alpha^3 \delta^8 m^{-4} Q^4
\sim \alpha^{1+2c+6d} \,.
\ee
This is to be applied in the region $-(1+c)/3 < d < -(1+3c)/9$.
In this region, $f$ varies from $\alpha^{-1}$, at $d=-(1+c)/3$, to
$\alpha^{\frac 13}$, at $d=-(1+3c)/9$.
Therefore we self-consistently find that new instabilities dominate
elastic scattering when they come to dominate overall $\qhat$,
throughout this region.

\subsection{$t>\tisosplit$: new thermal bath}

Now consider the region $(1-c)/3 > d > -(1+3c)/9$.  In this case, at the
time $\tisosplit$, the total $\qhat$ is still dominated by $\qhatinst$
arising from instabilities induced by the $p\sim Q$ excitations.
We want to determine whether split daughters can change this at times
$\tisosplit < t < \tbroaden$.  This range of momenta only exists if
$\tisosplit < \tbroaden$, which, using \Eq{tbroaden} and \Eq{tisosplit},
requires $d < -(1+3c)/7$.

Below the scale $\ksplit$, excitations undergo split-join processes on a
time scale shorter than the age of the system.  This allows the
excitations with $k<\ksplit$ to equilibrate into a nearly-thermal bath
with temperature $T$, in analogy with what we found in Sub-subsection
\ref{sec:bath}.  Let us find $T(t)$ as a function of
$t$.
The thermal bath is made by breaking down all daughters produced
with $k \lsim \ksplit$.
As in subsection \ref{sec:bottomup}, most of
the energy comes from the scale $\ksplit$; the number of daughters is
$\sim \nhard$, so
\be
\label{Tstar1}
\varepsilon \sim \ksplit \nhard \sim \alpha^{-1} \ksplit m^2 Q
\,, \qquad
T \sim \varepsilon^{\frac 14} \sim \alpha^{-\frac 14}
  \ksplit^{\frac 14} m^{\frac 12} Q^{\frac 14} \,.
\ee
Substituting in \Eq{ksplit},
\be
\label{Tstar2}
T \sim \alpha^{\frac 14} \delta^{-\frac 12} m^{\frac 54} Q^{\frac 14}
 t^{\frac 12} \,.
\ee
This grows more slowly with time than $\ksplit$.  We can also explicitly
check that at the earliest time we are considering, namely $\tisosplit$,
\be
\label{Tstar_compare}
T(\tisosplit) \sim \alpha^{-\frac{5}{12}} \delta^{-\frac{1}{6}}
                     m^{\frac 34} Q^{\frac 14}
                \sim \alpha^{\frac{-1-9c+5d}{24}} Q
\ee
whereas, from \Eq{kisosplit},
\be
\label{Tstar_2}
\kisosplit \sim \alpha^{-\frac 23} \delta^{-\frac 23} m
                    \sim \alpha^{\frac{-4-12c-4d}{24}} Q \,.
\ee
Since we are considering a region where $3c>-1-9d$, we find
$T(\tisosplit) \ll \kisosplit$.  Hence $T \ll \ksplit$ at all
subsequent times.  We also find that $T \ll \kiso$ at all times, since
$\kiso=\ksplit$ at time $\tisosplit$, and $T$ and $\kiso$ scale with
time in the same way, as $t^{\frac 12}$.

Most of the particles in the system are in the thermal bath.  To see
this, note that each particle at energy $\ksplit$ turns into
$\sim (\ksplit/T)\gg 1$ thermal bath particles.  Since
$\sim \nhard$ particles of energy $\ksplit$ get made, the bath
particles outnumber hard particles, $k\sim \ksplit$ particles, and (as
we will soon see) particles with $\ksplit \gg k \gg T$.  Hence we
expect that elastic scattering will be dominated by the thermal bath.
The value of $\qhat$ due to elastic scattering with thermal bath
particles is
\be
\qhatTelast \sim \alpha^2 T^3
\sim \alpha^{\frac{11}{4}} \delta^{-\frac{3}{2}} m^{\frac{15}{4}}
Q^{\frac 34} t^{\frac 32} \,.
\label{qhatTelast}
\ee

In addition, the thermal bath is generally not fully isotropic, and
neither are any of the excitations at scales $>T$.  Therefore there
can also be plasma instabilities due to the anisotropy of excitations at
any scale between $T$ and $\ksplit$.  First we need to estimate
occupancies in this range.
We know
$n(\ksplit) \sim \nhard$.  What about scales below $\ksplit$?
Below the scale $\ksplit$, the flux of energy density through a
logarithmic ``bin'' of momentum $k$ is $\sim \ksplit \nhard / t$.
The energy density equals this energy ``flux'' times the mean lifetime
of a particle to undergo democratic splitting,%
\footnote{%
    By democratic splitting we mean a splitting process where the
    daughters have comparable energy.  This is the main process which
    removes particles from some $k$-bin.}
which is $\tsplit(k) \propto k^{\frac 12}$ (see \Eq{tform},
\Eq{tsplit}).  Since the
energy density is $kn(k)$, we find $n(k) \propto k^{-\frac{1}{2}}$,
specifically $n(k) \sim \nhard (\ksplit/k)^{\frac 12}$.
The strength of screening from these particles is
$m^2(k) \sim \alpha^{-1} n(k)/k \propto k^{-\frac 32}$.

Recall that, for $k>\kisosplit$, particles split before their directions
are randomized.
Since $\ksplit > \kisosplit$, the largest $k$ values in the cascade will
be highly
anisotropic.  Specifically, in the time $\tsplit(k)$ during which a
particle resides at momentum $k$ before splitting further, it picks up a
mean-squared momentum $\qhat \tsplit(k)$.  This induces an angular range
of
\be
\theta^2(k) \sim \frac{\Delta k^2}{k^2} \sim \frac{\qhat \tsplit}{k^2}
 \propto k^{-\frac 32} \,.
\label{thetak}
\ee
There is also a contribution to $\theta^2$ from the angular distribution
of the parent, but since
we find $\theta^2$ increases with smaller $k$, the parent-distribution
is subdominant and we can use the above estimate.
The strength of plasma instabilities from a scale $k \gg \kisosplit$ is
\be
\qhatnew(k) \sim \theta^{-2}(k) m^3(k)
  \propto k^{\frac 32} k^{-\frac{9}{4}} \sim k^{-\frac 34}
\label{dontworry}
\ee
which is small $k$ dominated.

If $T < \kisosplit$, then there are also nearly-isotropic scales
between $\ksplit$ and $T$.  The time scale for a
particle of momentum $k<\kisosplit$ to be direction-randomized is
$\tiso(k) \sim k^2/\qhat$.  A fraction $\tiso(k)/\tsplit(k)$
of particles have not had time to randomize in direction, giving rise to
a residual (weak) anisotropy of size
\be
\epsilon(k) \sim \frac{k^2}{\qhat \tsplit(k)} \propto k^{\frac 32} \,.
\ee
The plasma instabilities arising from such excitations are of order
\be
\qhatnew(k<\kisosplit) \sim (\epsilon m^2)^{\frac 32} \propto k^0 \,,
\ee
essentially flat in $k$.  Therefore it is fair to say that, up to logs,
the size of new plasma instabilities is given by the contribution from
the softest scale in the game, which is $T$ itself.%
\footnote{%
    If $T < \kisosplit$, the argument is less clear.  In this case,
    the excitations may isotropize as they cascade from the scale
    $\kisosplit$ to the scale $T$, ``landing'' on the thermal bath
    already isotropic.  However, in this case the $\qhat$ of plasma
    instabilities from the $\OO(1)$ anisotropy at the scale $\kisosplit$
    turns out to be the same as the $\qhat$ we find for the thermal bath
    below, so our final answer remains correct.
}

At the scale $T$, the time it takes for an excitation to randomize its
direction is $\tiso(T) \sim T^2 / \qhat$.  Any excitation
which arrived longer than $\tiso(T)$ ago has had its direction
randomized; recent additions to the bath have not and are anisotropic.
Therefore the degree of anisotropy of the bath is
\be
\epsilon(t) \sim \frac{\tiso(T)}{t}
            \sim \frac{T^2}{\qhat t}
            \sim \frac{\alpha^{\frac 12} \delta^{-1} m^{\frac 52}
                       Q^{\frac 12} t}{\delta^{-2} m^3 t}
            \sim \alpha^{\frac 12} \delta^{1} m^{-\frac 12} Q^{\frac 12}
            \sim \alpha^{\frac{1+c+3d}{4}}\,,
\label{epsilon_again}
\ee
which turns out to be $t$ independent.

If we consider the thermal bath as a weakly-anisotropic system, in the
spirit of section~\ref{sec:weak}, then it has $c=0$.  In this case, we
saw there that instabilities dominate $\qhat$ from the thermal bath
if $\epsilon > \alpha^{\frac 13}$.  Otherwise elastic scatterings
dominate.  Therefore plasma
instabilities are more important provided
\be
\label{eps_tstar}
\epsilon > \alpha^{\frac 13} \quad \Rightarrow \quad
\alpha^{\frac{1+c+3d}{4}} > \alpha^{\frac 13} \quad \Rightarrow \quad
c < \frac{1-9d}{3} \,.
\ee
In this region, the value of $\qhat$ from $T$-bath plasma
instabilities is
\be
\label{qhatTinst}
\qhatTinst(t) \sim \epsilon^{\frac 32} m_{s}^3 \sim
            \epsilon^{\frac 32} \alpha^{\frac 32} T^3
   \sim \alpha^3 m^3 Q^{\frac 32} t^{\frac 32} \,.
\ee

Now we just need to check whether $\qhat$ from \Eq{qhatTelast} or
\Eq{qhatTinst} can exceed $\qhatinst \sim \delta^{-2} m^3$ before
the time $t=\tbroaden$.
New instabilities catch up with $\qhatinst$ at time $\tTinst$, defined as
\bea
\qhatTinst(\tTinst) & \sim & \qhatinst
\nonumber \\
 \alpha^3 m^3 Q^{\frac 32} \tTinst^{\frac 32} & \sim & \delta^{-2} m^3
\nonumber \\
\tTinst & \sim & \alpha^{-2} \delta^{-\frac 43} Q^{-1} \,.
\eea
Comparing with $\tbroaden$ from \Eq{tbroaden}, we find
\bea
\tTinst & \ll & \tbroaden    \qquad \quad \mbox{if}
\nonumber \\
\alpha^{-2} \delta^{-\frac 43} Q^{-1} & \ll & \delta^4 m^{-3} Q^2
\nonumber \\
\alpha^{-2-\frac{4d}{3}} Q^{-1} & \ll &
        \alpha^{\frac{-3+3c+5d}{2}} Q^{-1}
\nonumber \\
0 & > & 3+9c+23d
\eea
a line starting at $c=-1/3$ and with slope $-9/23$.
Therefore, new plasma instabilities from the scale $T_*$ come to
dominate the dynamics in the region with $9d>-1-3c$,
$23d<-3-9c$, and $9d<1-3c$.

For $9d>1-3c$, the thermal bath is too isotropic to have important
plasma instabilities but $\qhatTelast$ can grow to dominate.  This
happens at time
\bea
\qhatTelast(\tTelast) & \sim & \qhatinst
\nonumber \\
\alpha^{\frac{11}{4}} \delta^{-\frac{3}{2}} m^{\frac{15}{4}}
  Q^{\frac{3}{4}} \tTelast^{\frac{3}{2}} & \sim & \delta^{-2} m^3
\nonumber \\
\tTelast & \sim & \alpha^{\frac{-11}{6}} \delta^{\frac{-1}{3}}
                  m^{\frac{-1}{2}} Q^{\frac{-1}{2}} \,.
\eea
Again comparing with $\tbroaden$, we find
\bea
\tTelast & \ll & \tbroaden \qquad \qquad \mbox{if}
\nonumber \\
 \alpha^{\frac{-11}{6}} \delta^{\frac{-1}{3}}
                  m^{\frac{-1}{2}} Q^{\frac{-1}{2}}
& \ll & \delta^4 m^{-3} Q^2
\nonumber \\
\alpha^{\frac{-25+3c-7d}{12}} Q^{-1} & \ll &
       \alpha^{\frac{-18+18c+30d}{12}} Q^{-1}
\nonumber \\
0 & > & 7+15c+37d\,.
\eea
Therefore elastic scattering from a thermal bath dominates in a region
satisfying $37d<-7-15c$ but $9d>1-3c$.  This is a triangle in the $c,d$
plane with its apex at $c=-25/6,d=3/2$.

\begin{figure}
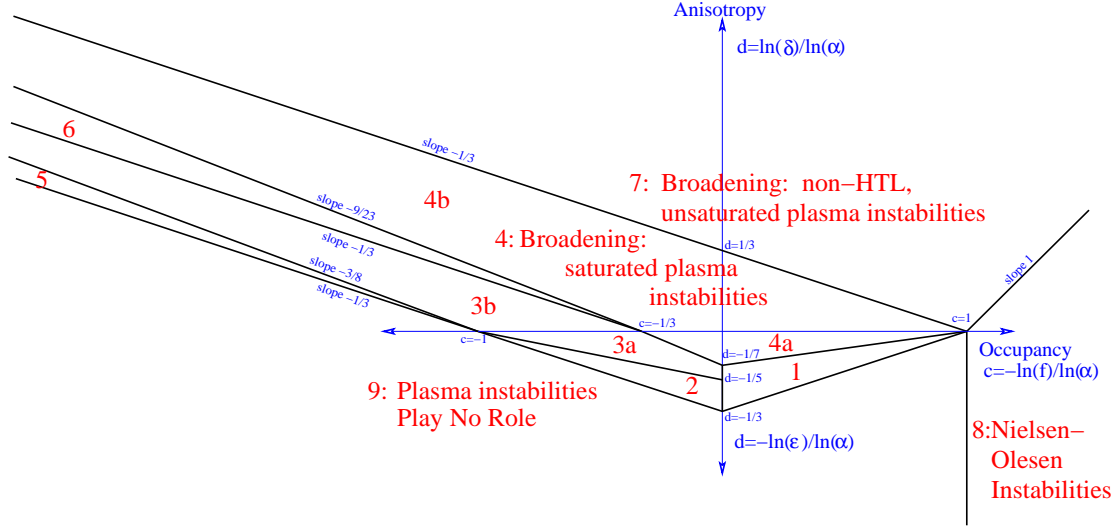

\centerbox{0.97}{oblate.eps}
\caption[Instability parameter space]
{Cartoon of $d$,$c$ plane indicating the
  regimes found for oblate distributions.
Plasma instabilities are important in regions 1 to 7.
In region 1, instabilities induce splitting which reduces anisotropy.
In 2, instabilities induce splitting, generating anisotropic,
occupancy-saturated particles which cause stronger plasma
instabilities.
Region 3 is similar but the new particles are not occupancy-saturated.
In regions 4 and 7, plasma instabilities directly lead to the broadening of
the primary-particle distribution.  In 4 they do this after
amplitude-saturation, in region 7 they do so before their amplitude
saturates and they are not well described by hard-loops.
In region 5, splitting creates new anisotropic
particles on a time scale of order the instability growth rate.
In region 6, splitting generates a thermal bath,
but it is incompletely isotropic and its plasma instabilities come to
dominate dynamics.
Region 8 is Nielsen-Olesen unstable.
In region 9, elastic scatterings play the dominant role.
Not shown is region 10, between regions 4,6 at very
negative $c$ and large $d$, where elastic scattering from a new thermal
bath comes to dominate dynamics.  The lower boundaries of regions 2,5
have not been completely studied and may require revision.
\label{fig:cartoon} }
\end{figure}

This completes our study of the region where plasma instabilities are
important and are well described by hard loops.  The regions we have
found and their properties are summarized in figure~\ref{fig:cartoon}.
The last region studied, where elastic scattering from a thermal bath
comes to dominate, lies off the edge of the figure.

\subsection{Highly anisotropic oblate distribution}
\label{sec:high}

Now we return to the case $c > 1-3d$.  Recall that, in this region, the
hard-loop approximation returned a range of unstable modes which is
broader than $\delta Q$.  This indicated an inconsistency in the
hard-loop approximation.  We also found for the marginal case
$c=1-3d$ that the time scale $\tbroaden$ for the momentum
distribution to broaden was the same as the $1/m$ time scale for the
instabilities to grow.
Let us try to understand the physics in this regime.

We claim that the dominant physics remains a Weibel-type instability,
with large magnetic fields deflecting the ``hard'' $p\sim Q$ excitations
and broadening the $p_\perp$ distribution of their momenta.
However, the range of plasma-unstable modes will be different than what
we found using the hard-loop approximation; and they will grow large
enough to broaden the $p\sim Q$ excitations' angular distribution
{\sl before} they grow large enough for nonlinear (nonabelian)
interactions to cut off the instability growth.  That is, the growth of
instabilities will be regulated by the angular broadening of the hard
excitation distribution, rather than nonabelian interactions between
unstable modes.

We will guess that the unstable modes still have
$k_z \gg k_\perp \sim m$ but $k_z \ll Q$.  But we now assume
$k_z > p_z \sim \theta Q$ in contrast to the previous section.  The
one-loop self-energy is still given by \Eq{eq:fullpi}, but now
$p\cdot k \sim Q k^0 + Q k_\perp + p_z k_z$ need not be large compared
to $k^2 \sim k_z^2$.
In order to find the range
of unstable $k_z$ available, take $k^0$ and $k_\perp$ to zero; then
$p\cdot k \sim p_z k_z$.  But we expect $k_z > p_z$, so
$k^2 \sim k_z^2$ is larger than $p \cdot k$.  So to find the range of
unstable $k_z$, drop the $p\cdot k$ terms in favor of $k^2$ terms.
In this approximation, \Eq{eq:fullpi} simplifies to
\be
\Pi^{\mu\nu}(k) \sim g^2 \int d^4 p \frac{p^2}{k_z^2} \delta(p^2) f(p)
  \sim \frac{Q^2}{k_z^2} g^2 \int d^4 p \delta(p^2) f(p)
  \sim  \frac{Q^2}{k_z^2} m^2 \,.
\ee
The propagator equation has instabilities when $k_z^2 \sim \Pi(k)$,
which we see requires
\be
k_z^2 \sim \frac{Q^2}{k_z^2} m^2 \qquad \Rightarrow \qquad
k_z \sim \sqrt{m Q} \,.
\ee
This is the range of unstable momenta.  In the parametric regime
considered, it is wider than $p_z\sim \delta Q$ (so the treatment is
self-consistent) but narrower than $\delta^{-1} m$.

What is the range of $k_\perp$ and the maximum growth rate $\gamma$?
They are determined as the values which significantly reduce the value
of $\Pi^{\mu\nu}$ estimated above.  This occurs when
$p\cdot k \sim k_z^2$.  Since $p^0 \sim Q \sim p_\perp$, this occurs
whenever $\gamma , k_\perp \sim m$.  So the transverse momentum range
and growth rate of the unstable modes remain $\sim m$.

To what amplitude do the plasma instabilities grow?  As before, they
grow until {\sl either} nonlinear physics interferes with the growth
mechanism, {\sl or} the unstable fields modify the dynamics in some
other way such that the calculation of the instability growth rate gets
amended.  In this region we will find that the latter occurs.  Our
previous estimate for the magnetic field strength where nonlinearities
such as color randomization shut off the instability can be extended in
a straightforward manner.  The fields would still saturate with
occupancy $f(k) \sim \alpha^{-1}$, with a magnetic field strength of
$B^2 \sim \alpha^{-1} k_z^2 m^2 \sim \alpha^{-1} m^3 Q$.  The maximum
value of $\qhat$ would then be $m^2 Q$.

The unstable fields grow exponentially, so the momentum randomization
will explore all values smaller than the above $\qhat$ before achieving
this value of $\qhat$.  However, as soon as
$\qhatinst \gsim \delta^2 m Q^2$, then the hard excitations undergo
significant broadening, $\Delta p_\perp > p_\perp \sim \delta Q$, in a
$1/m$ time scale.  In the region we are considering,
$m > \delta^2 Q$ and so $\delta^2 m Q^2 < m^2 Q$.  Therefore the
unstable modes reach a strength where they broaden the hard excitation
distribution {\sl before} they finish growing to become nonlinear.

This broadening of the hard excitation distribution changes which modes
are unstable, how large they can grow, and so forth.  In fact, it moves
where the system lies in the $c,d$ plane.  Therefore it constitutes an
important change to the properties of the system.

Hence we conclude that the physics of the region $c > 1-3d$ but
$c < 1+d$ is, that plasma instabilities grow in a time scale
$m \sim \alpha^{\frac{1-c+d}{2}}$ to become large enough that they
broaden the distribution of hard excitations.  The values of $c,d$ then
move along a line of fixed energy density, which is fixed $c-d$, towards
smaller $c,d$, reaching $c=1-3d$ in a time scale $t \sim 1/m$ up to
logs.  Thereafter they enter the $c<1-3d$ regime we have treated above.

\subsection{Final equilibration times}
\label{sec:teq}

In the previous subsections we have shown that plasma instabilities
generally dominate the dynamics at early times, and we have determined
what new phenomenon takes over from them and after what time scale.
The 10 regions we found can be broken into four broad categories:
\begin{enumerate}
\item
In region 9, elastic scattering dominates and the discussion of section
\ref{sec:iso} should be used.
\item
In region 8, N-O instabilities almost instantly change the system to be
one lying just off the boundary of region 8.  The equilibration then
proceeds as it would in that other region.
\item
In regions 1, 4, and 7 plasma instabilities cause the momentum
distribution to become more isotropic.  For $d>0$ this means $c,d$ each
decline with time; for $d<0$ only $d$ declines with time.  Therefore
after some amount of time the system enters region 3, 6, 9, or 10.
\item
In regions 2, 3, 5, 6, and 10, a bath of soft excitations develops and
comes to dominate the dynamics.
\end{enumerate}
Since we already solved the first case, and since the second and third
cases turn into either the first or fourth after some amount of time, we
only need to address what happens in the last case, when a bath of
soft excitations emerges to dominate the physics.  This bath may start
out nearly thermal (regions 6, 10) or far from thermal (regions 2, 3,
5).  But in every case we have that the energy density
$\varepsilon \sim \alpha^{-c} Q^4$ (if $d<0$) or
$\sim \alpha^{-c+d} Q^4$ (if $d>0$) is small compared to $Q^4$, so the
final temperature $\Tfinal \sim \varepsilon^{\frac{1}{4}} \ll Q$.

In every case we expect the late evolution to be controlled by the
physics of a bath of soft $p\lsim \Tfinal$ excitations, which steal
energy from the hard $p\sim Q$
modes and eventually become a thermal bath.  Since the thermalization
time of soft excitations is generally shorter than for hard excitations,
one expects that the soft bath becomes nearly thermal before it finishes
breaking down the $p\sim Q$ excitations.  It is safe to assume that, as
in the isotropic case, the most time consuming stage is the final
breakdown of the hard excitations when the bath temperature approaches
$T \sim \Tfinal$.

There are two possibilities.  Near the end, the thermal bath can either
be somewhat anisotropic, with $\epsilon > \alpha^{\frac 13}$; or it can
be more isotropic, $\epsilon < \alpha^{\frac 13}$.  In the former case
the dominant physics will be plasma instabilities; in the latter case it
will be elastic scattering.  In either case, what we need to compute is
the time scale at which the hard $p\sim Q$ excitations fragment
completely, that is, $\ksplit(\teq) \sim Q$.  Using our previous results
for $\ksplit$,
\be
\teq \sim \alpha^{-1} \qhat^{-\half} Q^{\half} \,.
\ee
It remains to self-consistently determine $\epsilon$ and $\qhat$.

First assume $\epsilon > \alpha^{\frac 13}$ so plasma instabilities
matter.  Then
$\qhat \sim \epsilon^{\frac 32} \alpha^{\frac 32} \Tfinal^3$.
We also have that
$\kisosplit \sim \alpha^{\frac{-1}{6}} \epsilon^{\half} \Tfinal$
which is $\gg \Tfinal$; so the fragmenting daughters are nearly
isotropic just above the scale $\Tfinal$.  However, because $\qhat$ is
caused by plasma instabilities, it is order-1 anisotropic, and so the
particles fragmenting into the thermal bath are order-1 anisotropic.%
\footnote{%
    This sounds like a contradiction; the particles are nearly isotropic
    as they fragment, but they land on the thermal bath in an
    anisotropic fashion.  The reason this can be true is that, at every
    stage, the fragmentation happens in an anisotropic fashion dominated
    by particles near the $p_x,p_y$ plane, but they then isotropize
    before the next step in the fragmentation.}
The level of anisotropy of the thermal bath $\epsilon$ is then set by
the ratio of the time scale for direction randomization,
$\tiso \sim \Tfinal^2/\qhat$, and $\teq$ -- since a fraction
$\sim \tiso/\teq$ of the thermal bath is made up of excitations which
landed less than $\tiso$ ago and have not isotropized.  That is,
\be
\epsilon \sim \frac{\tiso}{\teq}
  \sim \frac{\Tfinal^2 \qhat^{-1}}{\alpha^{-1} \qhat^{-\half} Q^{\half}}
  \sim \alpha^{\frac 14} \epsilon^{\frac{-3}{4}} \Tfinal^{\half}
  Q^{-\half}
\ee
and hence
\be
\epsilon \sim \alpha^{\frac 17} (\Tfinal/Q)^{\frac 27} \,, \qquad
\teq \sim \alpha^{\frac{-13}{7}} Q^{\frac 57} \Tfinal^{\frac{-12}{7}} \,.
\ee
This result only makes sense if $\epsilon > \alpha^{\frac 13}$,
which we see is $\Tfinal > \alpha^{\frac 23} Q$.  This is the region
$0 > c > -8/3$ for $d<0$ and $0 > c-d > -8/3$ for $d>0$.  It also only
makes sense if the resulting  $\teq \geq \alpha^{-2} \Tfinal^{-1}$.
Otherwise, the $p \sim Q$ excitations break down before the bath can
completely thermalize with itself, and this final thermalization is the
slowest process.

For $\Tfinal < \alpha^{\frac 23} Q$, thermalization takes so long that
the thermal bath is very nearly isotropic and elastic scattering
dominates.  In this case $\qhat \sim \alpha^2 \Tfinal^3$, and we find
\be
\teq \sim \alpha^{-2} Q^{\half} T^{\frac{-3}{2}} \,.
\ee
This case is the same as the isotropic case from subsection
\ref{sec:bottomup}.

There are two weaknesses in this analysis.  The first is that we assumed
that there is no long time-scale ``hang-up'' on the way to the final
thermalization we discuss.  This seems safe; we checked explicitly that
no such ``hang-up'' happens in isotropic systems, and anisotropic
systems should generally have faster dynamics due to plasma
instabilities.  Also, all the $\tchange$ time scales in Regions 1--7, 10
in table \ref{summary_table} are parametrically shorter than $\teq$.
The second weakness is that
we assumed that, for $\Tfinal > \alpha^{\frac 23} Q$, the thermal bath
will be anisotropic.  Once it is anisotropic, the anisotropic arrival of
daughters maintains this anisotropy, but it has to get that way to start
with.  However, in regions 2, 3, 5, and 6 the soft bath does start out
anisotropic; and region 10 (where it does not) is purely in the region
where $\Tfinal < \alpha^{\frac 23} Q$.  So this seems consistent.

It is possible that, even if a system starts out with $\qhatelast$
somewhat larger than $\qhatinst$, the thermal bath will still manage to
become anisotropic.  Therefore the boundaries between regions 2 and 9,
and between regions 5 and 9, are somewhat in doubt.  We will not try to
resolve this issue here but leave it for future work, if physical
systems which lie in this regime are found.

\subsection{Prolate distribution}

Now we will repeat the exercise of the last subsections, but for the
case of a highly {\sl prolate} anisotropic plasma, that is, one where
$f(\p) \sim \alpha^{-c}$ provided $p_z \lsim Q$ and
$p_x,p_y \lsim \delta Q$ with $\delta \sim \alpha^d \ll 1$.

The number density of typical excitations is $n \sim f \delta^2 Q^3$,
now with two powers of $\delta$ because momentum on two axes is more
constrained.  The energy density is
\be
\varepsilon \sim n Q \sim \alpha^{-c} \delta^2 Q^4
\label{eps_prolate}
\ee
and we will assume $\alpha^{-c} \delta^2 \ll \alpha^{-1}$,
or $c < 1+2d$ as otherwise there are Nielsen-Olesen instabilities.

The screening scale is
\be
\label{msq_prolate}
m^2 \sim \alpha \nhard Q^{-1} \sim \delta^2 \alpha^{1-c} Q^2
\sim \alpha^{1-c+2d} Q^2  \,.
\ee
Since the ``hard'' $p\sim Q$ modes all travel nearly in the $z$
direction, they remain coherently in a region of constant magnetic field
provided that $k_z \lsim m$ for the magnetic field; but {\sl both}
$k_x$ and $k_y$ may be larger by a factor of $\delta^{-1}$ and the
particles will remain in a region of coherent field.  Hence we find that
the range of unstable momenta and the growth rate are
\be
\label{kinst_prolate}
k_z \sim m \,, \qquad
k_\perp \sim m \delta^{-1} \,, \qquad
\gamma \sim m \,.
\ee
All unstable modes have polarization vectors in the $z$ direction.
The hard-loop approximation breaks down if $k_\perp$ for unstable modes
exceeds the typical transverse momentum of a ``hard'' excitation,
$k_\perp > \delta Q$, which is if $m > \delta^2 Q$ or
$c > 1-2d$.  The region $1+2d > c > 1-2d$ is analogous to the
case of subsection~\ref{sec:high}; in this region the growth of plasma
instabilities is cut off before amplitude saturation because the hard
momentum distribution gets broader.

The unstable mode amplitudes saturate when,
in traversing a distance $1/m$ along the $z$ axis, a
particle's color is rotated by an $\OO(1)$ angle, as previously
discussed.  This requires $A_z \sim m/g$,
$B \sim kA \sim m^2 \delta^{-1} \alpha^{-\half}$.  This is
equivalent to the condition that a Wilson loop of extent $k_z^{-1}$ by
$k_x^{-1}$ returns an order-1 phase.  The magnetic energy density
and $\qhat$ are
\be
\label{qhat_prolate}
B^2 \sim \alpha^{-1} m^4 \delta^{-2} \,, \qquad
\qhatinst \sim \alpha B^2/k_z \sim m^3 \delta^{-2} \,.
\ee
Unlike the oblate or weak-anisotropy cases,
saturation now occurs before the typical occupancy of
an unstable mode is $\sim \alpha^{-1}$;
$B^2 \sim f k_\perp^3 k_z$ so $f \sim \alpha^{-1} \delta$ for the
unstable modes.

The functional form of $\qhatinst$ is the same, when expressed in terms
of $m,\delta,\alpha,Q$, as for the oblate case, \Eq{qhat_extreme}.
The same will be true, for $c<0$, of $\qhatelast$, as we now show.
The momentum diffusion due to ordinary scatterings is
\be
\label{elast_prolate}
\qhatelast \sim \alpha^2 \int d^3 p f[1{+}f]
   \sim \left\{ \begin{array}{ll}
       \alpha^{2-2c} \delta^2 Q^3 & c > 0 \,, \\
       \alpha m^2 Q & c < 0 \,. \\
   \end{array} \right.
\ee
For $c>0$ instabilities always dominate $\qhat$ and we can ignore
$\qhatelast$.
For $c<0$, \Eq{elast_prolate} has the same form as \Eq{elast_msq}.
The region where plasma instabilities dominate is
\bea
\label{qhats_prolate}
\qhatinst & > & \qhatelast \nonumber \\
\alpha^{\frac{3}{2}(1-c)} \delta Q^3 & > & \alpha^{2-c} \delta^2 Q^3
    \nonumber \\
\delta^{-1} & > & \alpha^{\frac{1+c}{2}} \,,
\eea
or $2d > -1-c$.

We now go quickly through the prolate versions of the arguments we gave
in the previous sections for oblate systems.  We will find
that, when expressed in terms of $m^2,\delta,\alpha,$ and
$Q$, the results from the oblate case carry over to the prolate case.
Expressing in terms of $c,d$, the difference will just be the power of
$d$ appearing in $m^2$, \Eq{msq_prolate}.  Therefore the distinct
regions in the $c,d$ plane are the same as before but the boundaries are found
by replacing factors of $d-c$ in the oblate case with $2d-c$ in the
prolate case (since $m^2 \propto \alpha^{1+d-c}$ in \Eq{eq:msq} but
$m^2 \propto \alpha^{1+2d-c}$ in \Eq{msq_prolate}).

In particular, we find that
\Eq{tbroaden} for $\tbroaden$,
\Eq{tform} for $\tform$ and \Eq{tsplit} for $\tsplit$,
\Eq{kiso2} for $\kiso$,
\Eq{ksplit} for $\ksplit$, and
\Eq{tisosplit} and \Eq{kisosplit} for $\tisosplit$ and $\kisosplit$,
all still hold if one reads the versions expressed in terms of
$m^2,\alpha,\delta,Q$.  So do our arguments regarding the most important
scales for new daughters.  The temperature of the thermal bath of
daughters, $T_*$, is also still given by \Eq{Tstar2}, and the criteria
for its dominance are determined by the same expressions when we use
the $m^2,\alpha,\delta,Q$ variables.

\begin{figure}
\centerbox{0.9}{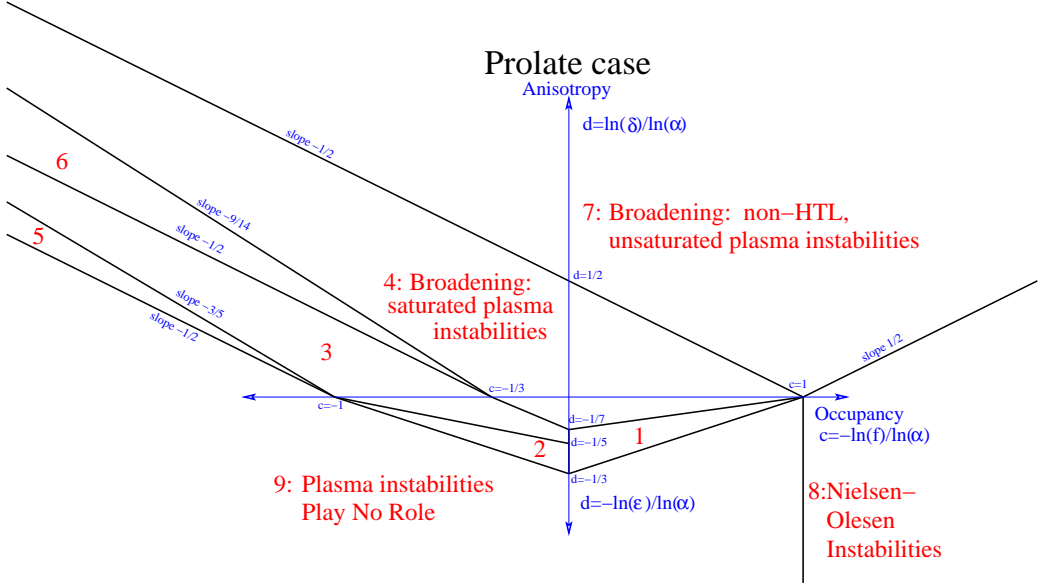}
\caption[Prolate case]
{ \label{fig:prolate}
The same as figure~\ref{fig:cartoon} but for the case of a prolate
anisotropic system.  The meaning of each region is the same as in the
previous figure; only the slopes of the lines separating regions have
changed.}
\end{figure}

Therefore we find the same regions as for an oblate distribution.  In
terms of $c,d$ the regions' boundaries are shifted, but in terms of $d$
and the energy density (which depends on $c-d$ for oblate and $c-2d$ for
prolate distributions) they are the same.  This is summarized in
figure~\ref{fig:prolate}.  Similarly we expect no parametric difference
in the final equilibration times.

\acknowledgments

We would like to thank Peter Arnold, for collaboration in some of the
preliminary work which developed into this paper, and Derek Teaney, for
pushing us into finally writing up these results.
This work was supported in part
by the Natural Sciences and Engineering Research Council of Canada and
the Institute of Particle Physics (Canada).

\appendix

\section{Daughters neglecting re-splitting}

\label{the_appendix}

In most of section~\ref{large_aniso}, when treating the split-off
daughters of ``hard'' $p\sim Q$ excitations, we made the simplifying
approximation that $\qhat(\theta)$ is uniform in angle and nearly
independent of an excitation's energy.  We justified this approximation
by arguing that we were really most interested in any new physics which
comes to supplant $\qhatinst$ as the dominant source of $\qhat$.  And in
every case we found that the new $\qhat$ has at most order-1
anisotropy.  Therefore, {\sl at the moment} when a new mechanism takes
over controlling $\qhat$, one is free to treat $\qhat$ as isotropic.
But what happens {\sl before} this moment?

We will present at least a partial answer to this question, by looking
at the evolution of particle occupancy and angular distribution, as a
function of momentum $p$ and time $t$ for $m < p < Q$.  We will treat
$\tbroaden > t > 1/m$, and we will also only try to understand
$p > \ksplit(t)$.  We will only consider the case of an oblate momentum
distribution.

Further, we will neglect elastic scattering.  We saw
previously that this is a good approximation for narrow angles and large
momenta provided that
$\qhatinst \sim \delta^{-2}m^3 \gg \qhatelast \sim \alpha m^2 Q$,
which requires $1+c+3d>0$.  For generic angles it requires
$\qhatinst(\theta\sim 1) \sim \delta^{-1} m^3 \gg \alpha m^2 Q$,
which requires $1+c+d>0$.  For the lowest-momentum excitations,
$p\sim m$, it requires $m^3 \gg \alpha m^2 Q$, which is true if
$1+c-d>0$.  These restrictions, and the restriction $p > \ksplit(t)$,
must be taken into account when applying the results of this appendix.

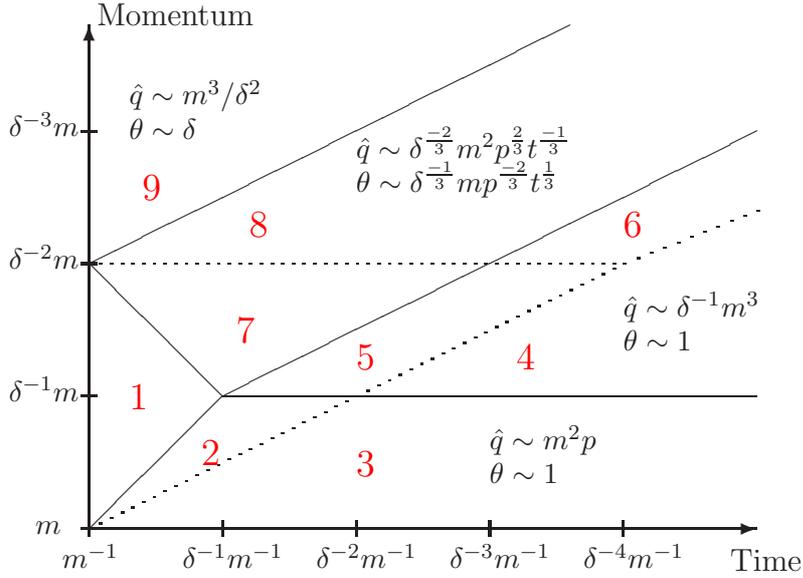
\begin{figure}
\centerline{\begin{picture}(300,225)
  \thicklines
  \put(20,20){\vector(1,0){250}}
  \put(20,20){\vector(0,1){190}}
  \multiput(20,17)(50,0){5}{\line(0,1){6}}
  \multiput(17,20)(0,50){4}{\line(1,0){6}}
  \put(260,4){\large Time}
  \put(23,210){\large Momentum}
  \put(10,5){$m^{-1}$}
  \put(55,5){$\delta^{-1} m^{-1}$}
  \put(105,5){$\delta^{-2} m^{-1}$}
  \put(155,5){$\delta^{-3} m^{-1}$}
  \put(205,5){$\delta^{-4} m^{-1}$}
  \put(0,18){$m$}
  \put(-10,68){$\delta^{-1} m$}
  \put(-10,118){$\delta^{-2} m$}
  \put(-10,168){$\delta^{-3} m$}
  \thinlines
  \put(20,20){\line(1,1){50}}
  \put(70,70){\line(-1,1){50}}
  \put(70,70){\line(1,0){200}}
  \put(70,70){\line(2,1){200}}
  \put(20,120){\line(2,1){180}}
  \multiput(20,20)(4,2){50}{\line(1,0){1}}
  \multiput(20,120)(4,0){50}{\line(1,0){1}}
  \multiput(220,120)(5,2){11}{\line(1,0){1}}
  \put(35,180){$\qhat\sim m^3/\delta^2$}
  \put(35,167){$\theta\sim \delta$}
  \put(120,160){$\qhat\sim \delta^{\frac{-2}{3}}m^2p^{\frac 23}
                t^{\frac{-1}{3}}$}
  \put(120,147){$\theta\sim \delta^{\frac{-1}{3}} m p^{\frac{-2}{3}}
                 t^{\frac 13}$}
  \put(170,50){$\qhat\sim m^2 p$}
  \put(170,37){$\theta\sim 1$}
  \put(220,100){$\qhat \sim \delta^{-1} m^3$}
  \put(220,87){$\theta \sim 1$}
  \put(35,65){\Large\red 1}
  \put(62,44){\Large\red 2}
  \put(120,40){\Large\red 3}
  \put(180,80){\Large\red 4}
  \put(120,80){\Large\red 5}
  \put(220,130){\Large\red 6}
  \put(75,90){\Large\red 7}
  \put(80,130){\Large\red 8}
  \put(40,144){\Large\red 9}
\end{picture}}
\caption[our favorite figure]
{\label{fig:app}
 The momentum-time plane for radiated daughters.  Regions separated by
 solid lines have daughters experience the indicated values of $\qhat$
 and fill the indicated angular ranges $\theta$; in region 1,
 $\qhat\sim  m^2 p$ and $\theta \sim m p^{\frac{-1}{2}} t^{\frac 12}$.
 Regions separated by solid and dotted lines are numbered to indicate
 how the particle occupancy varies, with $f(p,t)$ listed in
 \Eq{occ_early}.  The figure is only valid before further
 splittings occur, that is, above the line $\ksplit(t)$ (not shown,
 dependent on the ratio $\delta/\alpha$).  All lines have slope
 0, $\frac 12$, 1, or $-1$ except the dotted line between regions 4,6
 which has slope $\frac{2}{5}$.}
\end{figure}

Our results are presented in figure~\ref{fig:app}.  It is most
convenient to write $p = m \delta^{-a}$ and $t = \delta^{-b}/m$, that
is, to use $m$ as the characteristic scale and powers of $\delta^{-1}$
to distinguish how much larger than $m,1/m$ the momentum $p$ and time
$t$ are.  In these units, $Q \sim \delta^{-K} m$ with
$K = (1-c+d)/(2d)$.  Lines of equal $K$ are lines of constant slope
starting at $c=1,d=0$; the line $c=1-3d$ corresponds to
$Q\sim \delta^{-2} m$ and the $d=0$ axis is $K=\infty$.

First, we determine the range of angles which split-daughters will take,
as a function of $p$ and $t$.  The key facts we need are the following.
Radiated daughters are naturally born in the narrow angular range of the
hard parent population, and they then undergo transverse momentum
diffusion to gain mean-squared momentum $\Delta p^2 \sim \qhat t$.
The range of angles will be
\be
\theta^2 \sim \left\{ \begin{array}{ll}
  \frac{\qhat t}{p^2} & \mbox{if less than 1} \\
  \sim 1 & \mbox{if } \frac{\qhat t}{p^2} > 1 \\
\end{array} \right.
\ee
The value of $\qhat$ is dependent on both $p$ and $\theta$ (see
\Eq{qhat_theta} and \Eq{qhat_smallp}):
\be
\qhat(p,\theta) \sim {\rm min}\left( \frac{m^3}{\delta \theta} \,,\,
   m^2 p \right) \,.
\ee

First consider excitations with
$p < m/\delta$; then we always have $\qhat \sim m^2 p$.  At early times
the angular range is $\theta^2\sim \qhat t/p^2 \sim m^2 t/p$.
These particles
become isotropic on a time scale
\be
t \sim \frac{p^2}{\qhat} \sim \frac{p^2}{pm^2} \sim \frac{p}{m} m^{-1}
\,.
\ee
Next consider $m/\delta^2 > p > m/\delta$.  Initially $\qhat \sim m^2 p$
and again $\theta \sim m t^{\frac 12} p^{\frac{-1}{2}}$; but this
changes when $m^2 p \sim m^3/\delta \theta$, which is at
$t \sim \delta^{-2} p^{-1}$.
After this, $\qhat \sim m^3/\delta \theta$, and we find
\be
\theta^2 \sim \frac{\qhat t}{p^2} \sim \frac{m^3 t}{p^2 \delta \theta}
\qquad \Rightarrow \qquad
\theta \sim \delta^{\frac{-1}{3}} m p^{-\frac{2}{3}} t^{\frac 13}
\label{theta_middle}
\ee
until $\theta \sim 1$ at
the time scale $t \sim \delta^{-1} p^2 m^{-3}$.

Finally, if $p > \delta^{-2} m$, then initially $\theta \sim \delta$ the
angular range inherited from the radiating parent.  Enough momentum
broadening has accumulated to overcome this value when \Eq{theta_middle}
returns $\theta\sim \delta$, which is at $t \sim \delta^{-4} p^2 m^{-3}$.
After this the value of $\theta$ is given by \Eq{theta_middle} until
$t\sim \delta^{-1} p^2 m^{-3}$, when it is again order-1.
These time and momentum scales, $\qhat$ values and angular ranges are
summarized in figure~\ref{fig:app}.

Regarding the occupancy as a function of $p$ and $t$, we already saw
that the formation time for a radiated daughter is
$\tform^2 \sim p/\qhat$, which is $\tform \sim m^{-1}$
for $p<\delta^{-2} m$ and is
$\tform \sim \delta p^{\frac 12} m^{-\frac 32}$ for $p > \delta^{-2}
m$.  The rate of particle emission by a hard parent is
$\Gamma \sim \alpha[1{+}f(Q)]/\tform (dt dp/p)$ and the density of hard
parents is $\nhard \sim \alpha^{-1} m^2 Q$.  The powers of $\alpha$
cancel, so the number of soft excitations $n_p \equiv dn/d\ln p$ is
\be
n_p \sim \left\{ \begin{array}{ll}
    m^3 Qt [1{+}f(Q)] & p \ll \delta^{-2} m \,,\\
    \delta^{-1} p^{-\frac 12} m^{\frac 72} Q t [1{+}f(Q)] &
              p \gg \delta^{-2} m \,. \\
\end{array} \right.
\ee
Here $[1{+}f(Q)]$ is a stimulation factor which is important when the
hard particles are at large occupancy.

To convert these into an occupancy at the scale $p$, we use
$n_p \sim p^3 \theta f(p)$.  The factor of $\theta$ is because not all
of the angular range is filled.  When the resulting occupancy is larger
than $Q[1{+}f(Q)]/p$, then joining (re-absorption) will compete with
splitting and the occupancy will saturate.  This occurs for
$t > p^2 m^{-3}$ for $p<\delta^{-2} m$ and for
$t > \delta p^{\frac 52} m^{-\frac 72}$ for
$p > \delta^{-2} m$.
Therefore, in terms of the nine regions marked in figure~\ref{fig:app},
the typical occupancy will be:
\be
\label{occ_early}
\frac{f(p)}{1{+}f(Q)} \sim \left\{ \begin{array}{ll}
  m^2 p^{-\frac 52} Q t^{\frac 12}           &  \mbox{region 1}\,, \\
  m^3 p^{-3} Q t                             &  \mbox{region 2}\,, \\
  Q/p                                        &  \mbox{region 3}\,, \\
  Q/p                                        &  \mbox{region 4}\,, \\
  m^3 p^{-3} Q t                             &  \mbox{region 5}\,, \\
  \delta^{-1}m^{\frac 72} p^{-\frac 72} Q t  &  \mbox{region 6}\,, \\
  \delta^{\frac 13}m^{2} p^{-\frac 73} Q t^{\frac 23}
                                             &  \mbox{region 7}\,, \\
  \delta^{-\frac 23}m^{\frac 52} p^{-\frac{17}{6}} Q t^{\frac 23}
                                             &  \mbox{region 8}\,, \\
  \delta^{-2}m^{\frac 72} p^{-\frac 72} Q t  &  \mbox{region 9}\,. \\
 \end{array} \right.
\ee


\begin{thebibliography}{39}

\bibitem{inflaton_decay}
L.~F.~Abbott, E.~Farhi and M.~B.~Wise,
  Phys.\ Lett.\  B {\bf 117}, 29 (1982);
A.~Albrecht, P.~J.~Steinhardt, M.~S.~Turner and F.~Wilczek,
  Phys.\ Rev.\ Lett.\  {\bf 48}, 1437 (1982).

\bibitem{inflaton_resonant}
See for instance
L.~Kofman, A.~D.~Linde and A.~A.~Starobinsky,
  Phys.\ Rev.\ Lett.\  {\bf 73}, 3195 (1994)
  [arXiv:hep-th/9405187];
G.~N.~Felder, L.~Kofman and A.~D.~Linde,
  Phys.\ Rev.\  D {\bf 59}, 123523 (1999)
  [arXiv:hep-ph/9812289];
G.~N.~Felder, J.~Garcia-Bellido, P.~B.~Greene, L.~Kofman, A.~D.~Linde and I.~Tkachev,
  Phys.\ Rev.\ Lett.\  {\bf 87}, 011601 (2001)
  [arXiv:hep-ph/0012142];
J.~Garcia-Bellido, D.~G.~Figueroa and J.~Rubio,
  Phys.\ Rev.\  D {\bf 79}, 063531 (2009)
  [arXiv:0812.4624 [hep-ph]].

\bibitem{Grigoriev}
T.~Asaka, D.~Grigoriev, V.~Kuzmin and M.~Shaposhnikov,
  Phys.\ Rev.\ Lett.\  {\bf 92}, 101303 (2004)
  [hep-ph/0310100].


\bibitem{baryo_review}
J.~M.~Cline,
  [hep-ph/0609145].

\bibitem{AMY5}
P.~B.~Arnold, G.~D.~Moore and L.~G.~Yaffe,
  JHEP {\bf 0301}, 030 (2003)
  [hep-ph/0209353].

\bibitem{Weibel}
E. S. Weibel,
Phys.\ Rev.\ Lett.\ {\bf 2}, 83 (1959).

\bibitem{Mrow}
S.~\Mrowczynski,
Phys.\ Lett.\ B {\bf 214}, 587 (1988).
Phys.\ Lett.\ B {\bf 314}, 118 (1993).
S.~\Mrowczynski\ and M.~H.~Thoma,
Phys.\ Rev.\ D {\bf 62}, 036011 (2000)
[hep-ph/0001164].

\bibitem{RRS}
P.~Romatschke and M.~Strickland,
  Phys.\ Rev.\  D {\bf 68}, 036004 (2003)
  [hep-ph/0304092].

\bibitem{ALM}
P.~B.~Arnold, J.~Lenaghan and G.~D.~Moore,
  JHEP {\bf 0308}, 002 (2003)
  [hep-ph/0307325].

\bibitem{RRS2}
P.~Romatschke and M.~Strickland,
  Phys.\ Rev.\  D {\bf 70}, 116006 (2004)
  [hep-ph/0406188].

\bibitem{AMY7}
P.~B.~Arnold, G.~D.~Moore and L.~G.~Yaffe,
  Phys.\ Rev.\  D {\bf 72}, 054003 (2005)
  [hep-ph/0505212].

\bibitem{bottomup}
R.~Baier, A.~H.~Mueller, D.~Schiff and D.~T.~Son,
  Phys.\ Lett.\  B {\bf 502}, 51 (2001)
  [hep-ph/0009237].

\bibitem{Nielsen-Olesen}
N.~K.~Nielsen and P.~Olesen,
Nucl.\ Phys.\  B {\bf 144}, 376 (1978).

\bibitem{LPM}
L.~D.~Landau and I.~Pomeranchuk,
Dokl.\ Akad.\ Nauk Ser.\ Fiz.\  {\bf 92} (1953) 535;
Dokl.\ Akad.\ Nauk Ser.\ Fiz.\  {\bf 92} (1953) 735;
A.~B.~Migdal, Dokl.\ Akad.\ Nauk S.S.S.R.~{\bf 105}, 77 (1955);
Phys.\ Rev.\  {\bf 103}, 1811 (1956).

\bibitem{BDMPS}
R.~Baier, Y.~L.~Dokshitzer, A.~H.~Mueller, S.~Peigne and D.~Schiff,
  Nucl.\ Phys.\  B {\bf 484}, 265 (1997)
  [hep-ph/9608322];
  Nucl.\ Phys.\  B {\bf 483}, 291 (1997)
  [hep-ph/9607355].

\bibitem{Aurenche}
P.~Aurenche, F.~Gelis and H.~Zaraket,
  JHEP {\bf 0205}, 043 (2002)
  [arXiv:hep-ph/0204146].

\bibitem{AMY4}
P.~B.~Arnold, G.~D.~Moore and L.~G.~Yaffe,
  JHEP {\bf 0206}, 030 (2002)
  [arXiv:hep-ph/0204343].

\bibitem{Bethe}
H.~Bethe and W.~Heitler,
  Proc.\ Roy.\ Soc.\ Lond.\  A {\bf 146}, 83 (1934).

\bibitem{ArnoldDoganMoore}
P.~B.~Arnold, C.~Dogan and G.~D.~Moore,
  Phys.\ Rev.\  D {\bf 74}, 085021 (2006)
  [hep-ph/0608012].

\bibitem{anom_visc}
M.~Asakawa, S.~A.~Bass and B.~Muller,
  Phys.\ Rev.\ Lett.\  {\bf 96}, 252301 (2006)
  [hep-ph/0603092];
  Prog.\ Theor.\ Phys.\  {\bf 116}, 725 (2007)
  [hep-ph/0608270].

\bibitem{ALMY}
P.~B.~Arnold, J.~Lenaghan, G.~D.~Moore, L.~G.~Yaffe,
  Phys.\ Rev.\ Lett.\  {\bf 94}, 072302 (2005).
  [nucl-th/0409068].

\bibitem{RRS3}
A.~Rebhan, P.~Romatschke and M.~Strickland,
  JHEP {\bf 0509}, 041 (2005)
  [hep-ph/0505261].

\bibitem{others}
P.~Romatschke and R.~Venugopalan,
  Phys.\ Rev.\ Lett.\  {\bf 96}, 062302 (2006)
  [arXiv:hep-ph/0510121];
  Phys.\ Rev.\  D {\bf 74}, 045011 (2006)
  [arXiv:hep-ph/0605045];
A.~Dumitru, Y.~Nara and M.~Strickland,
  Phys.\ Rev.\  D {\bf 75}, 025016 (2007)
  [arXiv:hep-ph/0604149].

\bibitem{ArnoldMoore1}
P.~B.~Arnold and G.~D.~Moore,
  Phys.\ Rev.\  D {\bf 73}, 025006 (2006)
  [arXiv:hep-ph/0509206].

\bibitem{ArnoldMoore2007}
P.~B.~Arnold and G.~D.~Moore,
  Phys.\ Rev.\  D {\bf 76}, 045009 (2007)
  [0706.0490 [hep-ph]].

\bibitem{ASY}
P.~B.~Arnold, D.~Son and L.~G.~Yaffe,
  Phys.\ Rev.\  D {\bf 55}, 6264 (1997)
  [arXiv:hep-ph/9609481].

\bibitem{ArnoldMoore2}
P.~B.~Arnold and G.~D.~Moore,
  Phys.\ Rev.\  D {\bf 73}, 025013 (2006)
  [arXiv:hep-ph/0509226].

\end{thebibliography}
\end{document}